\documentclass[pre,twocolumn,superscriptaddress,floatfix]{revtex4-1}

\usepackage{etex} 

\usepackage{listings}
\usepackage[utf8]{inputenc}
\usepackage[text={7in,8.0in},centering]{geometry}
\usepackage{amsmath}
\usepackage{amssymb}
\usepackage{amsthm}
\usepackage[ampersand]{easylist}
\usepackage{bbm}
\usepackage{tabularx}
\usepackage{url}
\usepackage{fancyhdr}
\usepackage{algpseudocode}
\usepackage{algorithm}
\usepackage{empheq}
\usepackage{color}
\usepackage{graphicx}
\usepackage{hyperref}
\usepackage{tikz}
\usepackage[parfill]{parskip} 

\definecolor{lightyellow}{cmyk}{0,0,0.3,0}

\newcommand{\argmin}{\operatornamewithlimits{arg\,min}}

\newcommand{\vc}[1]{{\pmb{#1}}}
\newcommand{\ip}[2]{\langle{#1},{#2}\rangle}

\newcommand{\Av}{\vc{A}}

\newcommand{\ev}{\vc{e}}
\newcommand{\Ehat}{\widehat{E}}

\newcommand{\Iv}{\vc{I}}
\newcommand{\Hv}{\vc{H}}
\newcommand{\Jv}{\vc{J}}
\newcommand{\mv}{\vc{m}}

\newcommand{\sv}{\vc{s}}
\newcommand{\svtilde}{\tilde{\vc{s}}}

\newcommand{\tv}{\vc{t}}

\newcommand{\Vv}{\vc{V}}
\newcommand{\xv}{\vc{x}}
\newcommand{\xvtilde}{\tilde{\vc{x}}}
\newcommand{\Xv}{\vc{X}}
\newcommand{\yv}{\vc{y}}

\newcommand{\wv}{\vc{w}}
\newcommand{\Wv}{\vc{W}}
\newcommand{\vv}{\vc{v}}
\newcommand{\zv}{\vc{z}}

\newcommand{\etav}{\vc{\eta}}
\newcommand{\xiv}{\vc{\xi}}

\newcommand{\Sigmav}{\vc{\Sigma}}

\newcommand{\ftilde}{\tilde{f}}

\newcommand{\Jtilde}{\widetilde{J}}

\newcommand{\Jtildev}{\vc{\tilde{J}}}
\newcommand{\Vtildev}{\vc{\tilde{V}}}
\newcommand{\SigmaRootv}{\vc{\Sigma}^{\frac{1}{2}}}

\newcommand{\Vcal}{\mathcal{V}}

\newcommand{\Ncal}{\mathcal{N}}

\newcommand{\Scal}{\mathcal{S}}

\newcommand{\Reals}{\mathbb{R}}
\newcommand{\Integers}{\mathbb{Z}}
\newcommand{\Scube}{\mathbb{S}}

\newcommand{\OneVec}{\mathbf{1}}
\newcommand{\ZeroVec}{\mathbf{0}}
\newcommand{\Expect}[1]{\mathbb{E}_{#1}}

\newcommand{\dd}{\textrm{d}} 

\newcommand{\physExpect}[1]{\langle {#1} \rangle}

\DeclareMathOperator*{\median}{median} 

\usepackage{tikz-qtree}
\usepackage{tikz}
\usepackage{tikz-3dplot}
\usetikzlibrary{calc}
\usetikzlibrary{arrows.meta}
\usetikzlibrary{decorations.pathmorphing}

\begin{document}

\title{Wishart planted ensemble: \\
A tunably rugged pairwise Ising model with a first-order phase transition}

\author{Firas Hamze}
\affiliation{Microsoft Quantum, Microsoft, Redmond, Washington 98052, USA}
\affiliation{D-Wave Systems, Inc., Burnaby, British
  Columbia, Canada, V5G 4M9}

\author{Jack Raymond}
\affiliation{D-Wave Systems, Inc., Burnaby, British
  Columbia, Canada, V5G 4M9}

\author{Christopher A. Pattison}
\affiliation{Department of Physics, California Institute of
  Technology, Pasadena, California 91125, USA}
\affiliation{Department of Physics and Astronomy, Texas A\&M University,
College Station, Texas 77843-4242, USA}

\author{Katja Biswas}
\affiliation{Department of Physics and Astronomy, Texas A\&M University,
College Station, Texas 77843-4242, USA}

\author{Helmut G. Katzgraber}
\affiliation{Microsoft Quantum, Microsoft, Redmond, WA 98052, USA}
\affiliation{Department of Physics and Astronomy, Texas A\&M University,
College Station, Texas 77843-4242, USA}
\affiliation{Santa Fe Institute, Santa Fe,
New Mexico 87501 USA}

\date{\today}

\begin{abstract}
  We propose the Wishart planted ensemble, a class of
  zero-field Ising models with tunable algorithmic hardness and
  specifiable (or planted) ground state. The problem class
  arises from a simple procedure for generating a family of random
  integer programming problems with specific statistical symmetry
  properties but turns out to have intimate connections to a
  sign-inverted variant of the Hopfield model. The Hamiltonian
  contains only $2$-spin interactions, with the coupler matrix
  following a type of Wishart distribution. 
  The class exhibits a classical first-order phase
  transition in temperature. For some parameter settings the model
  has a locally stable paramagnetic state, a feature which correlates
  strongly with difficulty in finding the ground state and suggests an
  extremely rugged energy landscape. We analytically probe the
  ensemble thermodynamic properties by deriving the Thouless-Anderson-Palmer equations and
  free energy and corroborate the results with a replica and annealed
  approximation analysis; extensive Monte Carlo simulations confirm
  our predictions of the first-order transition temperature. The class
  exhibits a wide variation in algorithmic hardness as a generation
  parameter is varied, with a pronounced easy-hard-easy profile and
  peak in solution time towering many orders of magnitude over that of
  the easy regimes. By deriving the ensemble-averaged energy
  distribution and taking into account finite-precision
  representation, we propose an analytical expression for the location
  of the hardness peak and show that at fixed precision, the number of
  constraints in the integer program must increase with system size to
  yield truly hard problems. The Wishart planted ensemble is
  interesting for its peculiar physical properties and provides a
  useful and analytically transparent set of problems for benchmarking
  optimization algorithms.
\end{abstract}

\maketitle

\section{Introduction}
\label{sec:Introduction}

The interface between physics and computational complexity has yielded
fruitful insights over decades of research. Hard optimization
problems---which are ubiquitous throughout the natural sciences and
domains such as operations research---are of significant importance to
humans and are widely believed to admit no efficient algorithms for their
solution over all members of their class. It was recognized that such
problems show analogous features to those found in statistical
mechanical systems, for example the existence of algorithmic
phase transitions~\cite{hogg:96,monasson:99}, under which
typical problems show a dramatic increase in the difficulty faced by
known exact and heuristic algorithms.

In some cases, insights from the physics of spin glasses and
disordered systems have inspired remarkable new algorithms; for
example M\'ezard, Parisi, and Zecchina~\cite{mezard:02} studied
algorithmic hardness transitions in random 3-satisfiability (3-SAT)
problems using tools from statistical physics and subsequently
proposed survey propagation as a promising method for solving such
problems. In addition, physics-based approaches have suggested
ensembles of very hard problems; for example locked constraint
satisfaction problems~\cite{zdeborova:08} owe their difficulty to the
fragmentation of the solution space into widely separated sets. The NK
model~\cite{kauffman:87,kauffman:89} is a well-studied class of
tunably rugged cost functions proposed to capture the complexity of a
variety of physical and biological systems.

An important special class of hard problem ensembles are those whose
solutions are known to the constructor; these are often known as
optimization problems with planted solutions. Aside from their
theoretical interest, such problems are noteworthy for several
reasons. They may, for example, serve as candidates for cryptographic
one-way functions, that is functions whose outputs are cheaply
computable for any input but for which determining an input yielding a
given output is hard. Furthermore, they serve as useful benchmark
problems for evaluating heuristic or exact algorithms. In recent
years, the need for such benchmarks has increased with the advent of
physical devices implementing quantum annealing~\cite{johnson:11} and
related (e.g., Ref.~\cite{wang:13b}) algorithms. In such situations, it is
desirable to not only have access to a set of problems of tunable
hardness but to also be able to compare an algorithm's performance
with the correct answer.

A physics-based approach for generating hard 3-SAT problems with
planted solutions was proposed by Barthel \emph{et al.}~\cite{barthel:02}; more
recently, Krzakala and Zdeborov\'a presented a technique known as
quiet planting~\cite{krzakala:09} for devising graph
$q$-coloring problems with known solutions whose properties are
indistinguishable from those of a random ensemble. This concept can be
generalized to a variety of sparse problems, and has a close
connection to reconstruction on trees~\cite{riccitersenghi:19}.

While the aforementioned techniques and analyses have yielded numerous
elegant insights, they all share the undesirable property of
considering problems that are structurally far-removed from
contemporary physics-based optimization devices. $K$-satisfiability
problems, for example, require energy functions to include terms whose
value depends on groups of $K$ variables; in all the
previously mentioned work over binary variables, $K\geq 3$. Realistic spin system models and
optimization hardware on the other hand are typically restricted to
pairwise interactions; emulating higher-order interactions on
such systems can require tremendous overhead. In contrast,
$q$-coloring problems directly correspond to the antiferromagetic
Potts model of statistical physics and are thus expressible in terms
of $2$-body interactions, but $q$ must be larger than $2$ to yield
hard problems as $2$-coloring instances can be solved in linear time.
On devices natively encoding problems consisting of binary variables,
this can be problematic. Various techniques have recently been
published~\cite{hen:15a,wang:17,hamze:18,hen:19} for constructing planted
Ising instances on sparse graph topologies, which in some cases appear
to yield quite difficult problems~\cite{hamze:18}, but in common with
short-ranged disordered models in general, most of their known
properties are inferred from numerical simulation and much remains
unexplored about which features make them amenable as benchmarks for
given algorithms.

In this paper, we propose a simple randomized procedure for generating
systems of binary-constrained integer programs with a known
solution~\cite{pattison:19x}. The system coefficients are generated
according to a specific type of correlated multivariate Gaussian
distribution. When translated to an effective Ising Hamiltonian, we
obtain a novel type of disordered system with known ground state which
we call the Wishart planted ensemble; the name is inspired by the
distribution followed by the resultant random matrix of couplers.

The Hamiltonian includes interactions among all pairs of variables;
this aspect makes it a less-than-perfect fit for testing on devices
which implement short-range topologies. Unlike hard problem ensembles
based on $K$-SAT or graph coloring however, the variable domains are
binary valued and the interactions are pairwise. Vitally, the ensemble
displays an rich array of thermodynamic and computational properties
of considerable relevance to both classical and quantum algorithms.
Computationally, the problems can be tuned to range in difficulty from
very easy to extremely difficult at quite modest system sizes. We
emphasize that nearly all statements about problem ``hardness'' in
this paper refer to the empirically observed \emph{typical-case}
difficulty encountered by heuristics (and in all likelihood exact
algorithms) and have no bearing on theoretical computer science
questions concerned with \emph{worst-case} difficulty.

A striking physical property of the Wishart ensemble is the existence
of a first-order phase transition, a discontinuous jump in the free
energy derivative at some system-dependent critical temperature. A key
parameter in the generation procedure is $\alpha$, which specifies the
number-of-equations--to--number-of-variables ratio; $\alpha$ is
analogous to parameters such as the clause-to-variable ratio in the
satisfiability problems and exerts critical influence on the physical
and algorithmic complexity. By deriving the
Thouless-Anderson-Palmer~\cite{thouless:77} equations for the ensemble
with special care to account for the correlated couplers, we obtain
the mean-field free energy, from which we find that at any finite
$\alpha$, the internal energy drops discontinuously from some excited
value at some $\alpha$-dependent critical temperature $T_c$. We verify
this temperature and the nature of the transition with extensive
parallel tempering Monte Carlo \cite{geyer:91,hukushima:96}
simulations, which reveal that the system converges to its asymptotic
predicted properties quite rapidly. For large $\alpha>1$, the
magnitude of the discontinuity decreases monotonically and the
thermodynamics smoothly change character toward a traditional
second-order ferromagnetic transition. When $\alpha < 1$, the
paramagnetic state, i.e., the set of all configurations having no
correlation with the planted solution, is stable at any nonzero
temperature. This feature signals difficulty for classical heuristic
algorithms as it behaves as a deceptive dynamical ``trap.'' More
specifically, following free-energy gradients as done by methods like
simulated annealing~\cite{kirkpatrick:83} will overwhelmingly lead to
solutions far from the true optimum. Only by fortuitous initialization
within the ground-state basin of attraction will the problem be solved
with high probability. It turns out that $\alpha$ modulates the size
of the ground-state basin, with larger values increasing the
probability of solution by lucky initialization. When $\alpha\geq 1$,
on the other hand, the paramagnetic state becomes unstable for some
temperature $T_u<T_c$; at that point local algorithms can successfully
find the solution by ``rolling downhill'' and hence such problems are
typically easy. Remarkably, Barthel \emph{et al.}~\cite{barthel:02}
also argue that the hardness of their hard 3-SAT planted ensemble is
predicated on the existence of a first-order ferromagnetic transition.
We confirm the results of the Thouless-Anderson-Palmer (TAP) analysis
with two alternative approaches: the replica method~\cite{edwards:75}
and the annealed approximation.

First-order phase transitions are well known to exist in the $q$-state
Potts model for $q \geq 3$ and in some Ising-like systems such as the
Blume-Capel model~\cite{blume:66, capel:66} in which variables
nonetheless assume more than two states. They are, however, quite
unusual in the pairwise Ising model in zero field (though see Ref.~\cite{yedidia:90}); in
particular neither the Sherrington-Kirkpatrick~\cite{sherrington:75}
nor the Edwards-Anderson~\cite{edwards:75} spin glasses exhibit such a
transition. The Hopfield model~\cite{hopfield:82} displays a first-order transition 
between the spin glass and retrieval phases when a relatively small number of 
patterns ($\alpha < 0.05 $) are stored~\cite{amit:87}, though as discussed in 
Sec.~\ref{sec:Thermodynamics}, the Hopfield model is less appropriate as a 
class of problems with controllable hardness than the Wishart planted ensemble.
The puzzling presence of such a transition in our system is accounted for by
noting that the couplers are correlated to enforce the
existence of the planted solution rather than independently
disordered. Simulating systems with a first-order transition is widely
known to be challenging.

The Wishart planted ensemble is of particular interest because it
shares several features with models that have been shown by Nishimori
and Takada~\cite{nishimori:17} to be promising candidates for
exhibiting (limited) exponential speedup when simulated using
so-called nonstoquastic quantum driver Hamitonians; such
systems cannot be simulated classically and hence represent a
``strong'' type of quantum effect. The advantage of the Wishart
planted ensemble over the $p$-spin models considered in Ref.~\cite{nishimori:17} 
is once again that the interactions are naturally
pairwise rather than requiring terms of order $p \geq 3$.

Suitable nonstoquastic devices are not available as of the writing of
this paper, but the presented model can be used to explore
interesting problems on near-term stoquastic devices as well. In particular,
the combination of a rough multimodal landscape (replica symmetry
breaking) in the space orthogonal to the planted solution alongside
tunable control of the energy of the planted solution (at leading
order in $N$) opens up the possibility to explore the hard {\em
  population transfer} problem recently proposed by Smelyanskiy {\em
  et al}.~\cite{smelyanskiy:18}. This model is useful for
distinguishing physical quantum dynamics from classical dynamics such
as quantum Monte Carlo in transverse field Ising models due to the
multipath tunneling phenomena (miniband resonance). In principle one
can prepare a state in a planted mode, tune the mode energy to
equality (resonance) with other modes, and explore the rate of escape.

While this paper is mostly concerned with classical properties,
these intriguing connections to both types of quantum devices will
certainly be explored in future work.

Computationally, the Wishart planted ensemble emerges from our
procedure for generating a certain type of random integer linear
program~\cite{papadimitriou:98} and has numerous connections with
well-studied~\cite{fu:89,mertens:01,borgs:01,karmarkar:86,lagarias:85}
optimization problems such as the number partitioning and
subset sum~\cite{garey:79} problems. In common to these
problems, the allowable precision over the problem parameters has
important influence over combinatorial properties. There is also,
however a crucial distinction from these problems; when the parameter
$\alpha$ is fixed, the number of equations in the integer program
scales linearly with the number of variables rather than remain
constant (at unity, in the case of the previous two problems).

Random problem ensembles without a planted solution typically display
a parametrized ``easy-hard'' difficulty transition (e.g.,
Refs.~\cite{gent:96} and\cite{mertens:01}) in their optimization
variants. The Wishart ensemble, on the other hand, shows an
``easy-hard-easy'' character. One of the easy regimes is due to the
presence of a very large number of acceptable solutions coexisting
with the planted ground state; given the task of locating any one of
them, an optimization method has a relatively high likelihood of
success. The other is due to the planting procedure effectively
constraining the search space, providing ``hints'' to the algorithm
toward the solution. The hard regime, however, is seen to be
extremely difficult: Numerical experiments using a distributed,
state-of-the art parallel tempering implementation show a dramatic
hardness peak for small ($N=32$) system sizes. When $N=64$, parallel
tempering Monte Carlo fails to locate even an approximate solution
under lax and permissive target criteria within the allotted
simulation time of around 11 h on contemporary high-speed
hardware.

While we derive the Wishart planted ensemble in terms of a somewhat
abstract random integer program, the resultant model turns out to have
a remarkable structural similarity to the Hopfield
model~\cite{hopfield:82,amit:87} of biological neurons and more
particularly, a sign-inverted variant~\cite{nokura:98} proposed to
model neural ``unlearning''. As discussed in
Sec.~\ref{sec:Thermodynamics}, there turn out to be several important
differences between the models; nonetheless, it is exciting that such
completely different starting points as integer programming and
unlearning in neural networks result in models with close connections.

The rest of the paper is structured as follows. In Sec.~\ref{sec:WPE}
we describe our procedure to generate Wishart planted instances based
on random integer programming problems. We also discuss the ensemble's
computational properties and how to represent its members, whose
parameters are defined to take continuous values, with finite
precision. Section \ref{sec:Thermodynamics} analyzes the physical
properties of the class; the TAP free energy (derived in the Appendix)
is analyzed and shown to have global minima along a one-dimensional
subspace of the set of spin magnetizations. Furthermore, it has a
locally stable paramagnetic state for all temperatures when $\alpha<1$
and for $T > \alpha-1$ when $\alpha \geq 1$. These properties give
rise to the first-order transition between the paramagnetic and
planted states; we determine the transition temperature in terms of
$\alpha$. We show that as $\alpha$ grows, the system begins to
increasingly behave like a ferromagnet, i.e., with a second-order
transition. The predicted first-order transition temperature is
validated with extensive Monte Carlo simulation. In
Sec.~\ref{sec:hardnessPhase}, we turn our attention to empirical
algorithmic properties; under finite-precision representation, the
ensemble displays an easy-hard-easy relation with respect to parallel
tempering time to solution as $\alpha$ is varied. After showing that
the ensemble-averaged energies of the Wishart planted ensemble follow
a gamma distribution and introducing the notion of an intrinsic search
space, we analytically predict the location of the hardness peak for
any target energy threshold and confirm the prediction using optimized
parallel tempering simulations. The prediction is shown to be
precisely accurate even for approximate solution criteria. We show
that generating difficult problems under constant precision
restriction requires scaling the number of constraints in the integer
program approximately linearly. The Appendices \ref{sec:appendix}
contain most of our calculations, and confirm the TAP results through
replica analysis and an annealed approximation.

\section{The Wishart planted ensemble}
\label{sec:WPE}

\subsection{Generation procedure}
\label{sec:WPE:construction}

Our goal is to construct an ensemble of zero-field Ising Hamiltonians
over the $N$-spin complete graph with planted ground state
$\tv$, in other words, having the form
\[
H(\sv) = - \frac{1}{2}\sum_{i\neq j} J_{ij} s_i s_j ,
\]
where $\sv$ and $\tv$ refer to configurations on the $N$-spin Ising
model configuration space $\Scube^N\triangleq\{\pm 1\}^N$ and such
that \[
H(\pm \tv) = \min\limits_{\sv\in\Scube^N} H(\sv) .
\]
When not explicitly
stated $\tv$ will be taken to be the ferromagnetic ground state
$\tv = (+1,+1,\ldots,+1)$ and its $\Integers_2$ image, as the
minimizer to such a problem can be subsequently concealed by gauge
randomization.

Consider the $N\times M$ real-valued matrix
$\Wv \in \Reals^{N\times M}$, whose $M\geq 1$ columns are denoted by
$\wv^\mu$ for $\mu =1,\ldots,M$. The value of $M$ turns out to
modulate the ensemble properties such as thermodynamics and hardness;
in this work we are primarily concerned with the regime in which $M$
scales linearly as a constant factor of $N$, i.e., $M = \alpha N$ for
$\alpha > 0$.

Given a desired ground state $\tv$, our procedure seeks to construct a
consistent homogeneous Ising-constrained linear system with
$\sv = \pm\tv$ as a solution, in other words, to obtain $\Wv$ such
that \begin{equation}
  \Wv^T\tv = \ZeroVec .
  \label{eq:homogLinSys}
\end{equation}
This is because the positive semidefinite quadratic form
\begin{equation}
  G(\sv) = \frac{1}{2}\sv^T\Wv\Wv^T\sv = \frac{1}{2} \| \Wv^T\sv \|^2_2
\end{equation}
would then attain its minimum value of zero at
$\sv=\tv$, and hence if we define
\begin{equation}
  \Jtildev = -\frac{1}{N}\Wv\Wv^T = -\frac{1}{N}
  \sum_{\mu=1}^M\wv^{\mu}\otimes \wv^{\mu}
  \label{eq:JtildeDefn}
\end{equation}
and zero its diagonal to form
\begin{equation}
  \Jv = \Jtildev - \textrm{diag}( \Jtildev) ,
  \label{eq:JDefn}
\end{equation}
then the Hamiltonian 
\begin{equation}
  H(\sv) = -\frac{1}{2} \sv^T \Jv \sv
  \label{eq:isingHamiltonian}
\end{equation}
attains its ground state at $\sv=\tv$ with energy
\begin{align*}
  H(\tv) & =  -\frac{1}{2} \tv^T \Jv \tv \\
         & =  \frac{1}{2} \textrm{Tr}( \Jtildev
  ) ,
\end{align*}
where the property that $s_i^2 = 1$ for $\sv \in \Scube^N$ has
been used. The scaling by $1/N$ in the definition of $\Jtildev$ is to
make the energy extensive, i.e., scaling linearly with system size.

We obtain the linear system by individually generating the $M$ columns
$\{ \wv^\mu \}, i \in {1,\dots,M}$ of $\Wv$ such that
$\ip{\wv^\mu}{\tv} = 0$. We propose a simple projective method for
efficiently generating correlated Gaussian variates satisfying the
summation and other desirable properties. More precisely, the column
vectors are set to be distributed as
\begin{equation}
  \wv^\mu \sim \mathcal{N}( \ZeroVec, \Sigmav ) ,
  \label{eq:wDistrib}
\end{equation}
where the covariance matrix is given by
\begin{equation}
  \Sigmav = \frac{N}{N-1}\Big [ \Iv_N - \frac{1}{N}\tv \tv^T \Big]
\end{equation}
with $\Iv_N$ the $N$-dimensional identity matrix. In other words, for
each column vector $\wv$, all elements have unit variance, and for all
variable pairs $i \neq j$, the covariances are
\[
\Expect{}[ w_i w_j] = -\frac{t_it_j}{N-1} .
\]
Note that
$\textrm{rank}(\Sigmav) = N-1$ as expected; given any $N-1$ components
of $\wv$, the remaining one follows deterministically.
To generate the column vectors, we first determine the square root of
$\Sigmav$, i.e., $\SigmaRootv$ such that $\Sigmav = \SigmaRootv\SigmaRootv$,
to be
\begin{equation}
  \SigmaRootv = \sqrt{ 
  \frac{N}{N-1}} \Big[ \Iv_N - 
\frac{1}{N}
\tv\tv^T
\Big] .
\end{equation}
We then iterate over the loop described in Algorithm \ref{alg:WPEGen}:
\begin{algorithm}[H]
  \caption{Wishart Planted Ensemble Generator}
  \label{alg:WPEGen}
  \begin{algorithmic}[0] 
    \For{$\mu = 1, \ldots, M$}
      \State Sample uncorrelated Gaussian $\zv^\mu \sim \mathcal{N}(\ZeroVec, I_N)$
      \State $\wv^\mu \gets \SigmaRootv \zv^\mu$
    \EndFor
  \end{algorithmic}
\end{algorithm}
One can readily verify that $\ip{\wv^\mu}{\tv} = 0$ for all $\mu$
and that $\wv^\mu$ is distributed according to Eq.~(\ref{eq:wDistrib}).
While the components of $\wv$ are correlated, the Gaussian
nonetheless has strong structure that simplifies the subsequent
analysis. Following an appropriate gauge transformation to the
ferromagnetic state, the elements of $\Sigmav$ imply that the
distribution is \emph{exchangeable}, i.e., invariant to a permutation
of the components. Exchangeability is a stronger property than stationarity,
which merely requires the covariance of components $i$ and $j$
to depend only on $|i-j|$. The property is used in Sec.~\ref{sec:Thermodynamics:TAP} 
when deriving the TAP free energy and
again in Sec.~\ref{sec:hardnessPhase:energyHist} when obtaining
the ensemble energy
distribution.

The random matrix $\Wv\Wv^T$ follows a Wishart distribution, a
well-studied matrix generalization of the $\chi^2$ distribution; in
light of this we call our problem class the Wishart planted
ensemble~(WPE). When $M < N$, the support of the Wishart density lies
on a low-dimensional subspace of $N \times N$
matrices~\cite{uhlig:94}.  As we see in
Sec.~\ref{sec:Thermodynamics:Properties}, its spectral distribution is
a key feature underlying the phase behavior. We note also that while
sampling $\wv^\mu$ according to the Gaussian $\Ncal(\ZeroVec,\Sigmav)$
simplifies the analysis it is by no means necessary in practice for
our results to hold in the large $N$ limit.  For example, if
$\{\zv^\mu\}$ used to obtain $\{\wv^\mu\}$ were vectors of length $N$
consisting of independent and uniform $\{\pm 1\}$ variates rather than
uncorrelated Gaussians, then central limit arguments show that $\Wv\Wv^T$
is nonetheless asymptotically Wishart. We note however, that using
finite precision introduces the possibility that states other than
$\tv$ may attain the ground-state energy. The optimization problem in
this paper is defined to be that of locating any ground state; the
algorithmic implications of the presence of several solutions is
discussed in the next section and in detail in
Sec.~\ref{sec:hardnessPhase}.

\subsection{Computational properties}
\label{sec:WPE:computational}

Before proceeding to an examination of the WPE thermodyamics, we
discuss its properties from a computational perspective in light of
its interpretation as a constrained homogeneous linear system. Readers
familiar with related settings such as linear error correcting codes
should bear in mind that arithmetic here is over the real numbers
rather than a finite field such as ${\rm GF}(2)$.

Given a matrix $\Wv$, the task of finding the ground state of
Eq.~(\ref{eq:isingHamiltonian}) is equivalent to finding a solution to the
following NP-hard problem called \emph{integer programming
  feasibility}
\begin{align}
  \textrm{solve } \Wv^T \sv = \ZeroVec \nonumber \\
  \textrm{subject to } \sv \in \Scube^N .
  \label{eq:ILP}
\end{align}
Suspending for a moment the fact that $M$ scales with $N$ in the WPE,
we can obtain a sense for how it may impact problem difficulty. If
$\Wv$ consists of $M<N$ independent columns, then
$\textrm{dim}( \textrm{null} (\Wv^T) ) = N-M$. The dimension of the
nullspace of $\Wv^T$ implies the search space for potential Ising
state solutions to Eq.~(\ref{eq:ILP}), and so the larger it is, the
more difficult the problem may be guessed to be. In particular, when
$\Wv^T$ consists of a single row ($M=1$) and hence an $N-1$
dimensional nullspace, the problem may be surmised to be maximally
hard. When $\Wv$ is specified with relatively low precision, this
turns out to not be the case; exponentially many solutions other than
$\pm\tv$ overwhelmingly appear as $N$ increases and so locating any
such satisfying $\sv$ can be quite easy. This is reflected in the
Ising Hamiltonian (\ref{eq:isingHamiltonian}); when $M=1$, $\Jv$ can
be verified to be fully frustrated~\cite{comment:fullyFrustrated},
which gives rise to tremendous low-energy degeneracy.

At the other extreme in which $M$ assumes large values, the nullspace
of $\Wv^T$ becomes one-dimensional and hence the two Ising solutions to
Eq.~(\ref{eq:ILP}) are trivially recovered by inspection from a
vector spanning the nullspace. In the Hamiltonian
in Eq.~(\ref{eq:isingHamiltonian}), making $M$ large results in a
ferromagnetic system. To see this, we note that
\begin{align}
\Jtildev & = -\frac{1}{N}\Wv\Wv^T \\
        & = -\frac{\alpha}{M}\Wv\Wv^T\\
        & \to -\alpha \Sigmav ,
\end{align}
where the limit follows from the law of large numbers. From the
covariance matrix $\Sigmav$, the couplers thus uniformly
approach
\[
J_{ij} = \alpha \frac{t_it_j}{N-1} ,
\]
implying that the system reduces to a gauge-transformed and rescaled
Curie-Weiss ferromagnet, whose ground state is easy to find due to the
lack of frustration. Hence one may reasonably guess that for a given
allowable precision, the most difficult problems occur for some
intermediate value of $M$. This easy-hard-easy profile is shown to
indeed hold and is discussed in Sec.~\ref{sec:hardnessPhase}, in
which we make more explicit the role of $\textrm{null}(\Wv^T)$ and
develop the conjecture that the difficulty peak occurs at the
value of $M$ in which the fewest number of solutions occur relative to
the volume of the nullspace.

A prototypical special case of Eq.~(\ref{eq:ILP}), corresponding to $M=1$,
is the subset sum problem; its decision variant, that of
establishing whether a satisfying $\sv$ to Eq.~(\ref{eq:ILP}) exists, was
one of Karp's~\cite{karp:72} original 21 NP-complete problems. We note
that $\Wv$ must be specifiable to arbitrary precision for the
optimization variant in Eq.~(\ref{eq:ILP}) to avoid efficient solution via
the technique of dynamic programming as the complexity of this
algorithm scales exponentially in the number of bits needed to specify
the coefficients. This requirement holds more generally: when the
value of $M$ is fixed to any integer and the matrix elements belong to
a finite set, integer programming problems of the form of Eq.~(\ref{eq:ILP})
can be solved in pseudopolynomial time using the same
technique~\cite{papadimitriou:81}.

Subset sum problems with feasible solutions have been of particular
interest due to their complexity underpinning the security of an early
family of cryptographic systems~\cite{merkle:78}. Remarkably, a method
has been devised~\cite{lagarias:85} for solving with high probability a
family of low-density random subset sum problems, in which the
maximum element of the single-column $\Wv$ is large compared to $N$,
based on an idea known as lattice basis reduction~\cite{lenstra:82}.

A well-studied further specialization of the subset sum problem is
known as the number partitioning problem, corresponding to the
task of partitioning a base set of \emph{positive} integers into two
blocks with sums of minimum absolute difference (or ``discrepancy'').
When the integers are chosen independently from a uniform distribution
(the problem does not respect a planted solution), number partitioning
displays several interesting properties, in particular an algorithmic
easy-hard phase
transition~\cite{gent:96,borgs:01,mertens:01,mertens:06}. More
specifically, if the positive integers forming the base set are
bounded by $2^{\kappa N}$ for a fixed $\kappa > 0$, then when
$\kappa > \kappa_c$, partitions with a discrepancy of zero exist with
vanishing probability, and typical instances become difficult for
known heuristic and complete algorithms. Conversely, when $\kappa$
lies below this threshold, partitions with zero discrepancy are
abundant, and the problems are easily solved.

While the subset sum and number partitioning problems bear obvious
connections to the model presented in this work, it is apparent that
our assumptions and focus are different. Broadly speaking, in the WPE
a state must be found which now simultaneously satisfies the
$M$ relations in the integer program; the fact that $M$ scales as
$\alpha N$ rather than remaining fixed introduces consequences in the
algorithmic hardness properties.

\subsection{Representing a WPE instance}
\label{sec:WPE:representing}

The WPE construction presented in Sec.~\ref{sec:WPE:construction} was defined in
terms of Gaussian variables; in a practical implementation, however,
one must contend with finite precision and generally cannot represent
the required continuous-valued parameters. In the WPE, this can be
dealt with in one of two ways. The first, which will be examined in
this section, is to maintain the correlation structure defined by
$\Sigmav$ but replace the Gaussian variates $\zv$ with a discrete
zero-mean, unit variance ensemble; the simplest such choice is to have
$\zv$ uniformly and independently take the two values $\{ \pm 1\}$
(sometimes called a Rademacher distribution). This turns out to
allow an \emph{exact} representation of the problem parameters using a
logarithmic (in $N$) number of bits; further, as mentioned in Sec.~\ref{sec:WPE:construction}, 
the coupler matrix is nonetheless
asymptotically Wishart and the same physical properties derived in
this paper result.

The second and more heuristic way to represent a problem is to simply
round the parameters to the closest machine-representable number. This
introduces numerical errors; for example, the planted solution may no
longer have its theoretically intended energy. Consequently, one must
introduce a tolerance on what defines a ``solution.'' We use this
approach in our algorithmic hardness experiments because as discussed
in Sec.~\ref{sec:hardnessPhase:hardnessTransition}, it minimizes the
potential discrepancy between the finite-size statistical properties
and our analytical predictions, which may arise due to using
non-Gaussian generator variables on the small systems we were forced
to consider. Nonetheless, using the theoretically-derived energy
histogram, we are able to account for the observed hardness peak under
this approximate representation.

We now discuss the exact discretized WPE over a Rademacher distribution, i.e.,
where $\{z_i^\mu\}$ are independently and uniformly in $\{-1,1\}$
rather than drawn from $\Ncal(0,1)$ for $i \in \{1,\ldots,N\}$ and $\mu \in \{1,\ldots,M\}$.
We first rewrite
\begin{equation*}
  \SigmaRootv = \frac{1}{\sqrt{N(N-1)}} \Av ,
\end{equation*}
where the integer-valued matrix
\begin{equation*}
  \Av \triangleq N\Big[  I_N - \frac{1}{N} \tv\tv^T \Big] ,
\end{equation*}
i.e.,
\begin{equation*}
  A_{ij} =
  \left \{
    \begin{array}{ll}
      N-1 & \textrm{$i=j$}\\
      -1 & \textrm{$i\neq j$}
    \end{array}
  \right. .
\end{equation*}
Because
\[
\wv =\frac{1}{\sqrt{N(N-1)}} \Av \zv , 
\]
then up to the leading constant of $1/\sqrt{N(N-1)}$, the elements of
$\wv$ may assume values in the set of $2N-1$ equally spaced integers
\[
\Scal_\wv \triangleq \{ -2(N-1),
    \ldots, -2,0,2,\ldots, 2(N-1) \} .
\]
Thus, in the integer programming formulation, the
Rademacher-discretized WPE takes approximately $\log(2N-1)$ bits to encode
the parameters.

Obtaining the required precision in the Hamiltonian formulation (i.e.,
on $\Jv$) is more messy but analogous. Recalling that
\[
J_{ij} = -\frac{1}{N}\sum_{\mu=1}^M w_i^\mu w_j^\mu ,
\]
then using the previous restriction on the values of $\wv$ we can show
that
\[
N^2(N-1) J_{ij} \in \Scal_\Jv ,
\]
where
\[
\Scal_\Jv \triangleq \{ -4M(N-1)^2, \ldots, -4, 0, 4, \ldots 4M(N-1)^2 \} .
\]
Not all elements in this set of integers are actually attainable by
$J_{ij}$, but it provides a useful upper bound on the number of
possible values and shows that representing $J_{ij}$ in the
Rademacher-discretized WPE takes no more than on the order of
$\log(M) + 2\log(N)$ bits, which when $M = \alpha N$ is $O(\log N)$.

\section{Thermodynamic properties}
\label{sec:Thermodynamics}

Analyzing long-range disordered systems has a rich history in
statistical mechanics~\cite{nishimori:01}. The
Sherrington-Kirkpatrick~(SK) model~\cite{sherrington:75}, a
fully connected Ising model with independently sampled Gaussian bond
strengths, is a prototypical example of such systems. The replica
method~\cite{edwards:75} is a powerful framework for performing such
analyses, and has yielded great successes such as the solution to the
SK model~\cite{parisi:79} which have since been rigorously proved to
be correct (see, for example, Ref.~\cite{panchenko:13} and the
references therein). This approach is pursued in Appendix
\ref{sec:appendix:replicaMethod} were we are able to recover the
transition properties developed in this section, and through
connections with the anti-Hopfield model identify some additional
interesting transitions within the model.

A different and in many senses complementary approach to the analysis
of weakly coupled, fully connected disordered systems is due to Thouless,
Anderson, and Palmer~\cite{thouless:77}. The TAP equations are a
set of nonlinear relations satisfied by the local magnetizations for a
given instance. They can be arrived at in one of several
ways~\cite{opper:01}. For the SK model, they are often interpreted as
correcting the ``na\"ive'' mean field equations with a so-called
Onsager reaction term. The approach of Plefka~\cite{plefka:82} arrives
at the TAP equations by second-order expansion of the free energy at
constant magnetization, which turns out to have an appealing
information geometric interpretation~\cite{tanaka:00}. In this section
we use another approach called the cavity method~\cite{mezard:87}.

Determination of the TAP equations for the WPE is somewhat
complicated by the fact that $J_{ij}$ are not independent variates as
they are for the SK model. The TAP equations for systems with
correlated $\Jv$ have been determined in the past: A notable example,
which turns out to have a close connection to our ensemble, is the
Hopfield model~\cite{hopfield:82} of a biological neural network.
The couplings of the Hopfield model are given by
\[
J_{ij} = \frac{1}{N}\sum_{\mu=1}^p\xi_i^{\mu}\xi_j^{\mu} ,
\]
where the $p$ vectors $\{ \xiv^\mu \}$, known as patterns to be
stored for later retrieval, consist of $N$ \emph{independent}
zero-mean binary random variables. The physics of the Hopfield model
was first studied with the replica approach by Amit 
{\em et al}.~\cite{amit:85,amit:87}; the TAP equations were derived by
M\'ezard \emph{et al.}~\cite{mezard:87} using the cavity method, but
yielded results inconsistent with the replica analysis. TAP equations
consistent with Ref.~\cite{amit:85} were obtained by
Nakanishi and Takayama~\cite{nakanishi:97} using Plefka's
method, and subsequently by Shamir and
Sompolinski~\cite{shamir:00} via an elegant cavity
approach.

The connection of the WPE to the Hopfield model is apparent if we
express the elements of $\Jtildev$ defined in Eq.~(\ref{eq:JtildeDefn}) as
\[
\Jtilde_{ij} = -\frac{1}{N}\sum_{\mu=1}^M w_i^\mu w_j^\mu .
\]
An obvious difference from the Hopfield model is the presence of the
leading negation. Consequently, while the Hopfield Hamiltonian tends
to favor spin configurations aligned with the patterns
$\{ \xiv^\mu \}$, the WPE Hamiltonian penalizes configurations
overlapping with the directions $\{ \wv^\mu\}$. An additional
distinction, however, is that while the $M$ vectors are independently
drawn from the previously defined Gaussian distribution, the
components of each vector are now correlated, a property which emerged
due to the solution planting procedure. Remarkably, a model defined by
negating the sign of the Hopfield model Hamiltonian (only the first of
the two differences above) has been proposed as a model of
``unlearning'' paramagnetic configurations and thereby enhancing
learning for biological networks. A replica analysis of this
``anti-Hopfield'' model was undertaken by Nokura~\cite{nokura:98}. The
additional ``layer'' of correlations and the presence of a planted
solution in the WPE turns out to lead to very different behavior from
that of the anti-Hopfield model; in particular, the anti-Hopfield
model has no first-order transition comparable to the transition into
the planted solution, and the large-$\alpha$ regime is
Sherrington-Kirkpatrick-like rather than ferromagnetic.

Another Ising ensemble related to the WPE is the random orthogonal
model (ROM) proposed by Parisi and Potters~\cite{parisi:95a}, where
the $\Jv$ matrices are generated by uniformly sampling an orthogonal
matrix $\vc{O}$, forming diagonal matrix $\vc{D}$ whose elements
$D_{ii} \in \{\pm1\}$ (often additionally assumed to have a trace of
zero), and setting $\Jv = \vc{O} \vc{D} \vc{O}^T$. Having been devised
with different aims, this model is also quite different from the WPE.
First, no planting takes place, so the ground state is uncontrolled.
Further, the ground-state energy is only known if an Ising-feasible
state happens to be an eigenvector of $\Jv$, which become exceedingly
unlikely for even moderately sized systems. Finally, the eigenvalue
distributions of the $\Jv$ matrices are not the same, nor are the TAP
equations presented in Ref.~\cite{parisi:95a}. The systems thus have
quite differing thermodynamic properties; in particular, a first-order
transition analogous to the planting transition is absent in the ROM.

It is natural to wonder about the suitability of the original Hopfield
model to the task at hand. Unfortunately, it is not possible to
generate sufficiently rugged energy landscapes while maintaining
control over the ground states using this model. Hence, while the
Hopfield model is an appealing abstraction of an associative memory,
it is not as well suited to usage as a tunable planted optimization
objective. The Hopfield model does not exhibit, for any setting, a
persistent metastable paramagnetic state analogous to that of the
$\alpha < 1$ WPE. In contrast to the WPE in which the planted ground
state is undisturbed while a rough energy landscape is induced in the
orthogonal subspace, ruggedness in the Hopfield model arises due to
interference among the stored patterns, causing undesired spurious
minima~\cite{amit:87}. For values of the Hopfield $\alpha$, which
specifies the ratio of number of stored patterns to number of
variables, of less than around 0.05, a first-order transition between
a spin glass and "retrieval" phase takes place, but finding one of the
embedded patterns is quite easy in this case (as it should be for the
model's intended purpose). At somewhat larger values of $\alpha$, the
model continues to function as an associative memory, though the
patterns are only assured of being local optima rather than ground
states. When $\alpha > 0.138$, only a second-order paramagnetic to
spin glass transition takes place, and control over even the local
minima is completely lost.

\subsection{TAP equations}
\label{sec:Thermodynamics:TAP}

We derive the TAP equations for the WPE following the two-step cavity
approach of Shamir and Sompolinski~\cite{shamir:00}, making appropriate
adaptations to deal with the correlations among the components of each
$\wv$. The complete calculation is shown in Sec.~\ref{sec:appendix:cavityTAP}.  
Alternatives to the TAP method based on
the annealed approximation and replica method are considered in
Appendices \ref{sec:appendix:annealedApproximation} and
\ref{sec:appendix:replicaMethod}, yielding consistent results. In
these appendices the connection with the anti-Hopfield model is
explored, as is the notion of a planted solution with tuned energy
that might permit other uses, such as the population transfer
experiment discussed in the introduction.

Let $m_i = \physExpect{s_i}$ and $\mv$ be the vector of all
magnetizations. Define
\[
q = \frac{1}{N} \sum_{i=1}^N m_i^2
\]
and 
\[
V = \frac{
  \alpha (1-q)
}{
  1+\beta (1-q)
} .
\]
The TAP equations for the WPE are
\begin{equation}
  m_i = \tanh\Big[ \beta \Big(
  \sum_{i \neq j} J_{ij} m_j - \beta V m_i
  \Big) \Big] .
  \label{eq:TAP}
\end{equation}
Solutions to Eq.~(\ref{eq:TAP}) are stationary points of the following 
TAP free energy:
\begin{align}
    F_{\rm TAP}(\mv) & =
    -\frac{1}{2}\sum_{i \neq j} J_{ij} m_i m_j
    - \frac{1}{\beta} \sum_{i=1}^N S(m_i) \nonumber \\
    & - \frac{1}{2}\alpha N\Big[  (1-q) - \frac{1}{\beta}\log \Big( 1+
    \beta(1-q) \Big)  \Big] , \label{eq:FTAP}
\end{align}
where the local entropy terms are
\[
S(m_i) = - \frac{1+m_i}{2}\log
\frac{1+m_i}{2} - \frac{1-m_i}{2}\log \frac{1-m_i}{2} .
\]

To determine the quenched free energy one must consider a weighted sum
over minima~\cite{dedominicis:83}. In this section we present a more
limited analysis that is appropriate when only a single minimum
dominates the free energy. The result, which may be an upper bound to
the free energy, is consistent with the replica symmetric analysis of
Appendix \ref{sec:appendix:replicaMethod}, and in good agreement with
numerical results of Sec.~\ref{sec:hardnessPhase}. We expect the
approximation to be sufficient at higher temperature, and that in
other cases the consequences of multiple fixed points may be small
following the replica symmetry breaking analyses in related models as
discussed in Appendix
\ref{sec:appendix:replicaMethod}~\cite{parisi:95a,nokura:98}.

In general, determining the global minima of $F_{\rm TAP}$ is a difficult
task as $\mv$ lies in an $N$-dimensional space subject to the bound
constraints
\[
\mv \in [-1,1]^N .
\]
In our case, however, the existence of the planted solution
considerably simplifies things by assuring that the free-energy
minimum necessarily occurs along the ground-state direction:
\[
\min_{\mv \in [-1,1]^N}  F_{\rm TAP}(\mv) = \min_{m \in [-1,1]} F_{\rm TAP}(m \tv ) .
\]
To see this, consider the restriction of $\mv$ to the spherical shell
given by
\[
\Big \{ \mv \Big| \frac{1}{N} \sum_{i=1}^N m_i^2 = q_0 \Big \}
\]
for some
$q_0 \in [0,1]$ or, alternatively,
\[
\mv = \ev \sqrt{N q_0}
\]
for
arbitrary unit vector $\ev$. The claim is that
\[
\ev = \pm \frac{\tv}{\sqrt{N}}
\]
minimizes $F_{\rm TAP}$ on the shell;
from Eq.~(\ref{eq:FTAP}), the term
\[
-\frac{1}{2} \sum_{i \neq j} J_{ij} m_i m_j
\]
is minimized for
$\mv = \pm \sqrt{q_0} \tv$ due to the planting procedure. Further,
the term
\[
-\sum_{i=1}^N S(m_i)
\] is minimized on the shell at the
$2^N$ points at which the $\{m_i\}$ have the same magnitude, which 
holds when $\mv = \pm \sqrt{q_0}\tv$, and the final term in
$F_{\rm TAP}$ is independent of $\mv$ on the sphere. The global optimum,
which is the minimum over all the shells' minimizers, thus occurs
along \[
\mv = m \tv = \pm \sqrt{q} \tv .
\]
Recall that
\[
H(\tv) = -\frac{1}{2}\sum_{ij}J_{ij} t_{i} t_{j} = \frac{1}{2}
\textrm{Tr}(\Jtilde) 
\] and that
\[
\Jtilde_{ii} = -\frac{1}{N} \sum_{\mu=1}^M (w^\mu_{i})^2 .
\]
Using the fact that $S(m t_i) = S(m)$ for
$t_i \in \{\pm 1 \}$ we define the one-dimensional TAP free energy as
$F(m) \triangleq F_{\rm TAP}(m\tv)$, i.e.,
\begin{align}
  F(m) \triangleq & -\frac{1}{2N}m^2 \sum_{\mu=1}^M \sum_{i=1}^N  (w^\mu_{ii})^2
                    -\frac{N}{\beta}S(m) \label{eq:FTAP1D} \\
                  & -\frac{1}{2}\alpha N( 1-m^2) + \frac{\alpha N}{2\beta}\log(
                    1+\beta(1-m^2)). \nonumber
\end{align}
Consider the difference
\[
\frac{m^2}{2}\Big[   \alpha N -\frac{1}{N} \sum_{\mu=1}^M
\sum_{i=1}^N  (w^\mu_{ii})^2 \Big]
\]
present in Eq.~(\ref{eq:FTAP1D}). Using the higher-order moments of
correlated Gaussians, one can show that despite the
dependence among the components of $\wv^\mu$, the term scaling
$\frac{m^2}{2}$ is at most $O(\sqrt{M})$. Normalizing by $N$ to obtain
the free energy per spin, we obtain
\begin{align}
f(m) = & -\frac{m^2}{2N} O(\sqrt{\alpha N}) - \frac{1}{2}\alpha -
         \frac{1}{\beta}S(m) \label{eq:f} \\
       & + \frac{\alpha}{2\beta} \log(1+\beta(1-m^2)) . \nonumber
\end{align}
In the thermodynamic limit, the first term vanishes; neglecting the
constant term of $-\frac{1}{2}\alpha$, the global free-energy minima
of Eq.~(\ref{eq:FTAP}) can thus be found as those of the function
\begin{equation}
  \ftilde(m) \triangleq -\frac{1}{\beta}S(m) + \frac{\alpha}{2\beta}
  \log(1+\beta(1-m^2)) .
  \label{eq:ftilde}
\end{equation}
The mimina of $\ftilde$ must be stationary points of
Eq.~(\ref{eq:ftilde}), which are attained for solutions of the equation
\begin{equation}
  m = \tanh\frac{\alpha \beta m}{1+ \beta(1-m^2)} .
  \label{eq:ftildeStatPts}
\end{equation}

\subsection{Thermodynamic properties of the WPE}
\label{sec:Thermodynamics:Properties}

The WPE displays phase properties quite surprising in the Ising model.
Rather than undergo an SK-like spin-glass transition typical of many
disordered systems, the presence of a planted solution gives rise to a
thermal first-order phase transition. This transition in the WPE
occurs at an $\alpha$-dependent temperature $T_c$ such that two stable
states contribute equally to the free energy. As expected, at high
temperature the disordered paramagnetic state ($\mv = \ZeroVec$) is
the unique global minimizer to $F_{\rm TAP}$. At a critical
temperature $T_c$, this state remains globally minimizing but is no
longer unique; another state $\mv = m \tv$ with $m\neq 0$ is also
optimal. Below $T_c$, the paramagnet ceases to be the minimum, but
whether $\mv=0$ remains a local minimum (metastable state) to
$F_{\rm TAP}$ or an unstable state (saddle point or maximizer) depends
on $\alpha$ and the temperature. We find the fascinating result that
when $\alpha < 1$, the paramagnetic state is locally stable at all
nonzero temperatures. As will be discussed, this property has
considerable algorithmic implications.

The transition is first-order because of the discontinuity in the
minimizing $\mv$ and in the derivative of the free-energy minimum with
respect to $T$; consequently the internal energy also undergoes a jump
at $T_c$ as we will see. Remarkably, for any finite $\alpha$,
the transition is always technically first-order, in the sense
that there is a discontinuity in the log partition function
derivative. The magnitude of this discontinuity decreases with
$\alpha$, and the transition gradually segues from first to
second order.

\subsubsection{Stability of $\mv=\ZeroVec$}

The stability of $\mv=\ZeroVec$ can be ascertained from the Hessian of
$F_{\rm TAP}$, i.e., 
\[
\Hv(\mv) \triangleq \frac{\partial^2 F_{\rm TAP}}{\partial \mv^2} .
\]
The point $\mv=\ZeroVec$ is a local minimum of $F_{\rm TAP}$ if and only
if $\Hv(\ZeroVec)$ is positive definite. It turns out that the
spectral distribution of the random matrix $\Jv$ is key in determining
the relation between the definiteness of $\Hv(\ZeroVec)$ and $T$ at
given $\alpha$. The limiting eigenvalue distribution of $\Jv$ is
calculated in Sec.~\ref{sec:appendix:LSDJ}; somewhat surprisingly,
it is closely related to the Marchenko-Pastur~\cite{marchenko:67} law for the
spectrum of Wishart matrices constructed with \emph{uncorrelated}
Gaussians. The important aspect for now is that $-\alpha$ is always
(asymptotically) the smallest eigenvalue of $-\Jv$.

Computing the partial derivatives from the definition of $F_{\rm TAP}$ in
Eq.~(\ref{eq:FTAP}) we obtain the Hessian matrix elements
\begin{equation*}
  H_{ij} = \left \{
    \begin{array}{ll}
      \frac{1}{\beta(1-m_i^2)}
      - \frac{2\alpha\beta
      m_i^2}{N( 1+\beta(1-q))^2} +
      \frac{\alpha\beta(1-q)}{1+\beta(1-q)} & \textrm{$i = j$} \\[5pt]
      -J_{ij} -\frac{2\alpha\beta m_i
      m_j}{N(1+\beta(1-q)^2)} & \textrm{$i \neq j$}
    \end{array}
  \right. .
\end{equation*}
At $\mv=\ZeroVec$ we thus have
\begin{equation*}
  H_{ij} = \left \{
    \begin{array}{ll}
      \frac{1}{\beta} + \frac{\alpha\beta}{1+\beta} & \textrm{$i=j$} \\[5pt]
      -J_{ij} & \textrm{$i\neq j$}
    \end{array}
  \right. 
\end{equation*}
or compactly, the Hessian at $\mv=0$
\begin{equation}
  \Hv(\ZeroVec) = -\Jv + c(\alpha,\beta) \Iv
\end{equation}
with \[
c(\alpha,\beta) = \frac{1+\beta(1+\alpha \beta)}{\beta(1+\beta)} .
\]
$\Hv(\ZeroVec)$ is positive definite when $c(\alpha,\beta)$ is large enough to shift the eigenvalues
of $-\Jv$ to all be positive. From the spectral property mentioned
above, stability hence occurs when
\[
c(\alpha,\beta) \geq \alpha .
\]
For a given $\alpha$, this condition is equivalent to
\[
(T+1)(\alpha-T) \leq \alpha
\]
or
\[
T(T+1-\alpha)\geq 0 .
\]
Obviously, $T$ is non-negative; when $\alpha < 1$ the relation thus
holds for all $T$ while when $\alpha \geq 1$, it is satisfied
when $T \geq \alpha-1$.

To summarize, when $\alpha < 1$, the point $\mv=\ZeroVec$ is a locally
stable stationary point to the TAP free energy at any temperature; on
the other hand, when $\alpha \geq 1$, the Hessian becomes indefinite at
temperature $T_u = \alpha -1 $. For temperatures below $T_u$, note
that the eigenvalue of $\Hv$ corresponding to eigenvector $\tv$ (the
planted solution) becomes negative, which implies that the
paramagnetic solution becomes unstable along this
direction. Interestingly, the curvature of $F_{\rm TAP}$ along $\tv$ is of
largest magnitude for $\alpha\geq 1$ at the temperature such that
$c(\alpha,\beta)$ is minimal. In terms of $T$,
\[
c(\alpha,T) = T + \frac{\alpha}{T+1}
\]
which is smallest when $T = -1+\sqrt{\alpha}$. As local search
heuristics follow free-energy gradients in their quest for the
solution, it seems reasonable that this temperature is in some sense
optimal to start with when searching for the ground state in the
$\alpha \geq 1$ (easy) regime. A further special interpretation of
this temperature relates to the anti-Hopfield model, as discussed in
Appendix \ref{sec:appendix:replicaMethod}.

\begin{figure}
  \centering
  \includegraphics[width=1.\columnwidth]{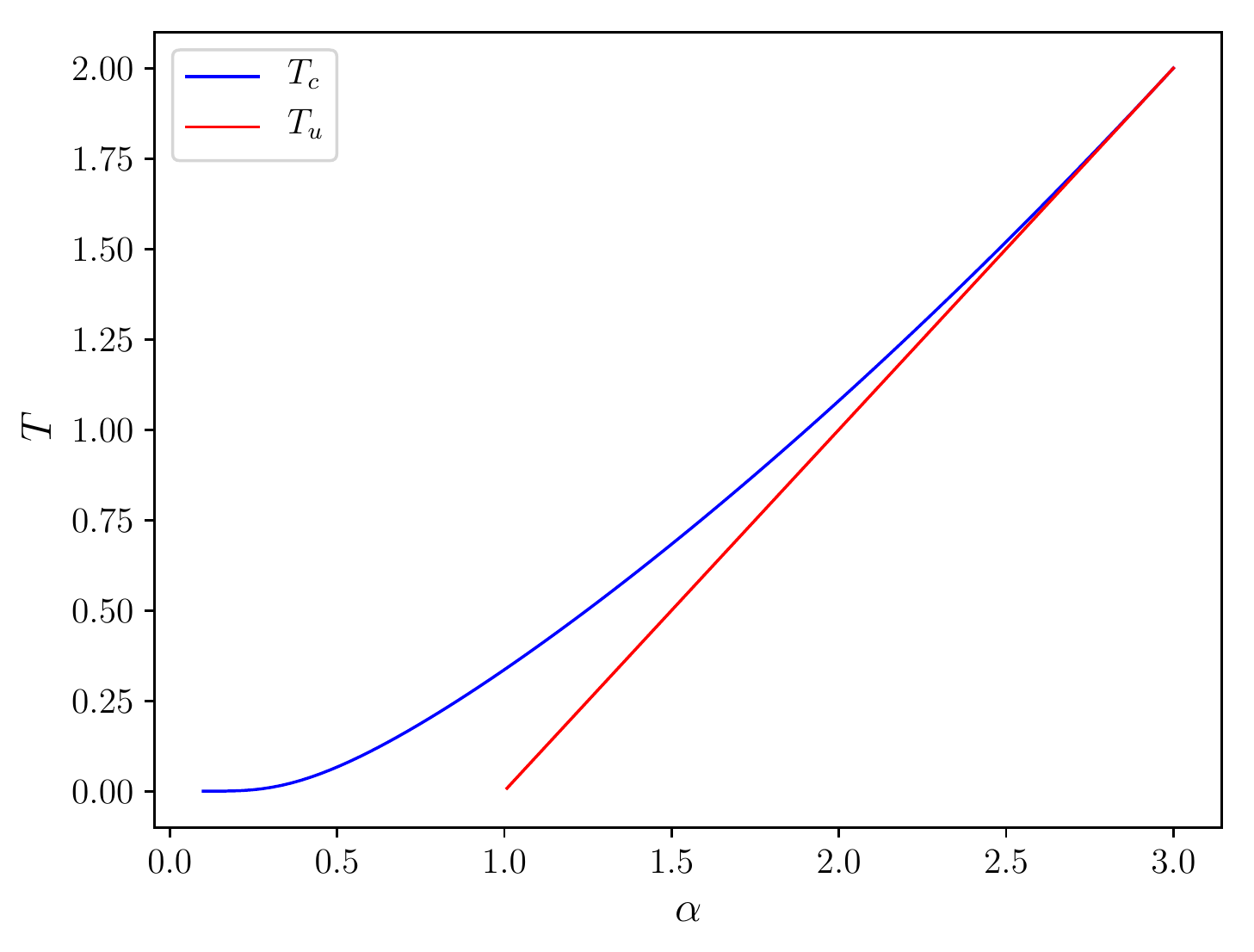}
  \caption{Transition temperatures for the Wishart planted ensemble as
    a function of $\alpha$. $T_c$ (solid blue) refers to the
    first-order (coexistence) phase transition temperature, while for
    $\alpha\geq 1$, $T_u$ (solid red) marks the instability of the
    paramagnetic phase. Note that as
    $\alpha$ increases past $1$, $T_u$ and $T_c$ converge, which is
    expected from the construction procedure. The values of $T_c$ in
    the low-$\alpha$ portion of the function are small but nonzero;
    they are well-approximated there by $T_c \approx 2^{2/\alpha}-1$.}
  \label{fig:transitionTemps}
\end{figure}

\subsubsection{First-order phase transition}

The first-order transition for given $\alpha$ occurs at a $T_c$ such
that $\ftilde(0) = \ftilde(m)$ for $m \neq 0$. For any $T$, the
stationary points of $\ftilde$ are determined numerically by
iteratively solving the saddle point equation [Eq.~(\ref{eq:ftildeStatPts})];
using binary search in $T$, we can then localize the transition
temperatures at any $\alpha$. Figure \ref{fig:transitionTemps} shows
the relation of $T_c$ with $\alpha$, and also shows for
$\alpha \geq 1$ the paramagnetic instability point $T_u = \alpha - 1$.
It is clear that $T_c > T_u$ uniformly but as $\alpha$ increases the
two temperatures converge; this in turn constrains the
jump magnitude between the two sides of the transition. While
$T_c$ has no closed-form expression in $\alpha$, we can obtain a lower
bound $\widehat{T_c}$ as
\begin{equation}
T_c \geq \widehat{T_c} \triangleq 2^{2/\alpha}-1 \label{eq:TcApprox}
\end{equation}
which serves as a surprisingly accurate approximation for small
$\alpha$ but deteriorates as $\alpha$ gets larger than $1$.

\begin{figure}
  \centering
  \includegraphics[width=0.85\columnwidth]{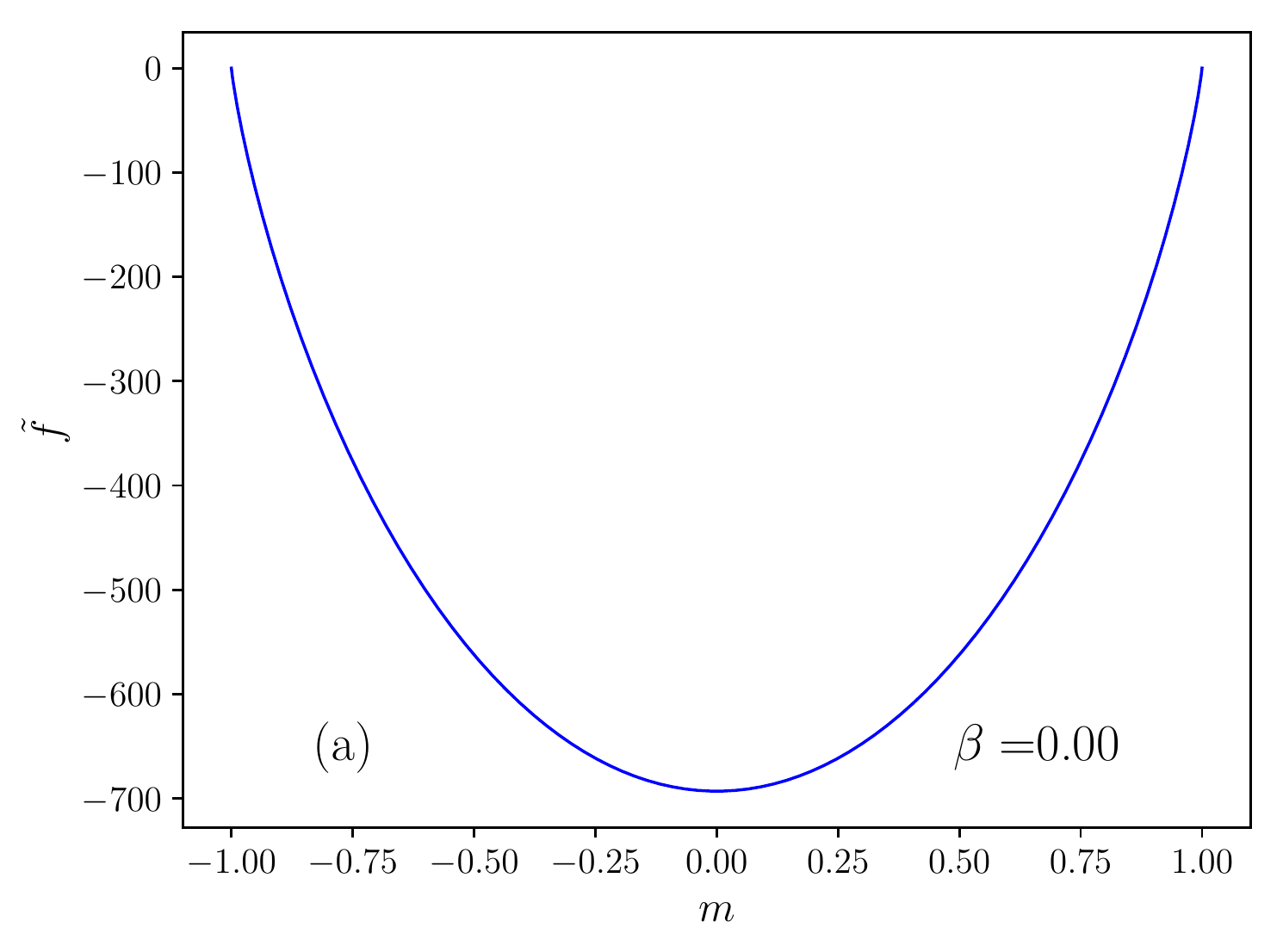}\\
  \includegraphics[width=0.85\columnwidth]{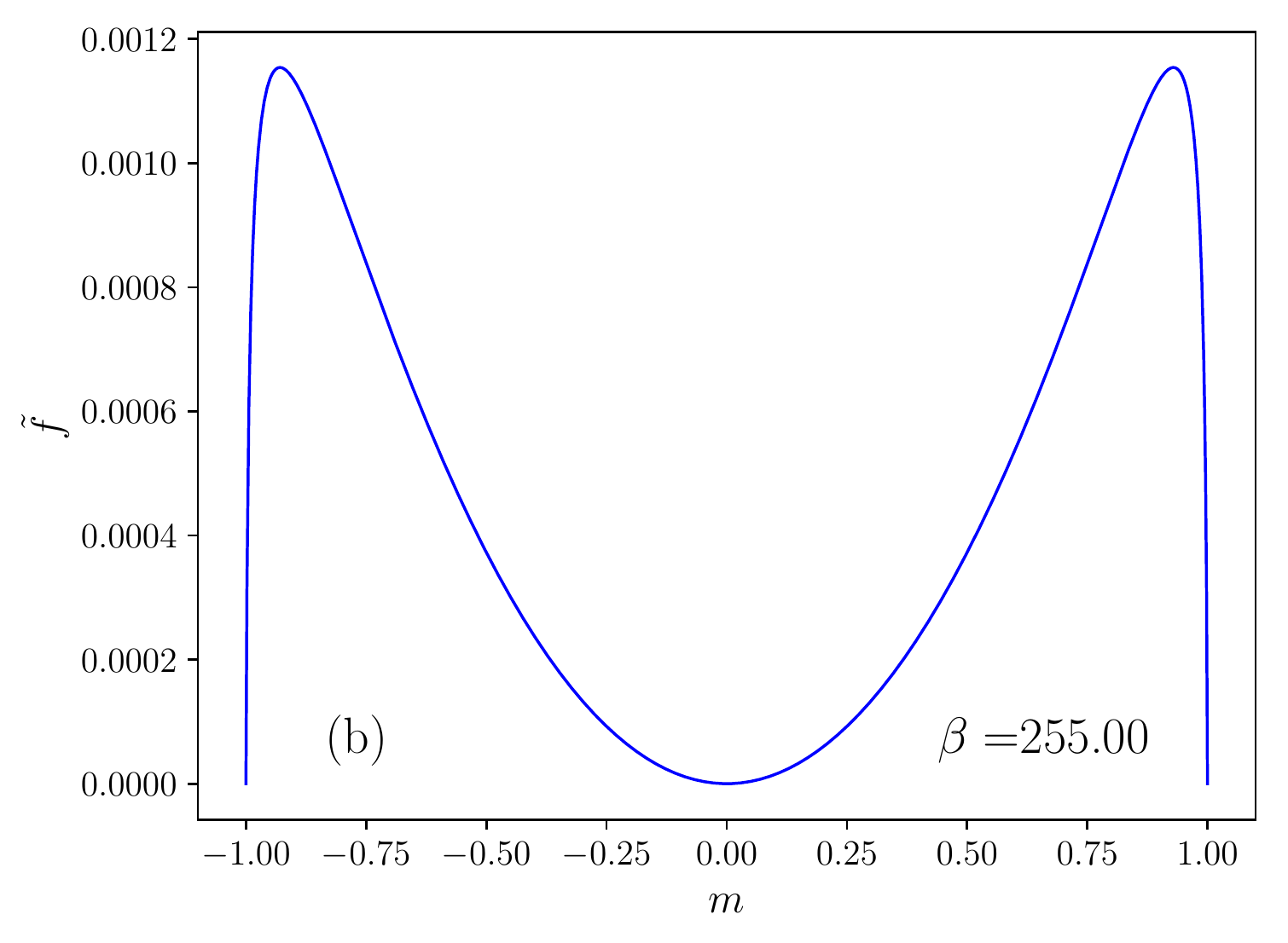}\\
  \includegraphics[width=0.85\columnwidth]{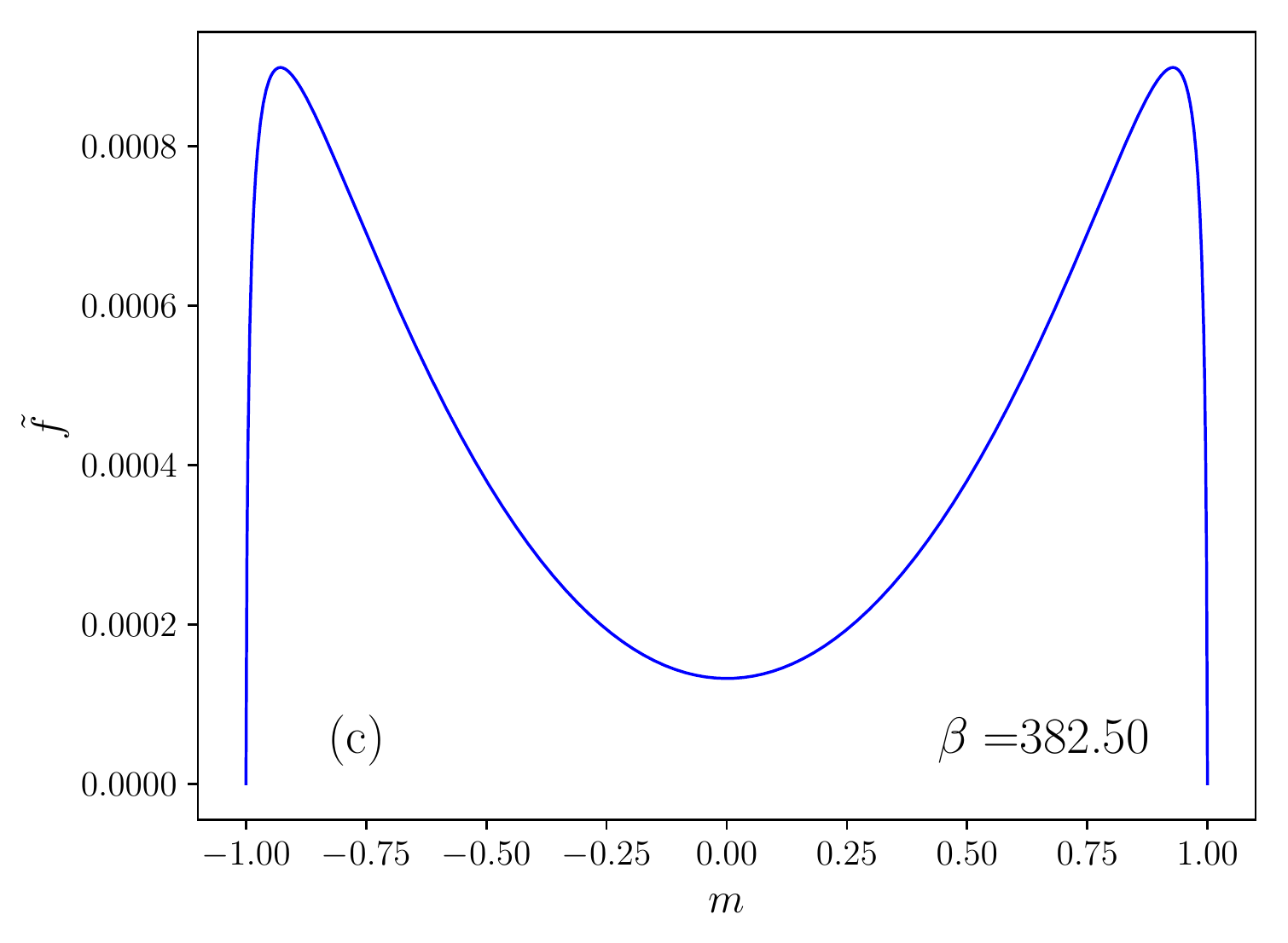}
  \caption{Function $\ftilde$, whose global minima correspond to
    those of the TAP free energy, for a Wishart planted problem with
    $\alpha=0.25$. In (a), we see the high-temperature regime, with a
    single minimum at $m=0$. Panel (b) shows the first-order
    transition, occurring at $T_c= 0.004$ and characterized by equal
    contribution of $m=0$ and the minima close to $m=\pm 1$, while 
    panel (c) displays the low-temperature phase ($T\approx 0.003$). Note that $m=0$
    remains metastable in spite of the low temperature, while
    the free-energy minima lie around $m=\pm 1$ corresponding to the
    planted solution.}
  \label{fig:fTildeAlpha0_25}
\end{figure}

We illustrate these results by plotting $\ftilde(m)$ at representative
temperatures for a few values of $\alpha$. Consider first the
relatively small value of $\alpha=0.25$, with landscape illustrated in
Fig.~\ref{fig:fTildeAlpha0_25}. The first-order transition at
$T_c=0.004$ is clearly visible in the middle panel as $m=0$ is now a
coexisting global minimum with the other minima close to the
end points~\cite{comment:ftildeMinima}. Remarkably, when
$T \approx 0.003$, shown in the bottom panel, $m=0$ remains locally
optimal.

\begin{figure}
  \centering
  \includegraphics[width=0.85\columnwidth]{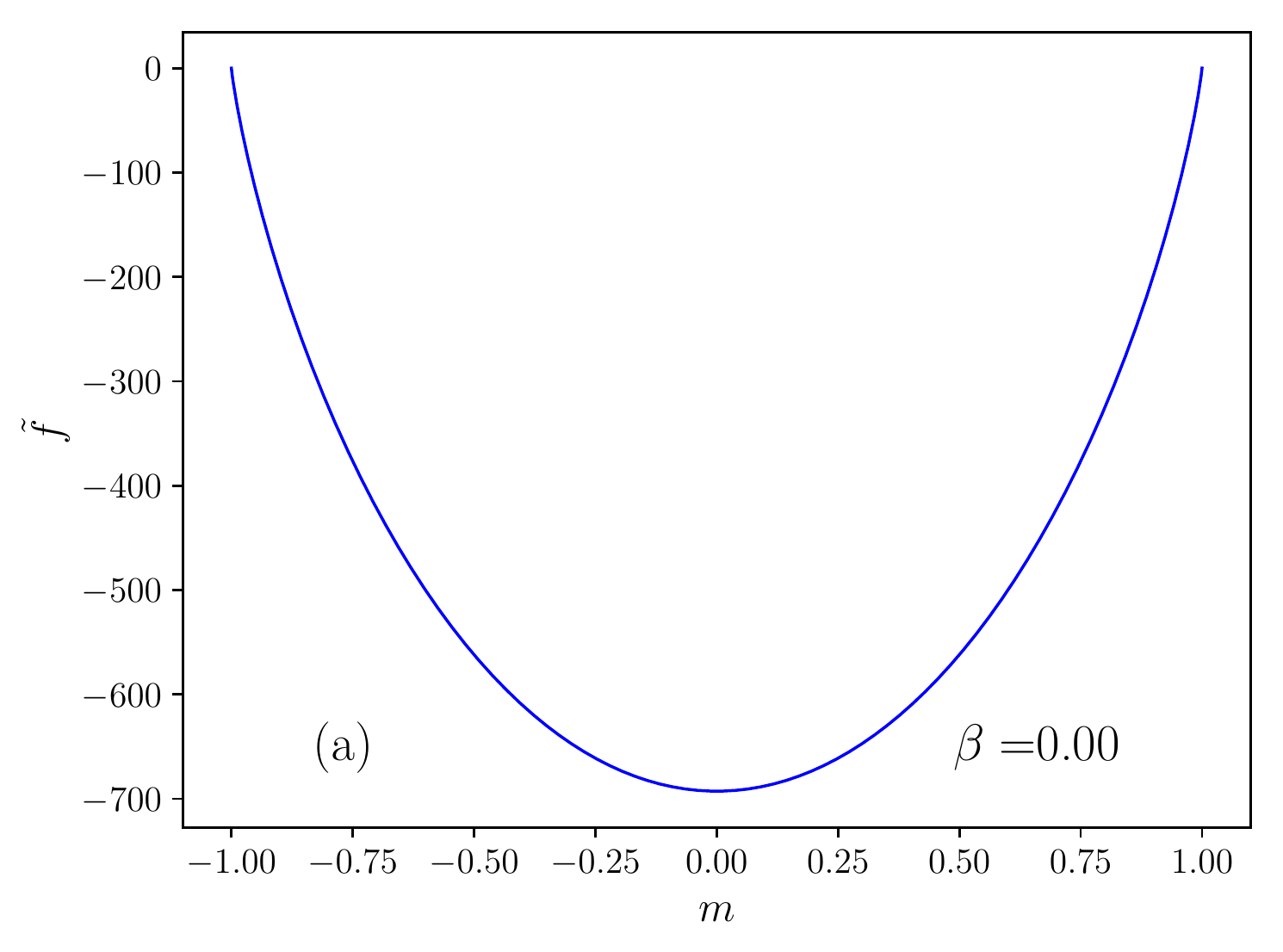}\\
  \includegraphics[width=0.85\columnwidth]{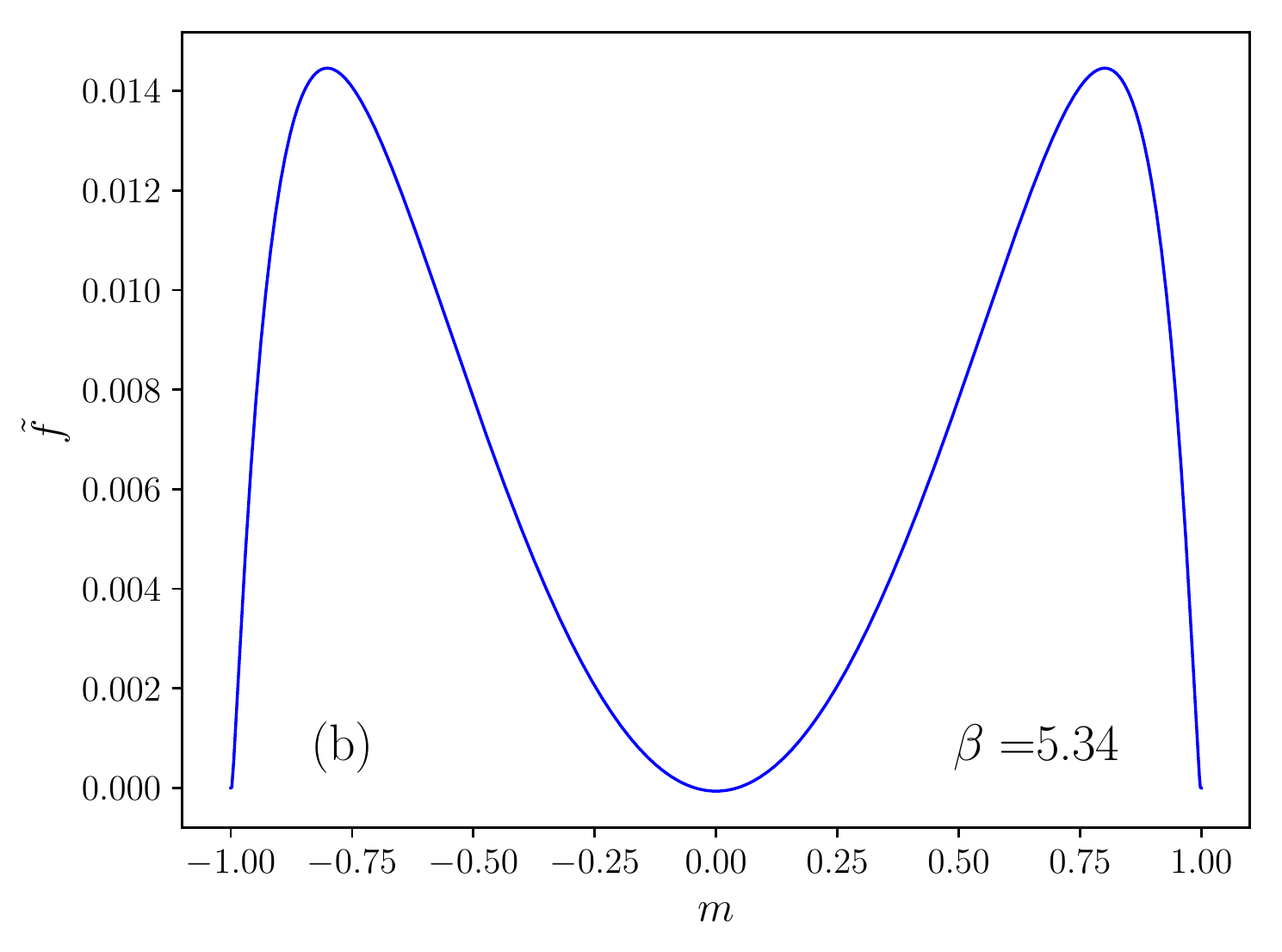}\\
  \includegraphics[width=0.85\columnwidth]{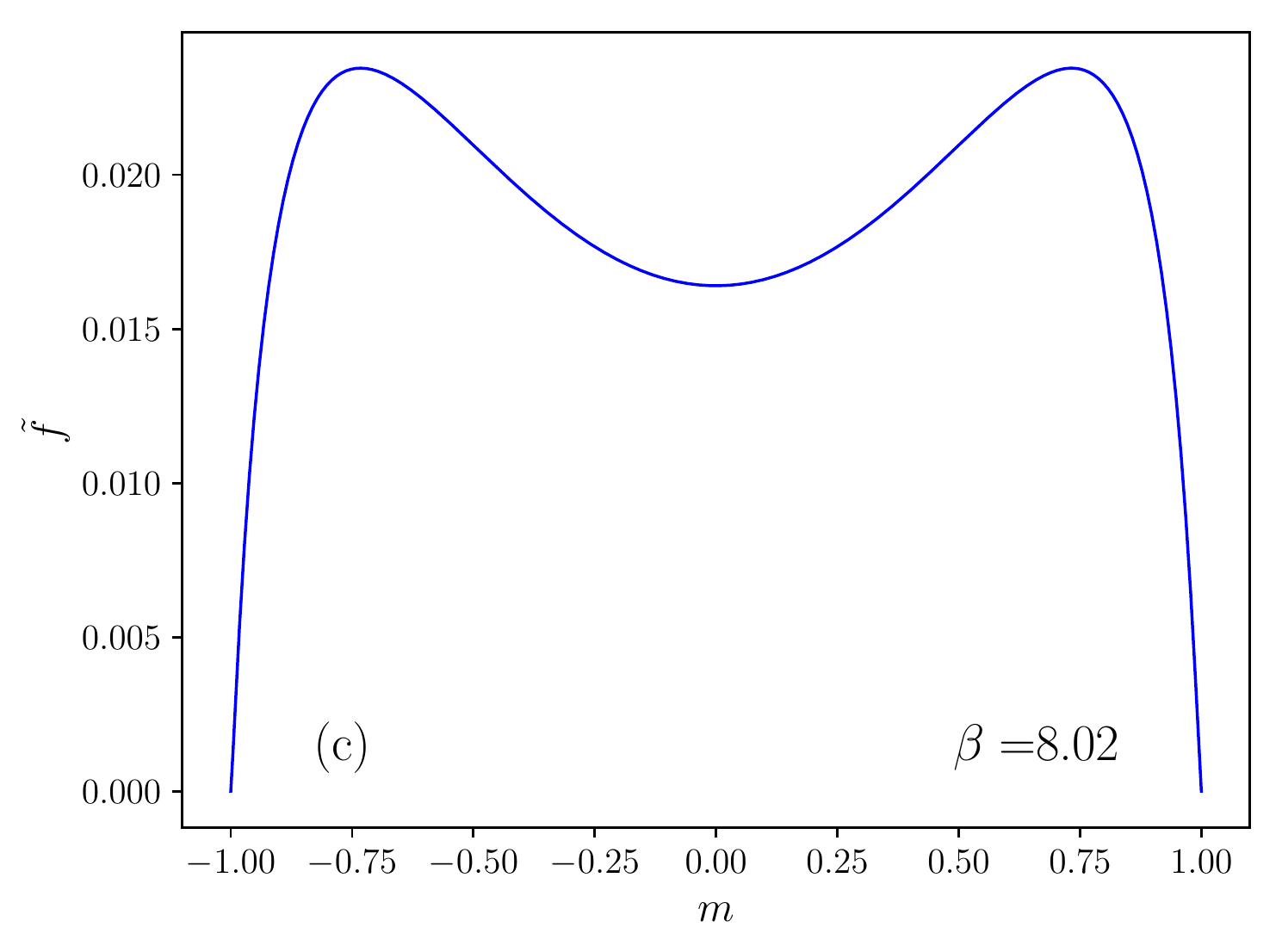}
  \caption{Function $\ftilde$ for a Wishart planted problem with
    $\alpha=0.75$. In panel (a) we show the high-temperature regime. 
    Panel (b) shows the first-order transition at $T_c= 0.187$.
    Panel (c) illustrates the low-temperature ($T\approx 0.125$) phase, with
    $m=0$
    remaining metastable as it did for $\alpha=0.25$ shown in Fig.~\ref{fig:fTildeAlpha0_25}. 
    Note further that the sizes of the regions
    between the \emph{maxima} and the boundaries are larger than they
    were for $\alpha=0.25$, suggesting that the ground state has a
    wider basin of attraction, resulting in this being a
    computationally easier problem.}
  \label{fig:fTildeAlpha0_75}
\end{figure}

In Fig.~\ref{fig:fTildeAlpha0_75} we see similar behavior with an
increased $T_c \approx 0.187$ and a persistent metastable $m=0$ state
at low temperature for $\alpha = 0.75$.

From an algorithmic perspective, the low-temperature stability of $m=0$ is
quite intriguing as it is widely believed to correlate with genuine
combinatorial hardness, sounding the death knell for heuristic
approaches attempting to locally optimize trial configurations. This
category of algorithms includes workhorses such as simulated
annealing and parallel tempering Monte Carlo, which exploit correlations
in the energy landscape to search by performing biased random walks.
Paramagnetic stability suggests that a problem is hard because it
implies that cooling an initially disordered configuration will
overwhelmingly lead to states that are also disordered. There is no
exploitable information within the landscape guiding the dynamics to
the correct region of the state space, and the only hope is to begin
the search from the appropriate basin. Disordered states form the vast
bulk of the configuration space, however, so the probability of a
correct initialization decreases exponentially with system size. While
all problems with $\alpha<1$ exhibit this feature, the specific value
of $\alpha$ turns out to modulate the size of the planted solution
basin, i.e., all states from which the ground state can be reached at
reasonable cost, which of course exerts a critical effect on the
observed performance of heuristics. This aspect can be observed by
comparing the intervals between the end points and \emph{maxima} of
$\ftilde$ in Figs.~\ref{fig:fTildeAlpha0_25} and
\ref{fig:fTildeAlpha0_75} and noting them to be wider for
$\alpha=0.75$. Broadly speaking, small values of $\alpha$ lead to
smaller basins and hence, one would surmise, more difficult problems.
This assertion assumes, however, that the vectors $\{\wv^\mu\}$ are
composed of real numbers and specifiable to arbitrary numerical
precision. When the precision is bounded, the interpretation is more involved
and is discussed in detail in Sec.~\ref{sec:hardnessPhase}.

\begin{figure*}
  \centering
  \includegraphics[width=0.85\columnwidth]{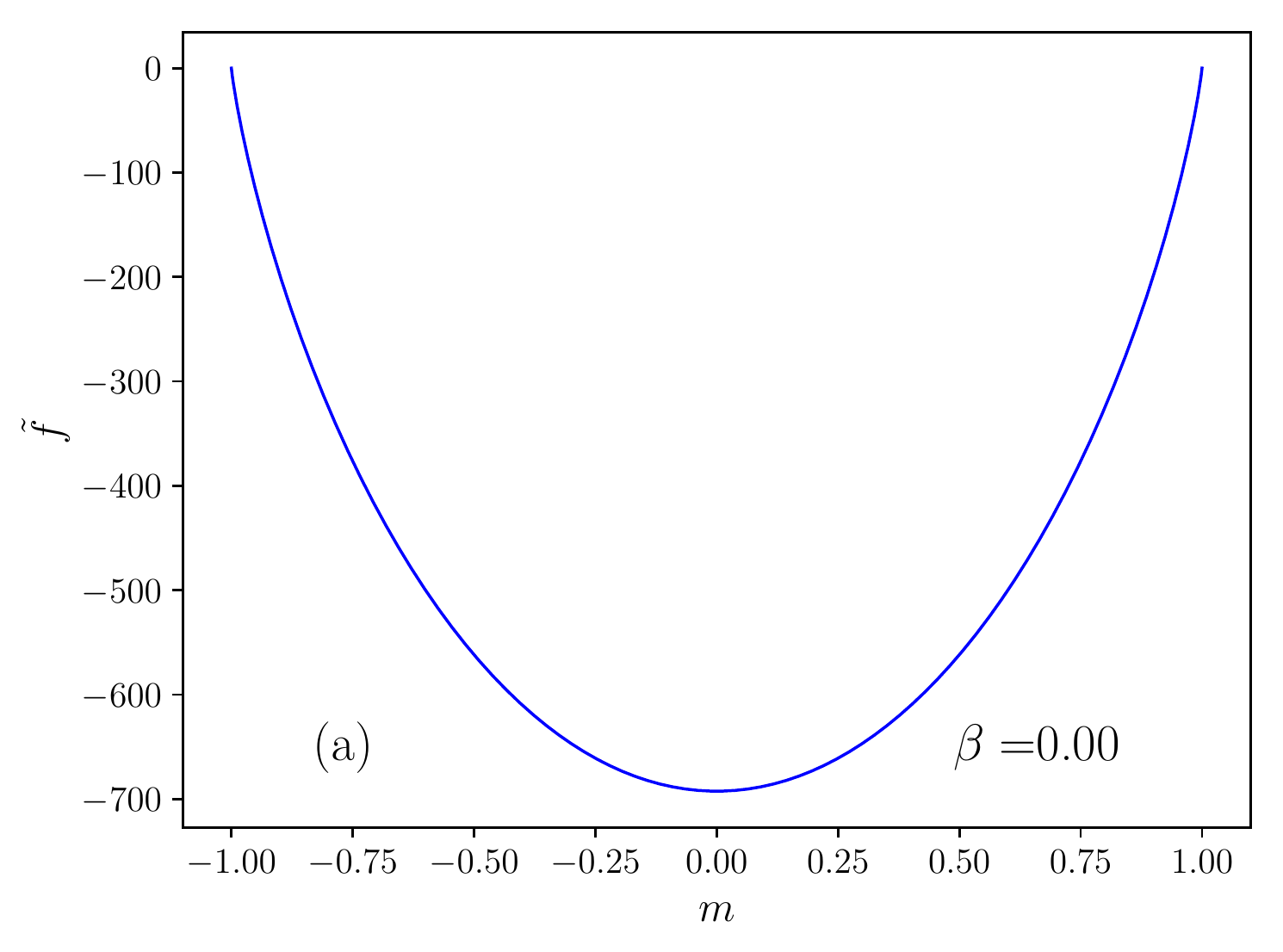}
  \includegraphics[width=0.85\columnwidth]{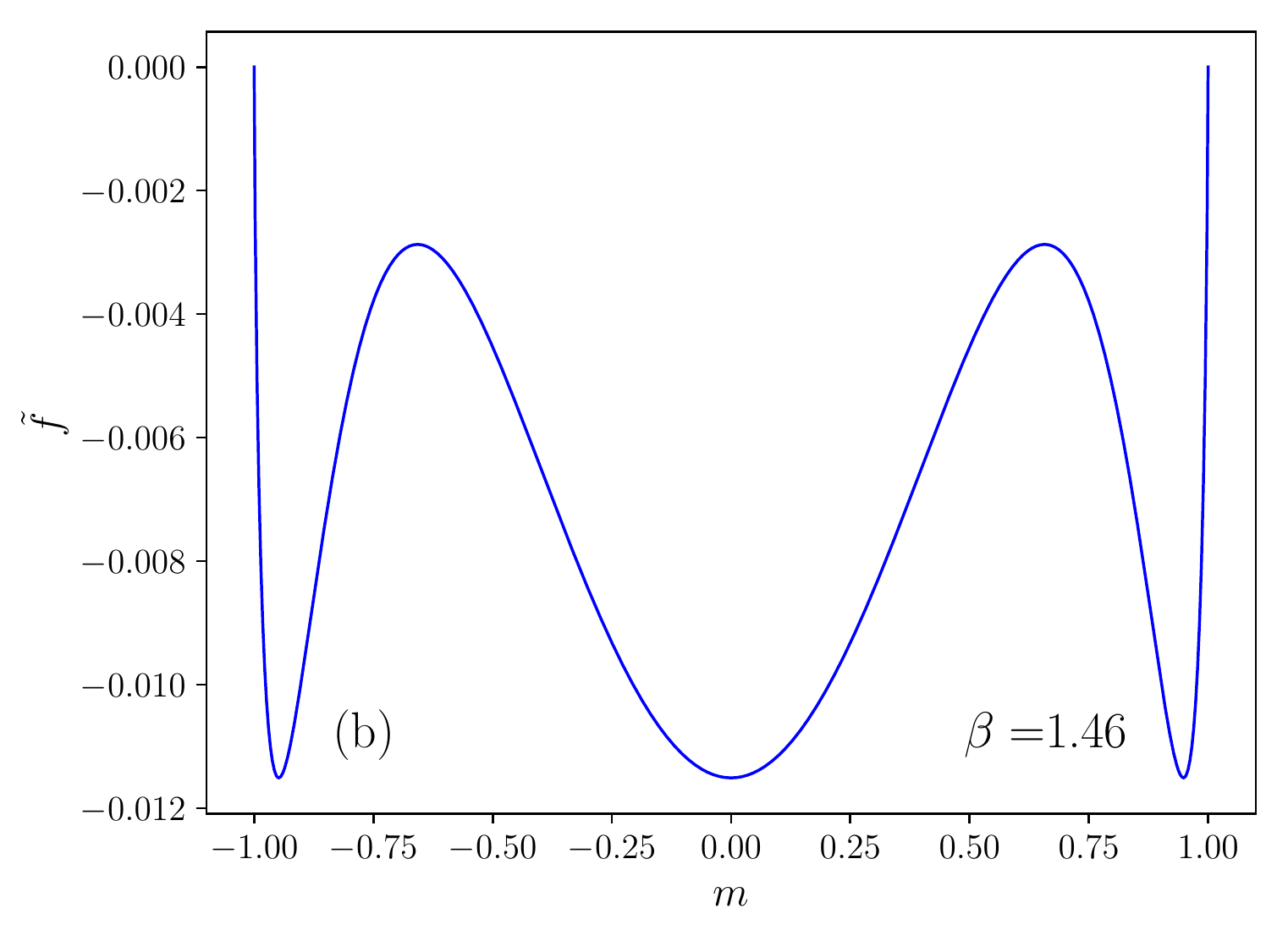}\\
  \includegraphics[width=0.85\columnwidth]{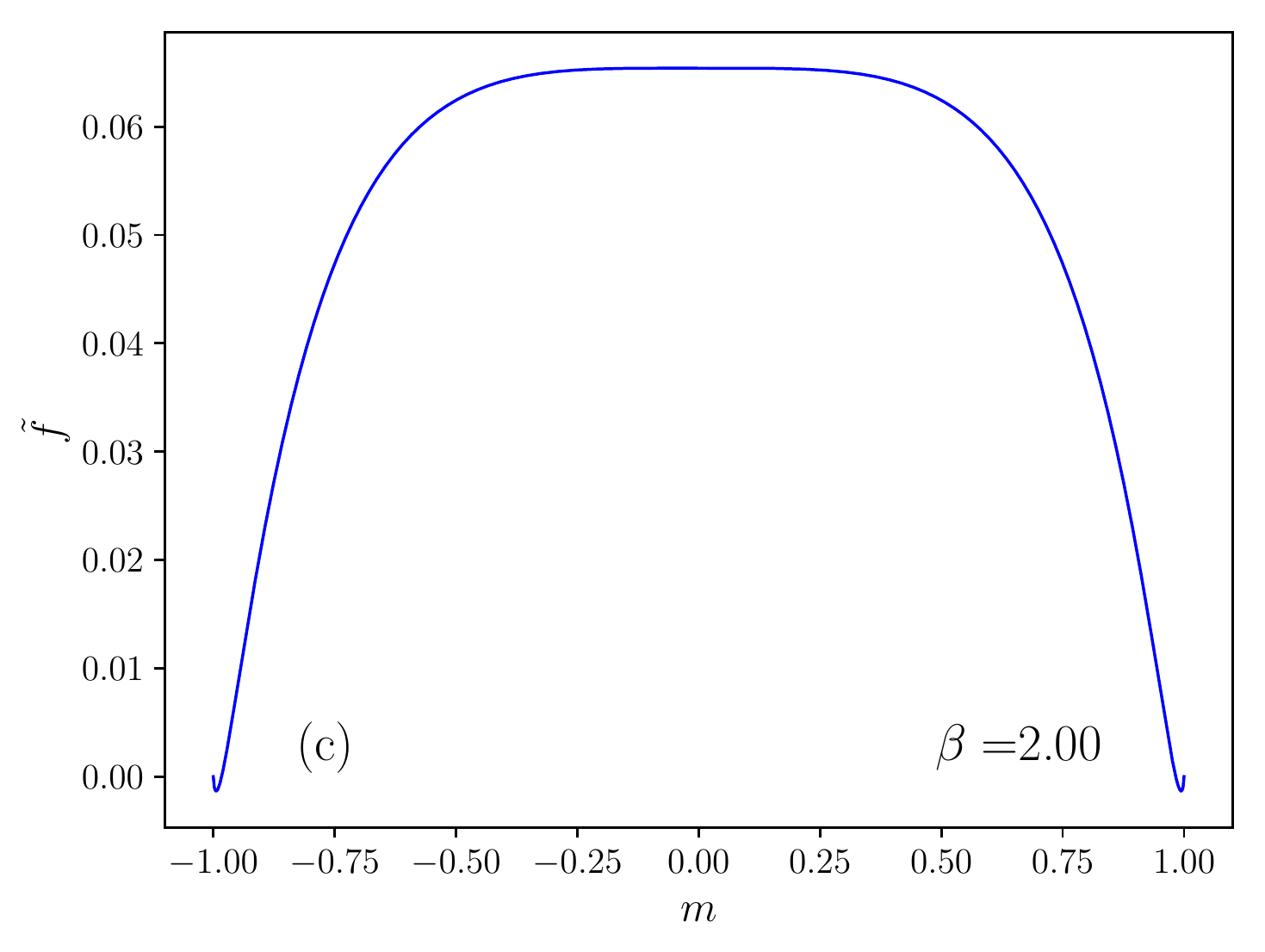}
  \includegraphics[width=0.85\columnwidth]{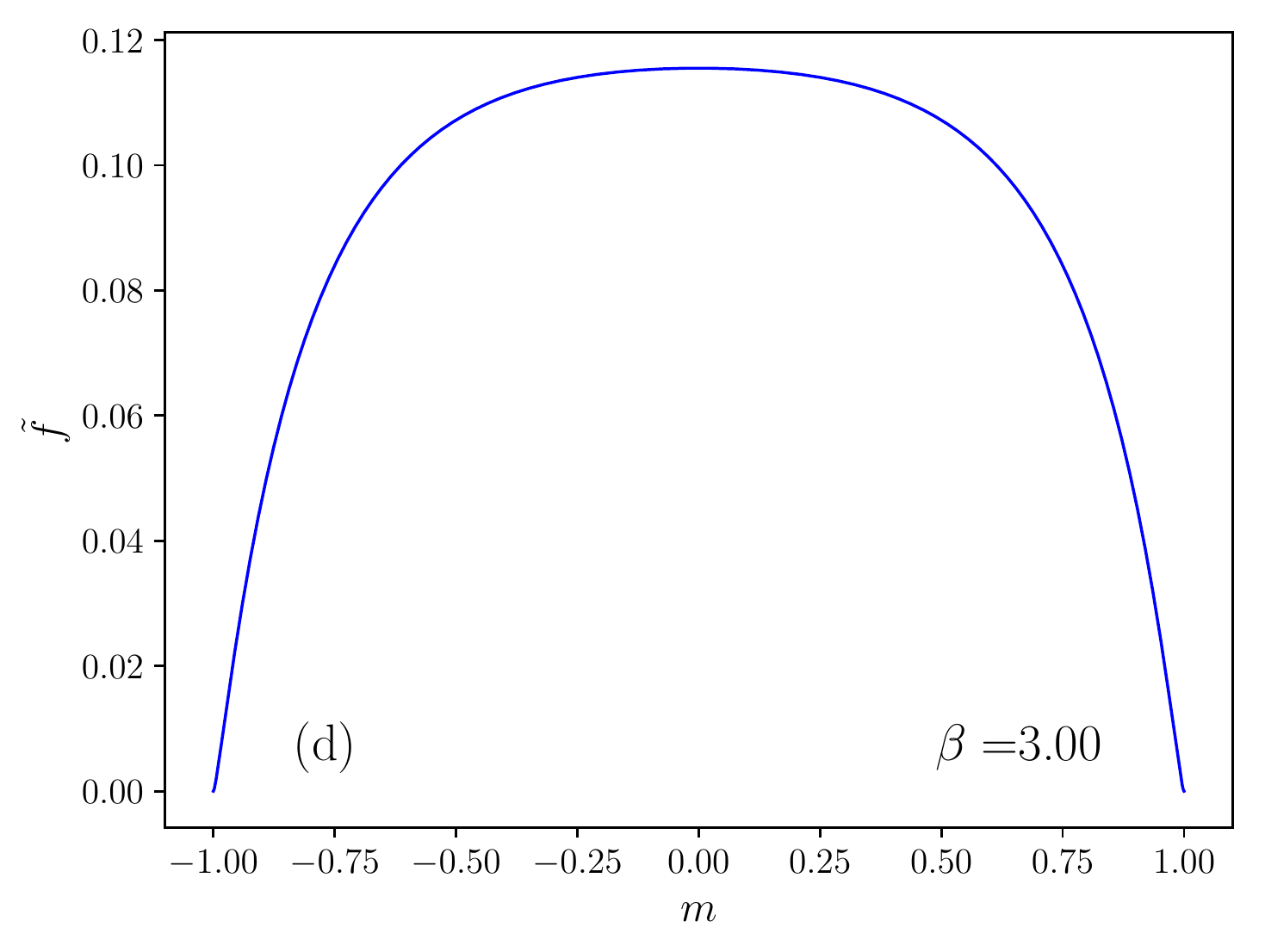}
  \caption{$\ftilde$ for $\alpha=1.5$. Panel (a) shows the  high temperature
    phase, whereas panel (b) shows the first-order phase transition at $T_c = 0.68$.
    The paramagnet at $m=0$ remains stable until $T_u = 0.5$,
    shown in panel (c), at which point it becomes unstable.
    For $T<T_u$ [panel (d)], the
    ground states are easily attainable.}
  \label{fig:fTildeAlpha1_5}
\end{figure*}
Figure \ref{fig:fTildeAlpha1_5} shows how $\ftilde$ is affected by $T$
for a value of $\alpha=1.5$, a regime in which we expect $m=0$ to
become unstable. The first-order transition is still readily apparent
at $T_c=0.68$. For temperatures lower than $T_c$ but higher than
$T_u = 0.5$, the paramagnet remains a local minimum. Finally, when
cooled to $T_u=0.5$, $m=0$ becomes a maximum along the planted ground
state direction. From a computational perspective, we anticipate that
a well-designed local algorithm will be likely to succeed below this
temperature as it follows the free-energy gradients leading it to the
planted solution.

To further appreciate the first-order transition, we illustrate the
discontinuity at $T_c$ of some key thermodynamic quantities for a
few representative $\alpha$ values. A key observation is that the
nature of the transition changes ``gracefully'' from first to
second order as $\alpha$ increases past unity.

First, we recall (see, for example, Ref.~\cite{mezard:09}) that the ensemble 
partition function is
related to the mean-field free-energy density $f(m)$ as
\[
Z_N(\beta) \doteq \int_{-1}^{1}\exp\Big[ -N\beta f_\beta(m) \Big] \dd m .
\]
Using Laplace's method, we obtain the ``log partition density''
\[
\lim\limits_{N\to\infty} \frac{1}{N} \log Z_N(\beta) = -\beta \min\limits_{m\in [-1,1]} f_\beta(m) ,
\]
where an additive term arising from Laplace's formula vanishes for
large $N$. We define the minimum and minimizer of the free-energy
density, respectively, as
\[
f^*(\beta) \triangleq \min\limits_{m\in [-1,1]} f_\beta(m)
\]
and
\[
m^*(\beta) \triangleq \argmin\limits_{m\in [-1,1]} f_\beta(m) .
\]
The limiting internal energy density is
\begin{align*}
  e(\beta) \triangleq & \lim\limits_{N\to\infty} \frac{1}{N}E_N(\beta) \\
  = & \lim\limits_{N\to\infty} -\frac{1}{N}\frac{\dd}{\dd\beta} \log Z_N(\beta)
\end{align*}
so that
\[ e(\beta) =
\frac{\dd}{\dd\beta} \beta f^*(\beta) .
\]

\begin{figure*}
  \centering
  \includegraphics[width=0.85\columnwidth]{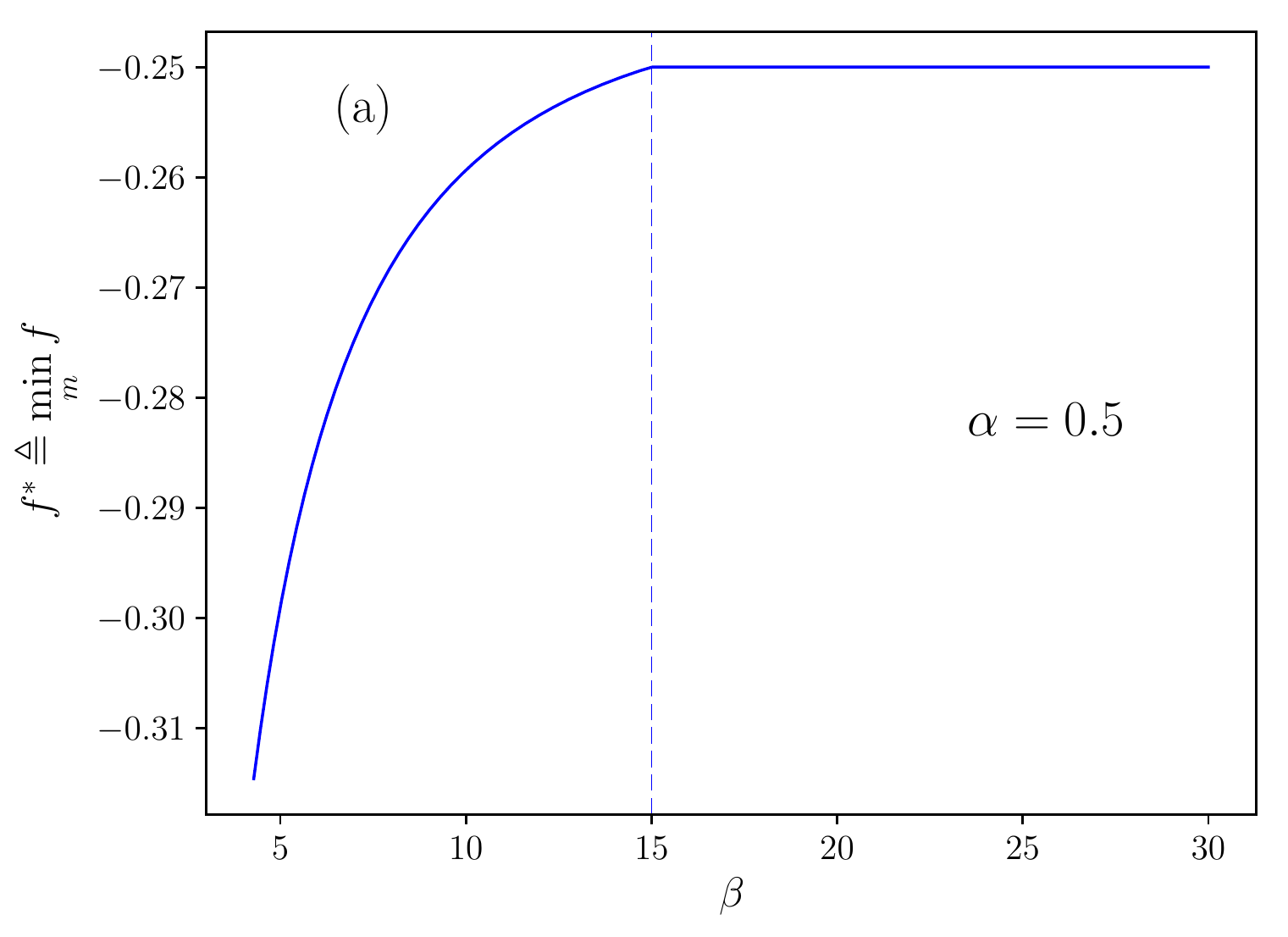}
  \includegraphics[width=0.85\columnwidth]{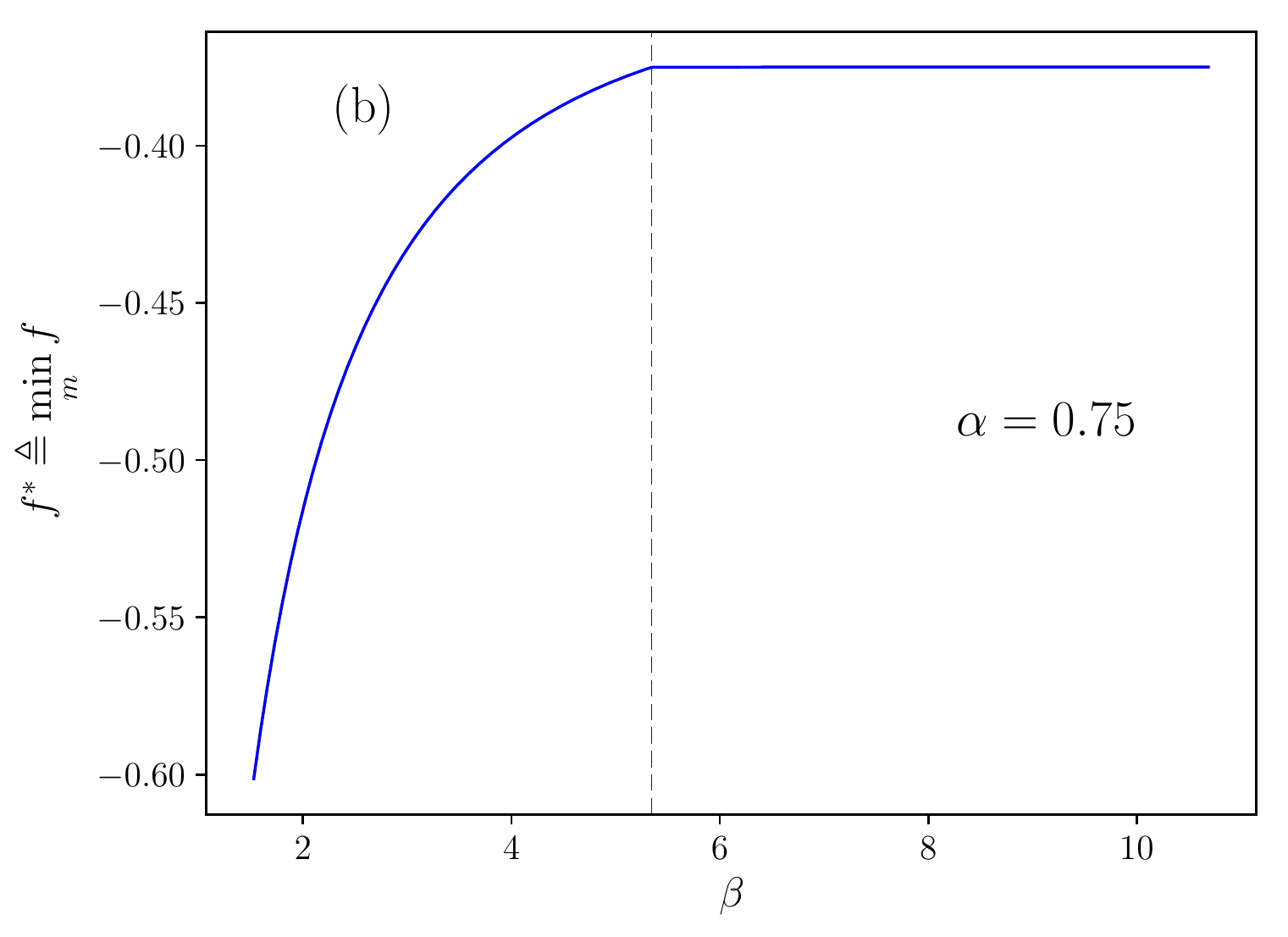}\\
  \includegraphics[width=0.85\columnwidth]{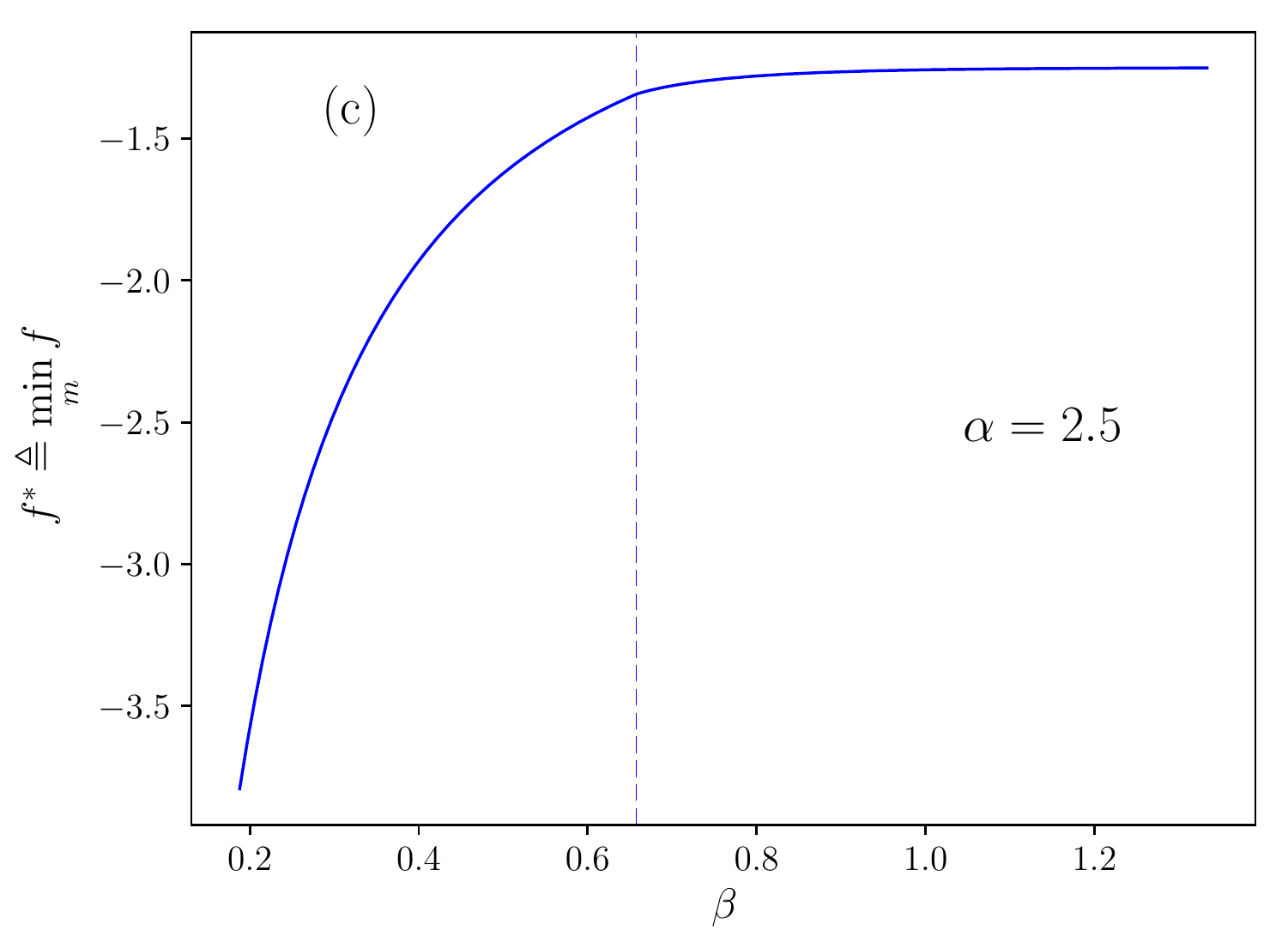}
  \includegraphics[width=0.85\columnwidth]{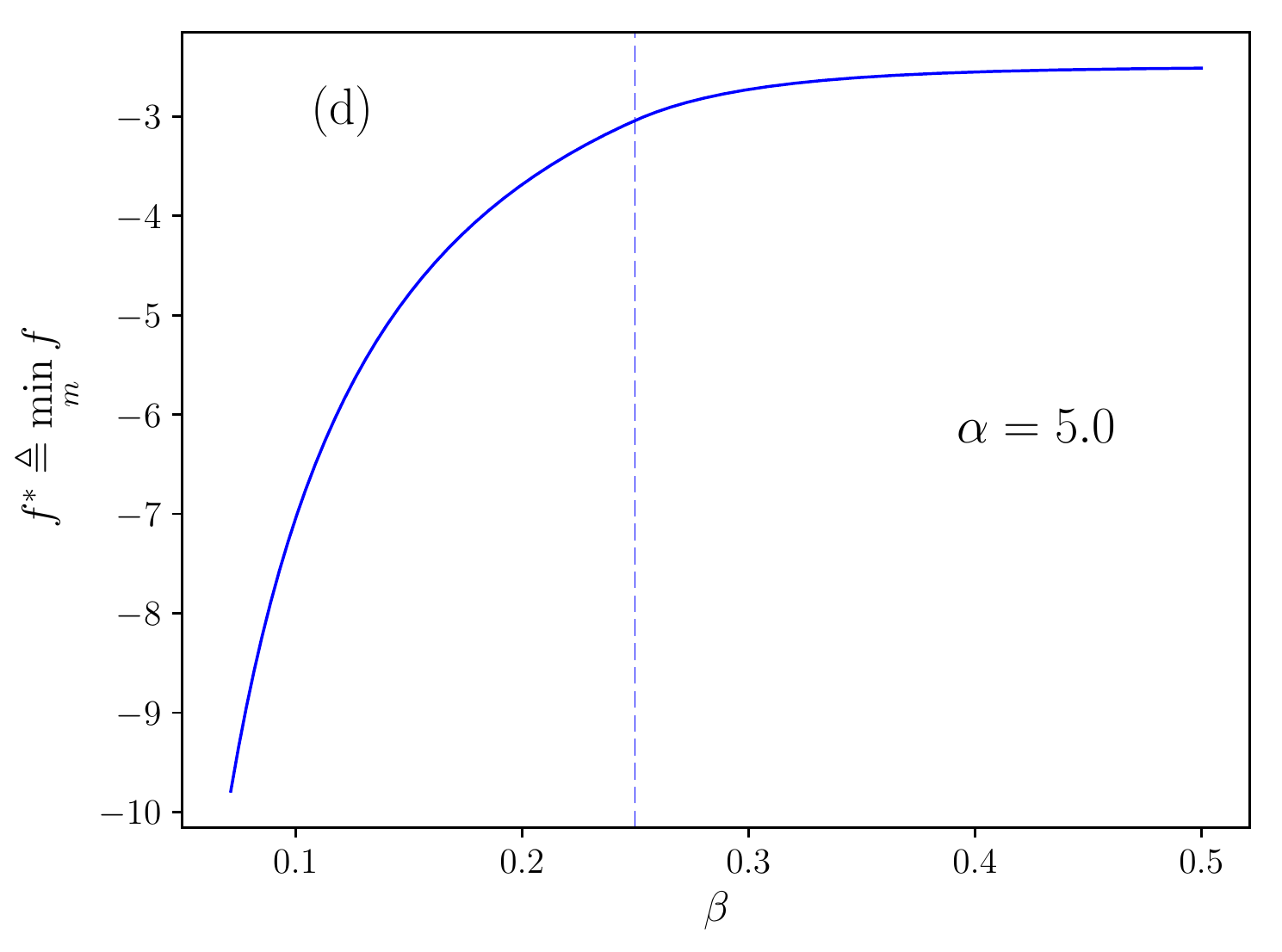}
  \caption{$f^*(\beta)$ corresponding to the asymptotic log partition
    function density for four values of $\alpha$. The first-order
    transition occurs at $\beta_c = 1/T_c$ denoted by the dashed vertical line in
    each panel. When $\alpha=0.5$, shown in panel (a), and $\alpha=0.75$, shown in panel (b), the
    function is clearly not analytic at $\beta_c$. It remains so for
    $\alpha=2.5$ [panel (c)], but the discrepancy in the derivatives on the two
    sides has lessened. When $\alpha=5$ the function, plotted in panel (d)
    visually appears smooth at $\beta_c$.}
  \label{fig:fStarPlots}
\end{figure*}

In Fig.~\ref{fig:fStarPlots}, we plot $f^*(\beta)$ for four values
of $\alpha$, with the corresponding $\beta_c = 1/T_c$ located at the vertical
dashed line. For the first two values of $\alpha$ ($0.5$ and $0.75$),
the log partition function is clearly a smooth function on either side
of the transition but is not differentiable at $\beta_c$. At
$\alpha=2.5$, the two sides still have different derivatives but the
discrepancy is diminished, and when $\alpha=5$ it is no longer visibly discernible. 

\begin{figure*}
  \centering
  \includegraphics[width=0.85\columnwidth]{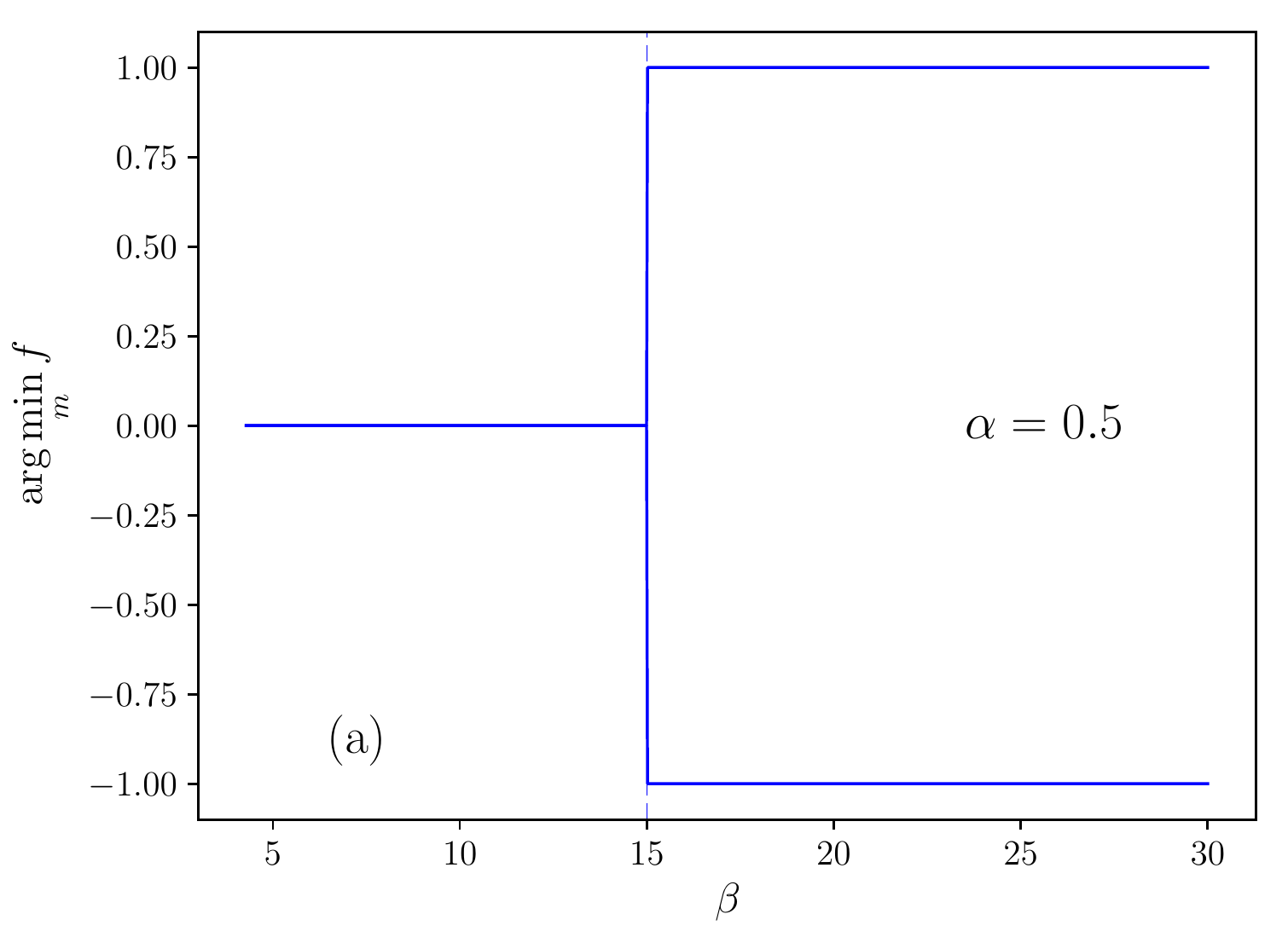}
  \includegraphics[width=0.85\columnwidth]{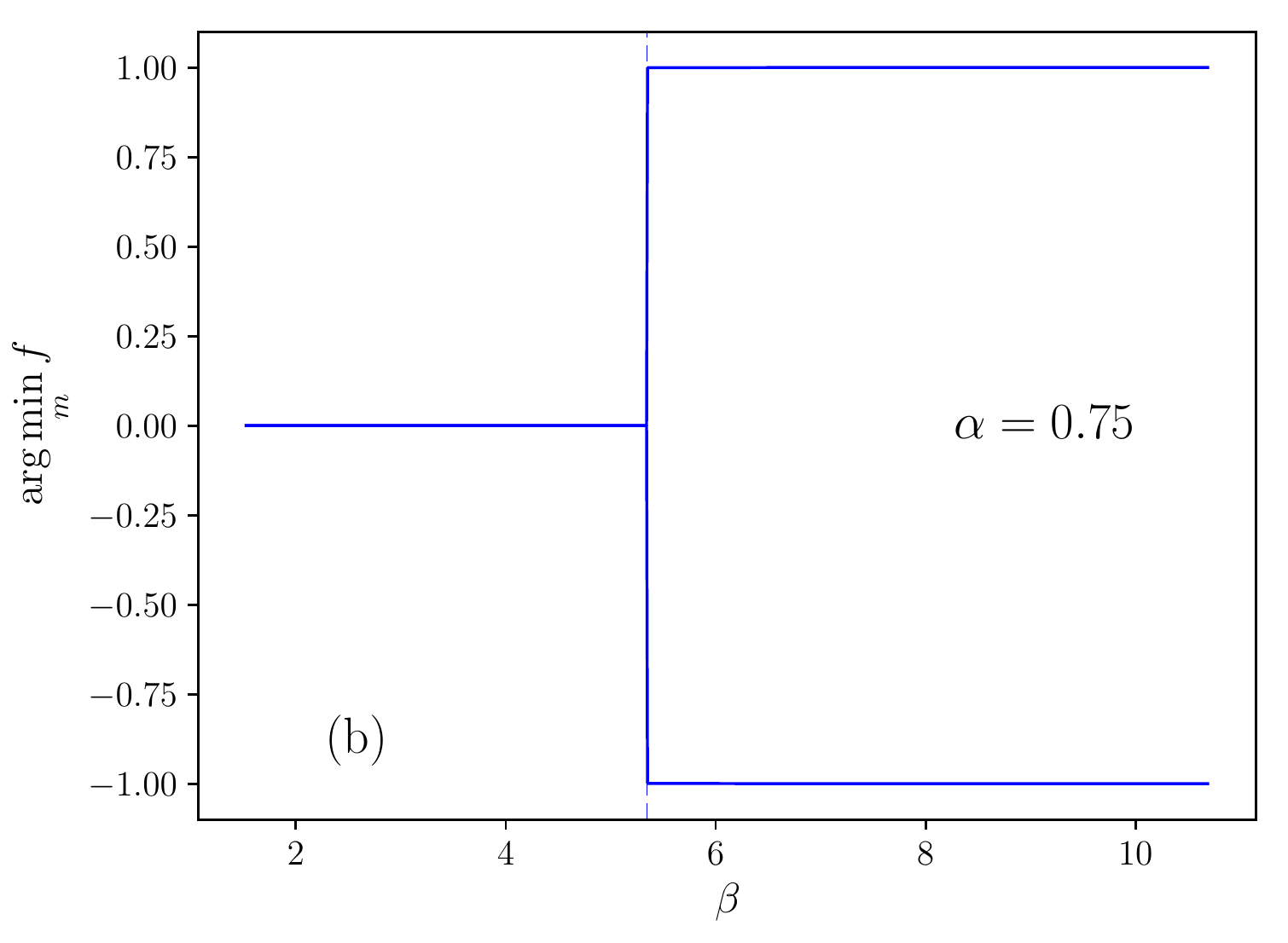}\\
  \includegraphics[width=0.85\columnwidth]{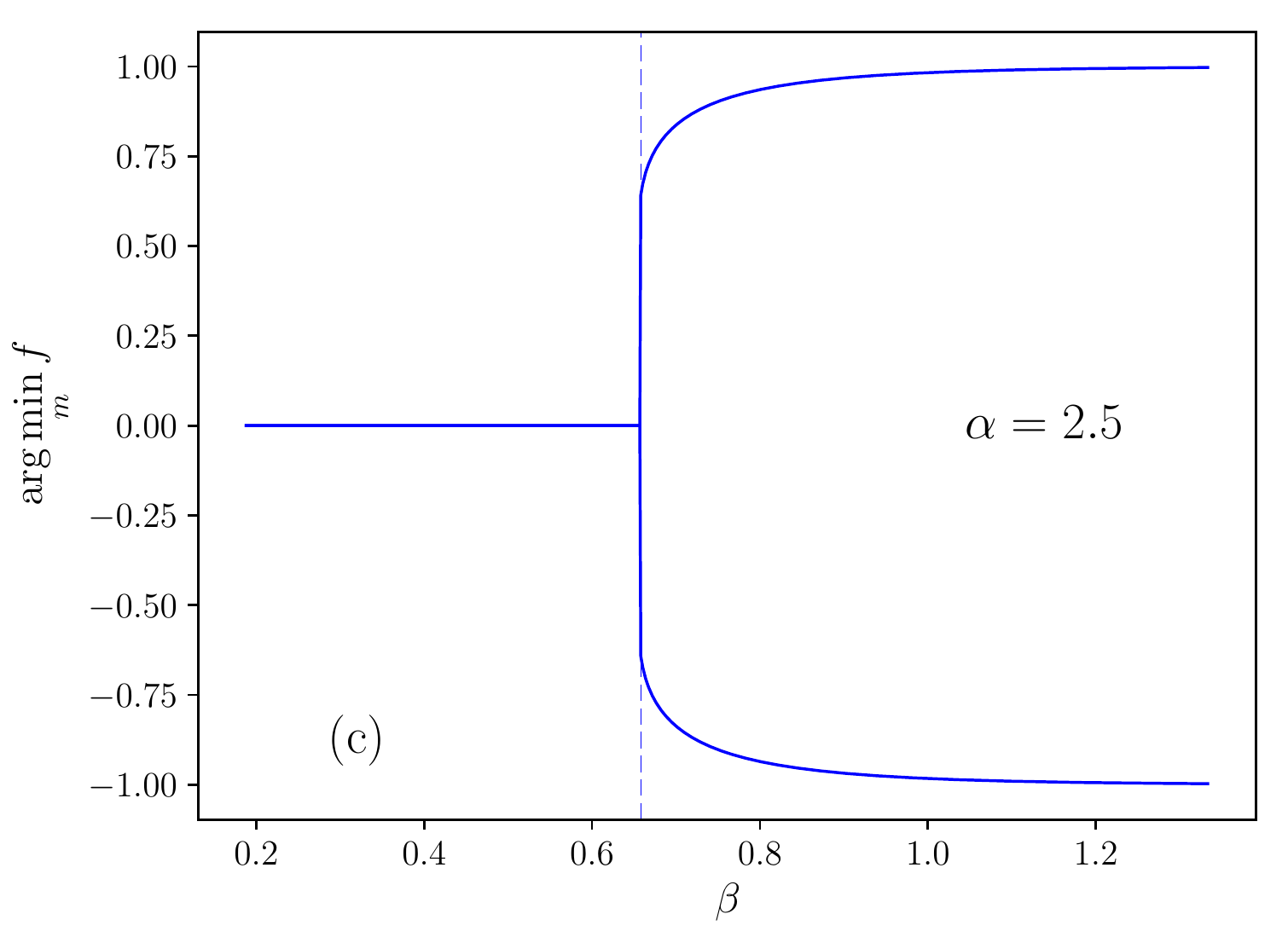}
  \includegraphics[width=0.85\columnwidth]{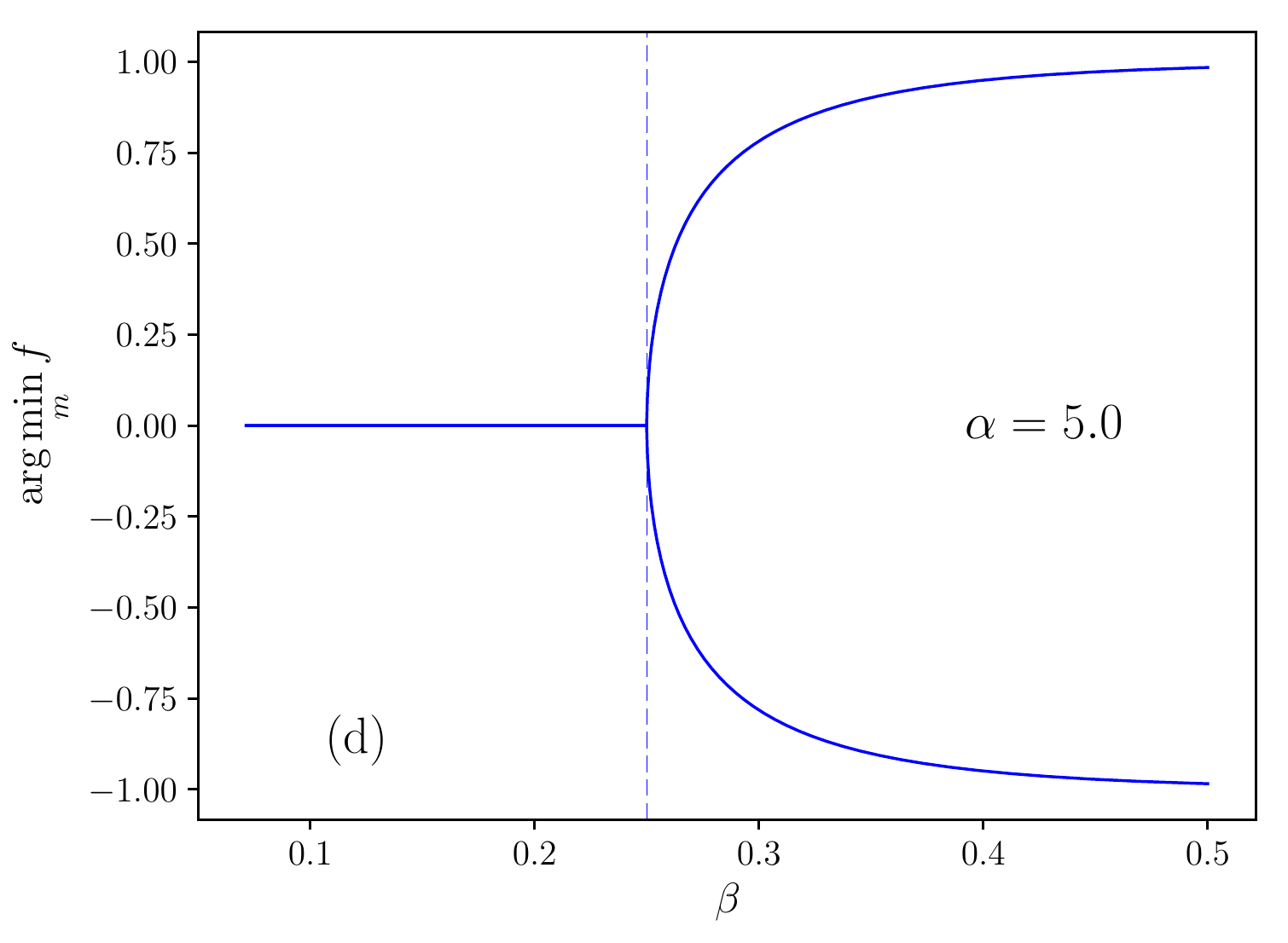}
  \caption{Magnetizations $m^*(\beta)$ for the same values of $\alpha$
  used in Fig.~\ref{fig:fStarPlots}. The first-order transition
  appears as an abrupt bifurcation at $\beta_c$ (dotted vertical
  lines) most apparent for
  $\alpha=0.5$ (a) and $\alpha=0.75$ (b). When
  $\alpha=2.5$ (c) the discontinuity is diminished,
  while for $\alpha=5$ (d) the bifurcation appears to have
  changed to second-order.}
  \label{fig:mStarPlots}
\end{figure*}

Figure \ref{fig:mStarPlots} shows the magnetization $m^*(\beta)$ for
the same four values of $\alpha$; the smaller $\alpha$ values show a
sharp, discontinuous bifurcation at $\beta_c$ at which point the
system abruptly shifts from being paramagnetic to being closely
aligned with the planted solution. The jump is still apparent but
attenuated for $\alpha=2.5$, while for $\alpha=5$ the bifurcation
starts to resemble a second-order ferromagnetic transition.

\begin{figure*}
  \includegraphics[width=0.85\columnwidth]{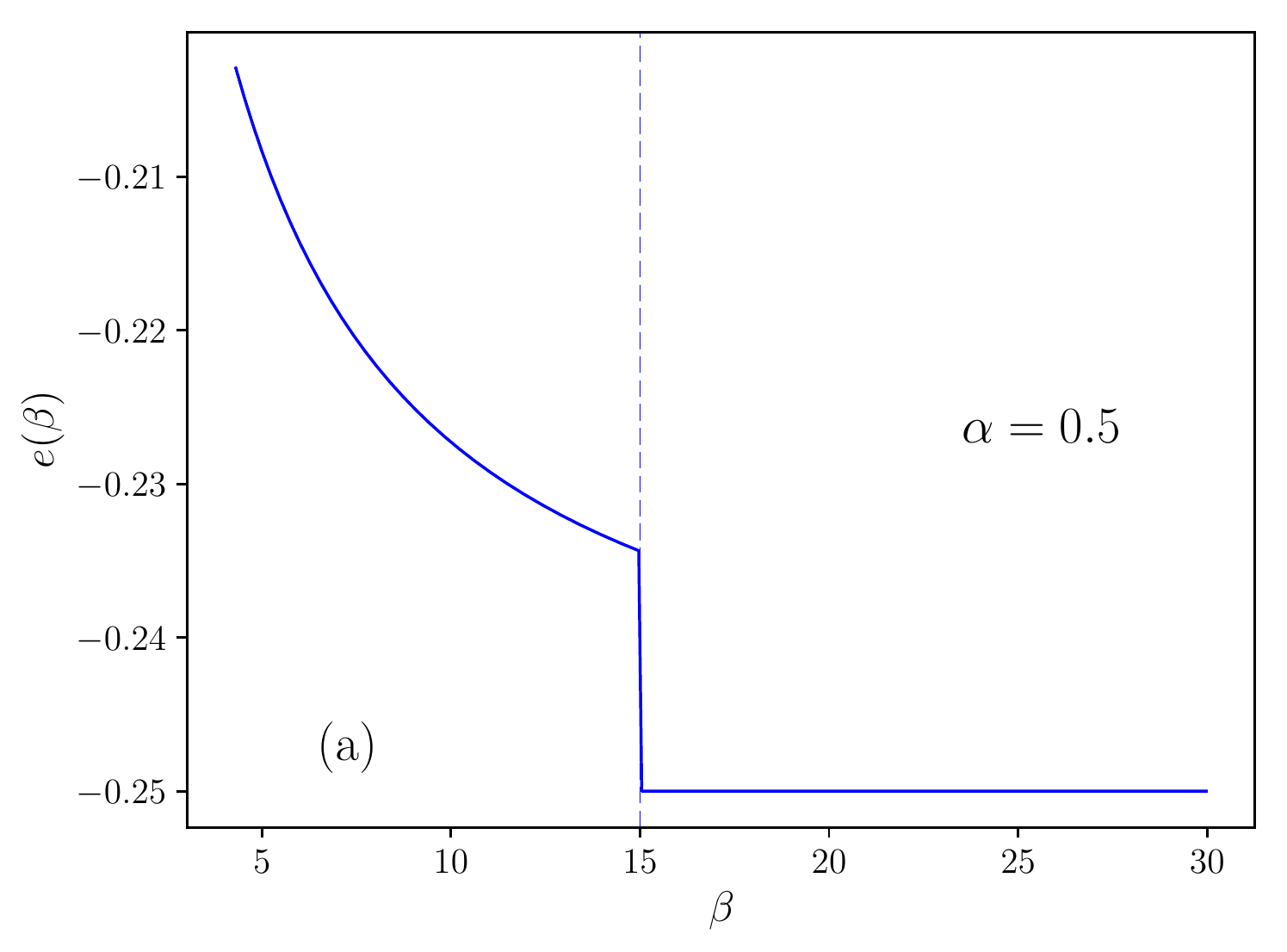}
  \includegraphics[width=0.85\columnwidth]{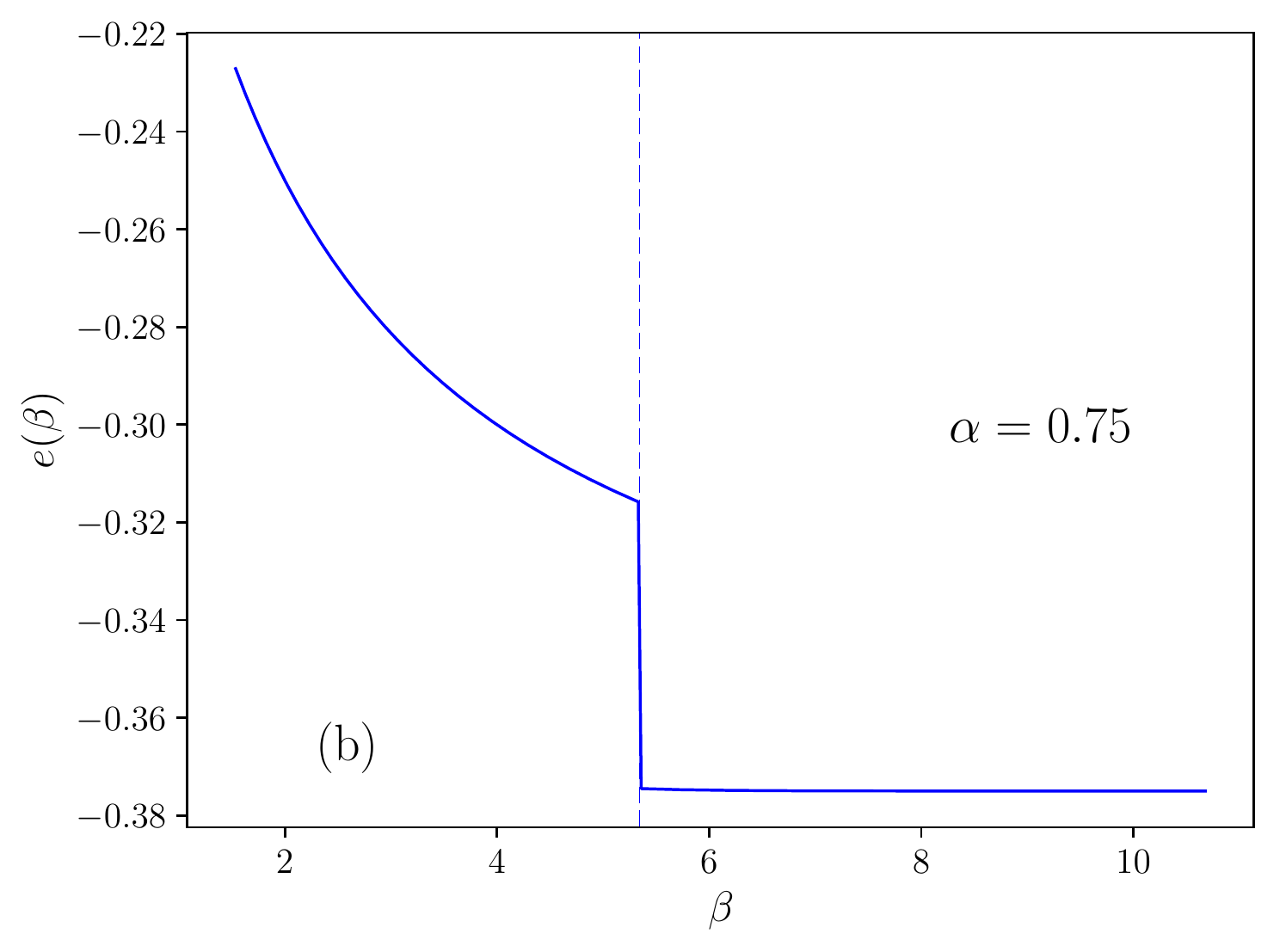}\\
  \includegraphics[width=0.85\columnwidth]{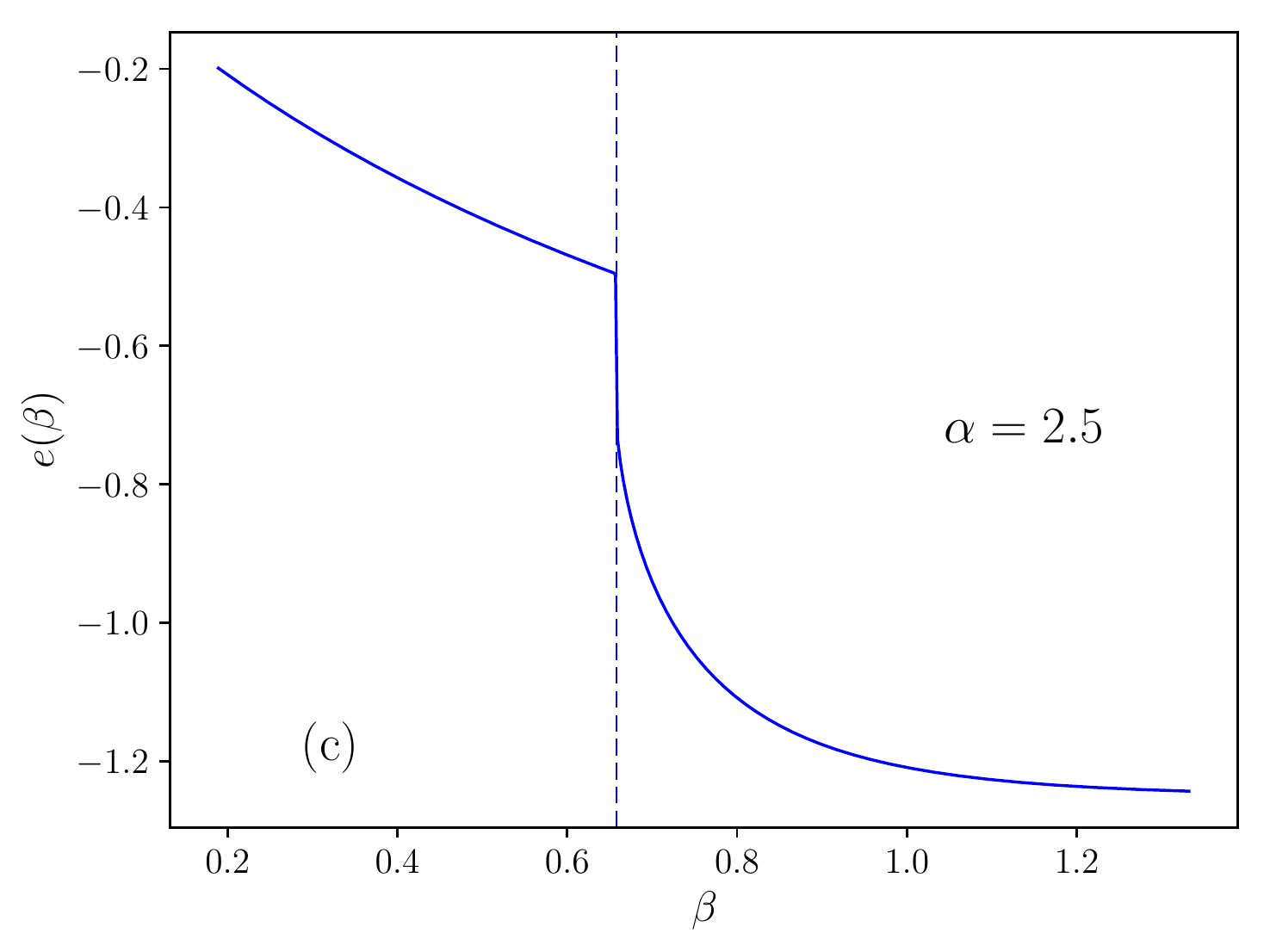}
  \includegraphics[width=0.85\columnwidth]{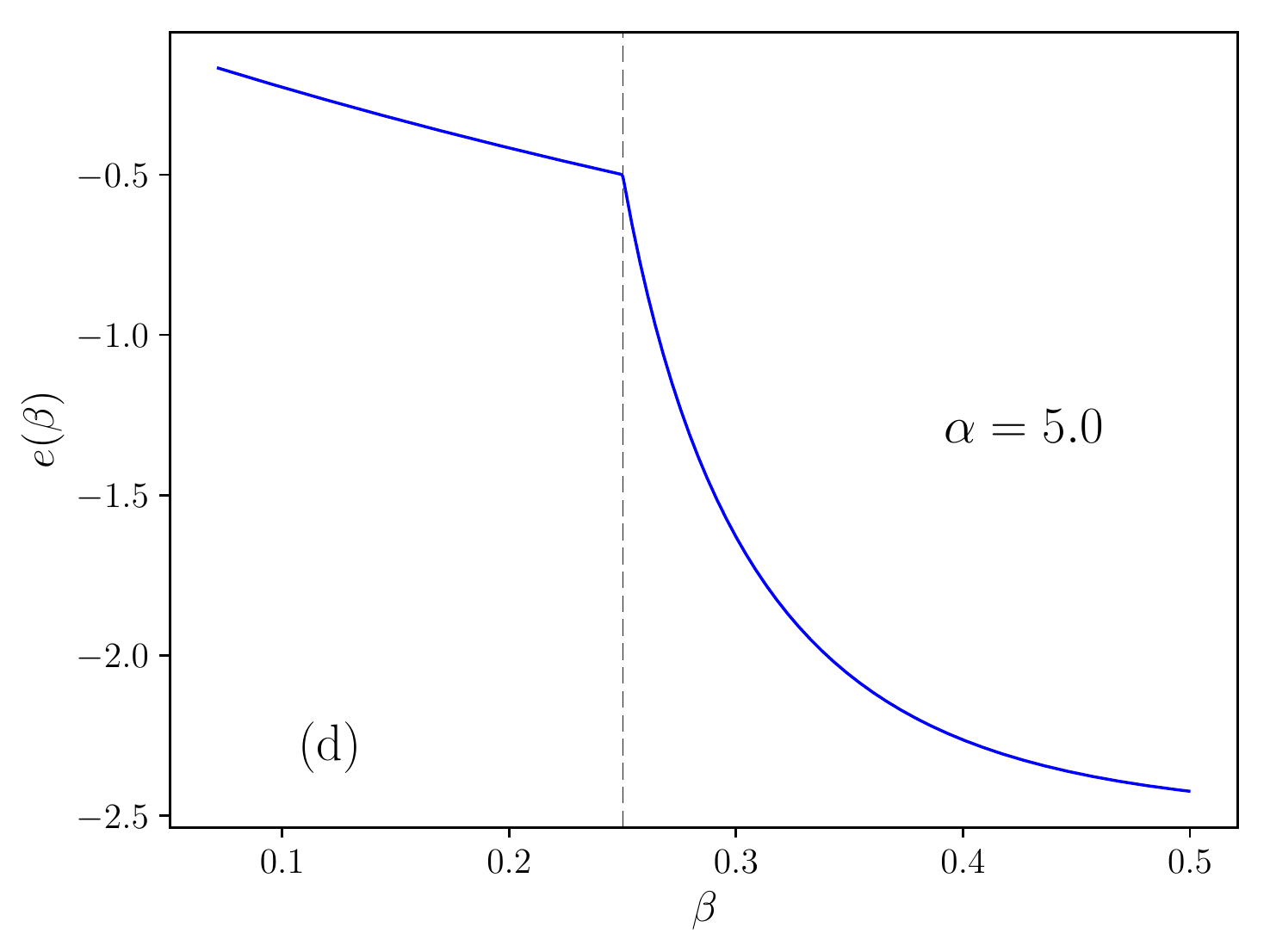}
  \caption{Internal energy densities $e(\beta)$ for the values of
    $\alpha$ used in Figs.~\ref{fig:fStarPlots} and
    \ref{fig:mStarPlots}. For
    $\alpha=0.5$ (a) and $\alpha=0.75$ (b) the
    energy density drops discontinuously from an excited value to the
    planted solution ground-state energy density of $-\alpha/2$. For
    $\alpha=5$ (d) the discontinuity appears to have
    disappeared and the internal energy drops gradually with
    temperature.}
  \label{fig:eDensPlots}
\end{figure*}

Finally, the internal energy densities are plotted in Fig.~\ref{fig:eDensPlots}. 
Once again, the energies drop discontinuously at
$\beta_c$ from some excited values to the planted solution ground
state energies (per spin) of $-\alpha/2$, which is most evident for
the smaller two values of $\alpha$, diminished but still apparent when
$\alpha=2.5$, and essentially invisible for $\alpha=5$. Note that the
energy gap between the two sides of $\beta_c$ is larger when
$\alpha=0.75$ than when $\alpha=0.5$, in accordance with its higher
transition temperature. The gap closes again for large $\alpha$ but
the energies thereafter progress continuously to their ground-state values.

The gradual reversion to a continuous transition is sensible in light
of the construction procedure; when $M$ is very large, the matrix
$\Jtildev$ converges to $-\alpha\Sigmav$, where $\Sigmav$ is the $\wv$
covariance matrix. Since $\Sigma_{ij} = -\frac{1}{N-1}$, this implies
that in the large $\alpha$ limit, the system becomes a (scaled)
 Curie-Weiss ferromagnet with a transition temperature of
$T \sim \alpha$.

\subsection{Monte Carlo simulation}
\label{sec:Thermodynamics:Simulation}

In this section we perform finite-temperature Monte Carlo simulation
of fully connected
spin glasses with couplers drawn from a Wishart distribution. To detect the existence
of a finite-temperature phase transition we measure the Binder cumulant~\cite{binder:81} 
given by
\begin{equation}
g = \frac{1}{2}
\left(
3 - \frac{[\langle m^4\rangle_T]_{\rm av}}{[\langle m^2\rangle_T]_{\rm av}^2}
\right) \; ,
\label{eq:binder}
\end{equation}
where $\langle \cdots \rangle_T$ represents a thermal average, $[\cdots]_{\rm av}$ 
represents a disorder average, and
\begin{equation}
m = \frac{1}{N}\sum_{i = 1}^N s_i
\label{eq:mag}
\end{equation}
is the magnetization per spin. Although the model is disordered, it orders into a 
ferromagnetic phase because the ferromagnetic ground state is planted by 
construction. In general, the Binder ratio scales as
\begin{equation}
g = \widetilde{G}\left(N^{1/\nu}[T - T_c] \right)  \; ,
\label{eq:g_scale}
\end{equation}
where $\widetilde{G}(.)$ is polynomial for small values of its argument,
$T$ is the temperature, and $T_c$ the critical temperature. In
general, $g(T = T_c)$ is independent of $N$ and, as such, on can
determine $T_c$ by the point where data for different $N$ cross.
However, because as we shall see the transition is first order, the
shape of the Binder ratio as a function of temperature is somewhat
different to the commonly known shape in second-order phase
transitions (see, for example, Ref.~\cite{katzgraber:06}). When a
first-order phase transition is present, the Binder ratio starts at
$g(T \to 0) \to 1$, dips into $g(T) < 0$ and then plateaus to
$g(T \to \infty) \to 0$. One can determine the critical transition
temperature by either extrapolating $g(T_{\rm min},L)$ for
$L \to \infty$, where $T_{\rm min}$ is the minimum value of $g$.
Alternatively, one can study the crossing of data for different system
sizes $N$ in the range $0 \le T \le T_{\rm min}$. The former is
usually easier to use to detect a transition, because the latter is
often close to $T = 0$ where data for different $N$ do not splay
enough to see a clean crossing. However, as shown in
Refs.~\cite{binder:92a,vollmayr:93},
$g(T_c)-g(T_{\rm min},N) \sim 1/N$, whereas the corrections to scaling
in Eq.~\eqref{eq:g_scale} decrease proportional to
$g(T_c) - g(T,N)\sim 1/N^2$. As such, here we determine the position
of the critical temperature by studying the crossing of the Binder
ratio.

To further corroborate the existence of a first-order transition to a 
ferromagnetic phase we study the distribution $P(m)$ of the magnetization
[Eq.~\eqref{eq:mag}]. Close to the transition temperature where latent
heat is present the order parameter should signal two competing phases, 
i.e., peaks at $|m|\to 1$, as well as a competing peak at $m =0$.

The simulations are done using parallel tempering Monte Carlo
~\cite{geyer:91,hukushima:96}. Thermalization is verified by ensuring
that all measured observables are independent of simulation time.  We
do this by analyzing how the results for all observables vary when the
simulation time is successively increased by factors of $2$
(logarithmic binning). We require that the last three results for all
observables agree within error bars. Simulation parameters are shown
in Table~\ref{tab:simparams}.  Error bars are determined via a
jackknife analysis over the disorder.

\begin{table}
\caption{
Parameters of the simulations. 
$N_{\rm sa}$ is the number of samples, $N_{\rm sw} = 2^b$ is the total number 
of Monte Carlo sweeps for each of the $N_T$ replicas for a single sample,
$T_{\rm min}$ [$T_{\rm max}$] is the lowest [highest] temperature simulated, 
and $N_T$ is the number of temperatures used in the parallel tempering scheme 
for each system size $N$ and $\alpha$.
\label{tab:simparams}
}
\begin{tabular*}{\columnwidth}{@{\extracolsep{\fill}} c r r r r r r }
\hline
\hline
$\alpha$ & $N$ & $N_{\rm sa}$ & $b$  & $T_{\rm min}$ & $T_{\rm max}$ & $N_{T}$  \\
\hline
$0.50$   &  $48$ &  $1000$   & $22$ & $0.0650$      & $1.40000$     & $150$ \\
$0.50$   &  $64$ &  $1000$   & $22$ & $0.0650$      & $1.40000$     & $150$ \\
$0.50$   &  $96$ &  $1000$   & $26$ & $0.0650$      & $1.40000$     & $150$ \\ \\
$0.75$   &  $48$ &  $1000$   & $21$ & $0.1150$      & $1.40000$     & $130$ \\
$0.75$   &  $64$ &  $1000$   & $21$ & $0.1150$      & $1.40000$     & $130$ \\
$0.75$   &  $96$ &  $1000$   & $21$ & $0.1150$      & $1.40000$     & $130$ \\
$0.75$   & $128$ &  $1000$   & $21$ & $0.1150$      & $1.40000$     & $130$ \\ \\
$1.00$   &  $48$ &  $1000$   & $21$ & $0.1150$      & $1.40000$     & $130$ \\
$1.00$   &  $64$ &  $1000$   & $21$ & $0.1150$      & $1.40000$     & $130$ \\
$1.00$   &  $96$ &  $1000$   & $21$ & $0.1150$      & $1.40000$     & $130$ \\
$1.00$   & $128$ &  $1000$   & $21$ & $0.1150$      & $1.40000$     & $130$ \\
$1.00$   & $192$ &  $1000$   & $21$ & $0.1150$      & $1.40000$     & $130$ \\
\hline
\hline
\end{tabular*}
\end{table}

\begin{figure*}
\includegraphics[width=0.85\columnwidth]{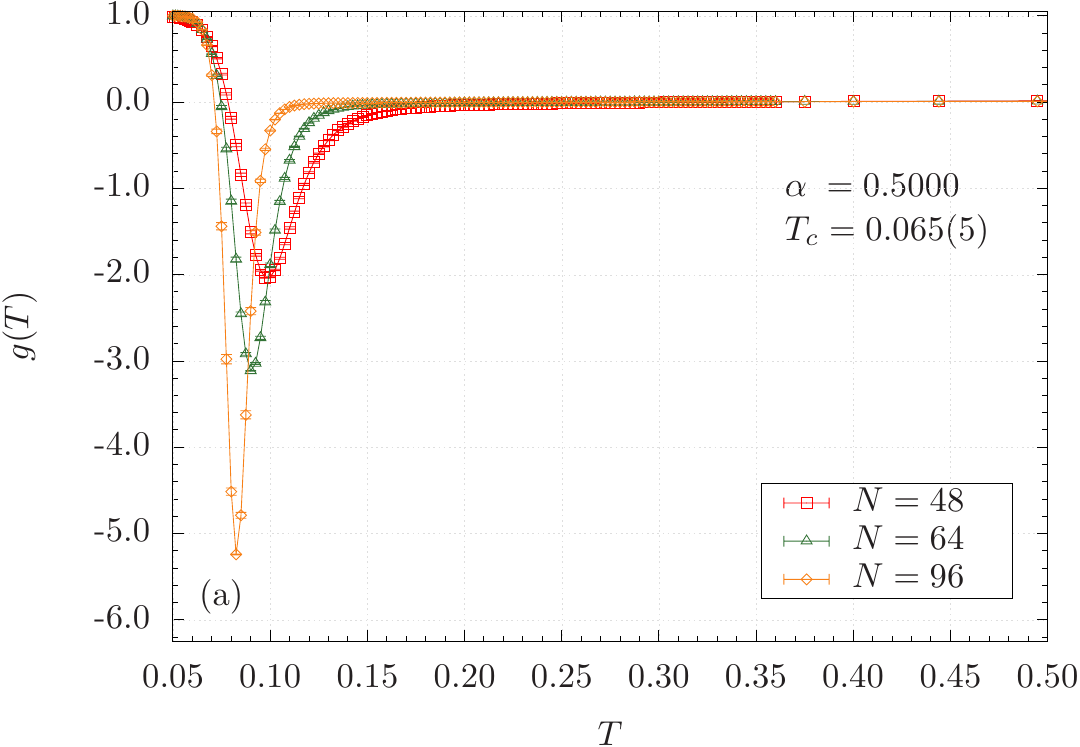}
\includegraphics[width=0.85\columnwidth]{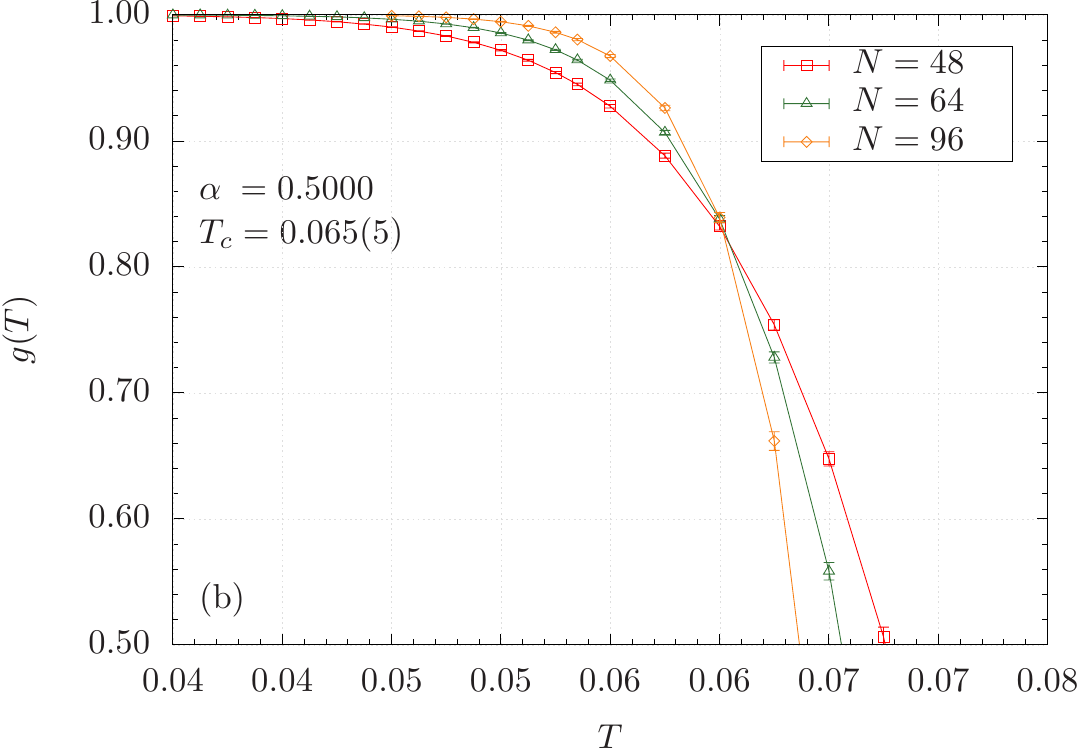}\\[2em]
\includegraphics[width=0.85\columnwidth]{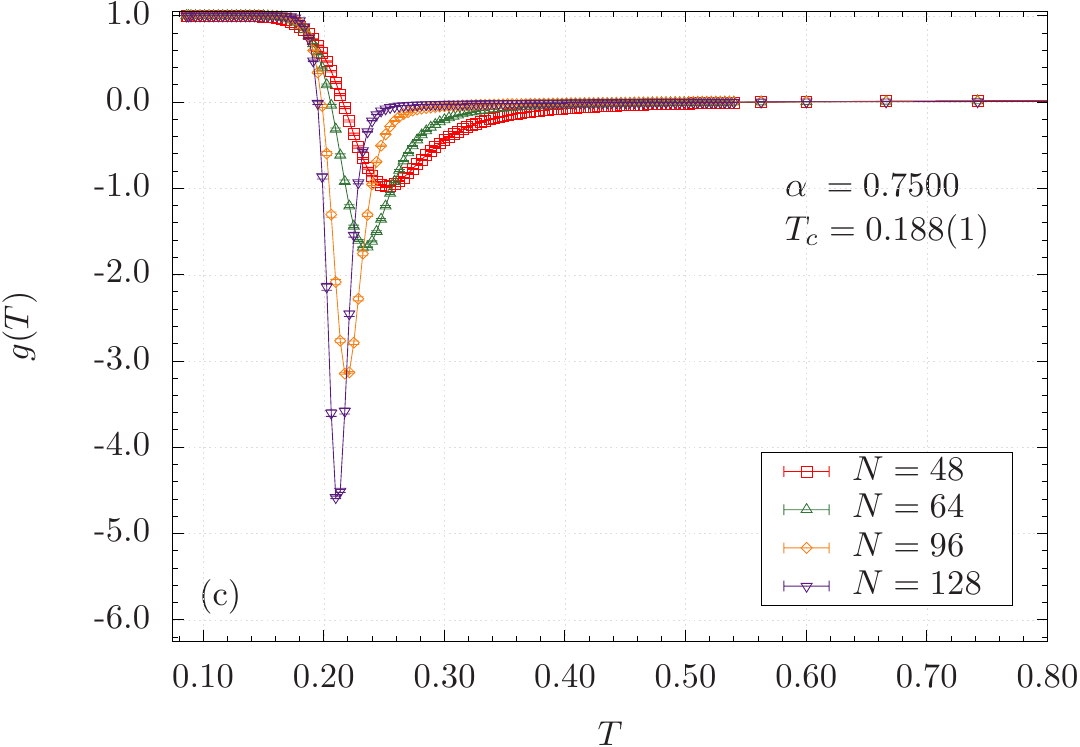}
\includegraphics[width=0.85\columnwidth]{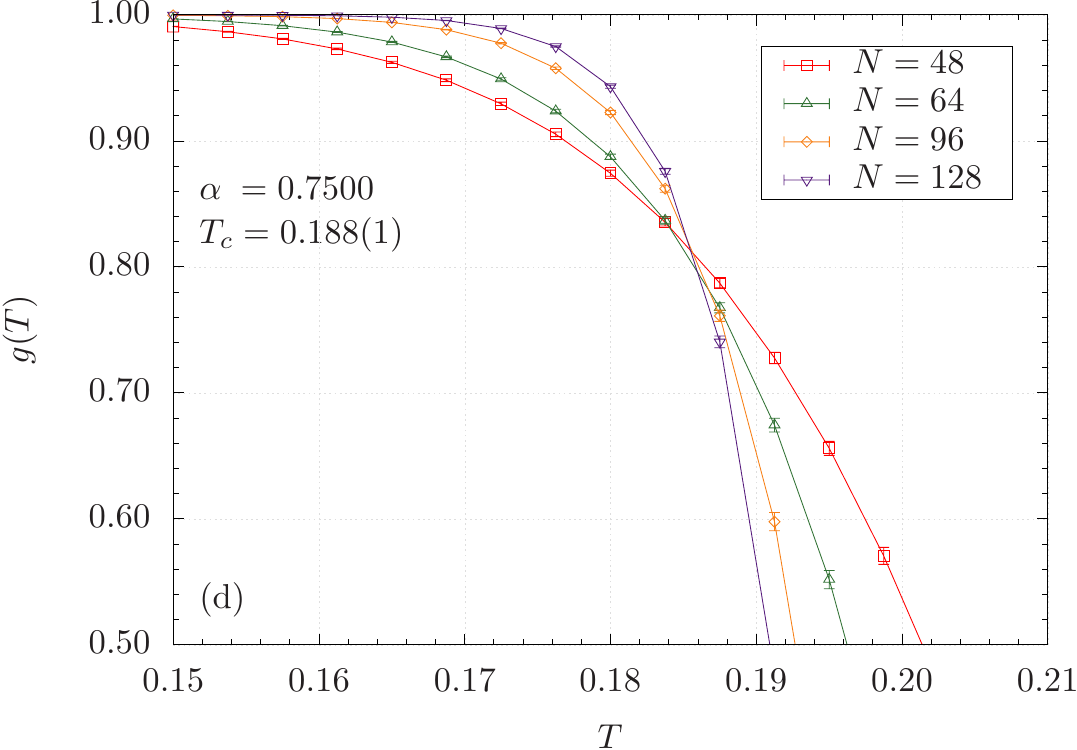}\\[2em]
\includegraphics[width=0.85\columnwidth]{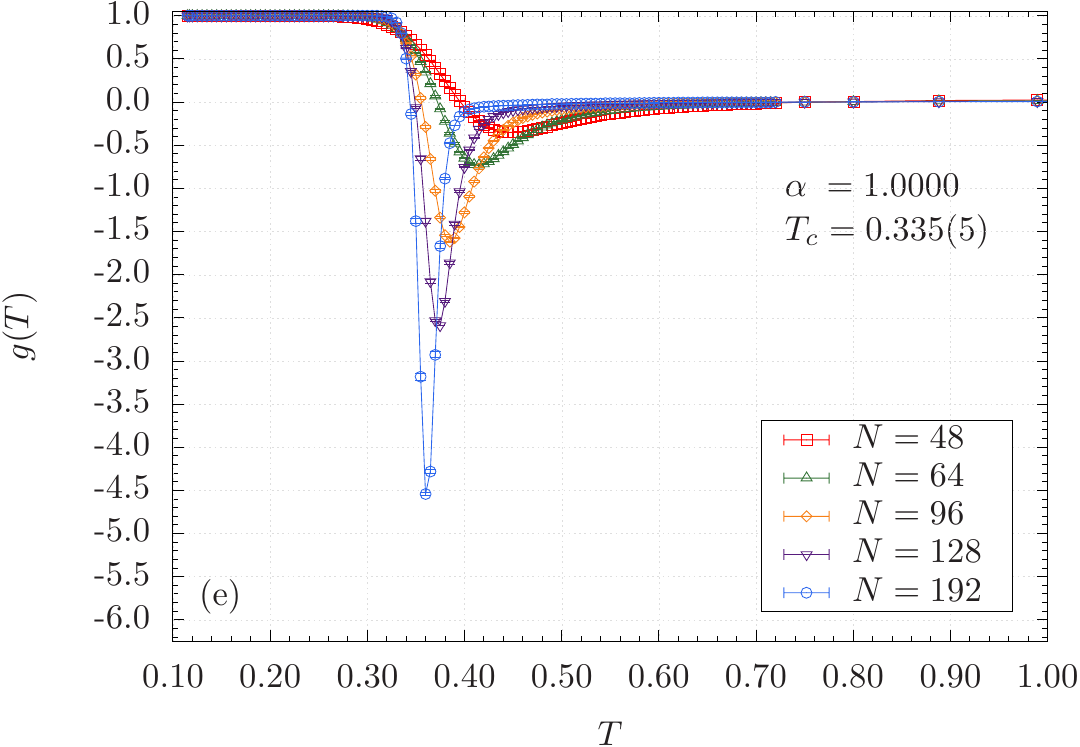}
\includegraphics[width=0.85\columnwidth]{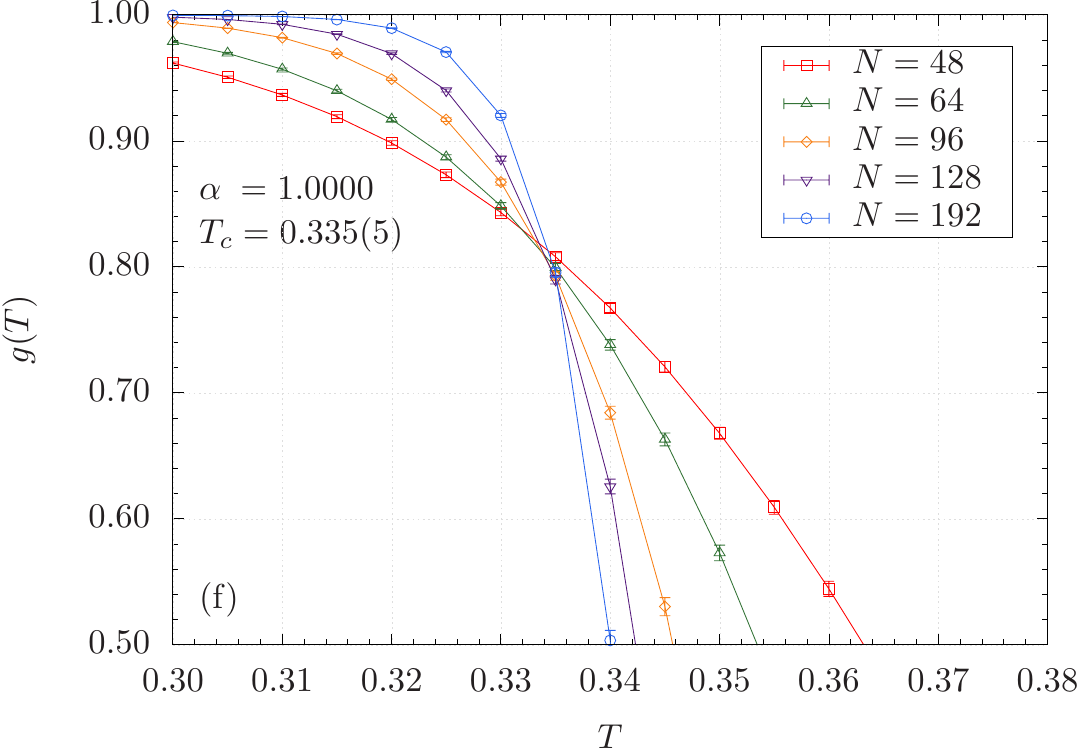}
\caption{
Binder ratio as a function of temperature for $\alpha = 0.50$, $0.75$, and $1.00$.
Panels (a), (c),  and (e) show data for the complete temperature range, whereas
panels (b), (d), and (f) zoom into the region where $T \in [0,T_{\rm min}]$. 
A first-order transition is clearly visible.
}
\label{fig:binder}
\end{figure*}

Figure \ref{fig:binder} shows data for the Binder ratio as a function of
temperature. Figures \ref{fig:binder}(a), \ref{fig:binder}(c), and \ref{fig:binder}(e)
show the Binder ratio for $\alpha = 0.50$, $0.75$, and $1.00$, respectively,
over the whole temperature range. A negative dip signaling a first-order transition
is clearly visible. Figures \ref{fig:binder}(b), \ref{fig:binder}(d), and \ref{fig:binder}(f)
zoom into $T \in [0,T_{\rm min}]$. As can be seen, the data cross for all three
$\alpha$ values. From the crossing points we estimate $T_c(\alpha = 0.50) = 0.065(5)$,
$T_c(\alpha = 0.75) = 0.188(1)$, and $T_c(\alpha = 1.00) = 0.335(5)$, in perfect
agreement with our analytical estimates. Note that $T_c(\alpha) \to 0$ for 
decreasing $\alpha$, i.e., the problems become increasingly harder numerically for smaller  
values of $\alpha > \alpha_c$.

\begin{figure}
\includegraphics[width=0.85\columnwidth]{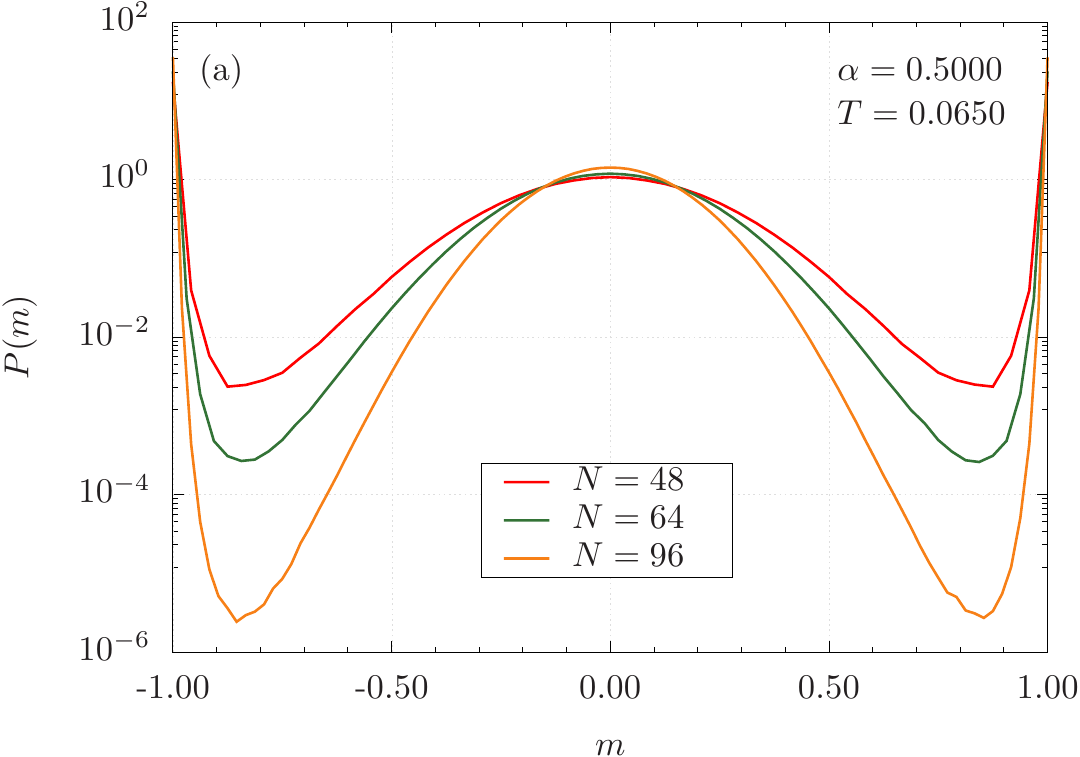}\\[2em]
\includegraphics[width=0.85\columnwidth]{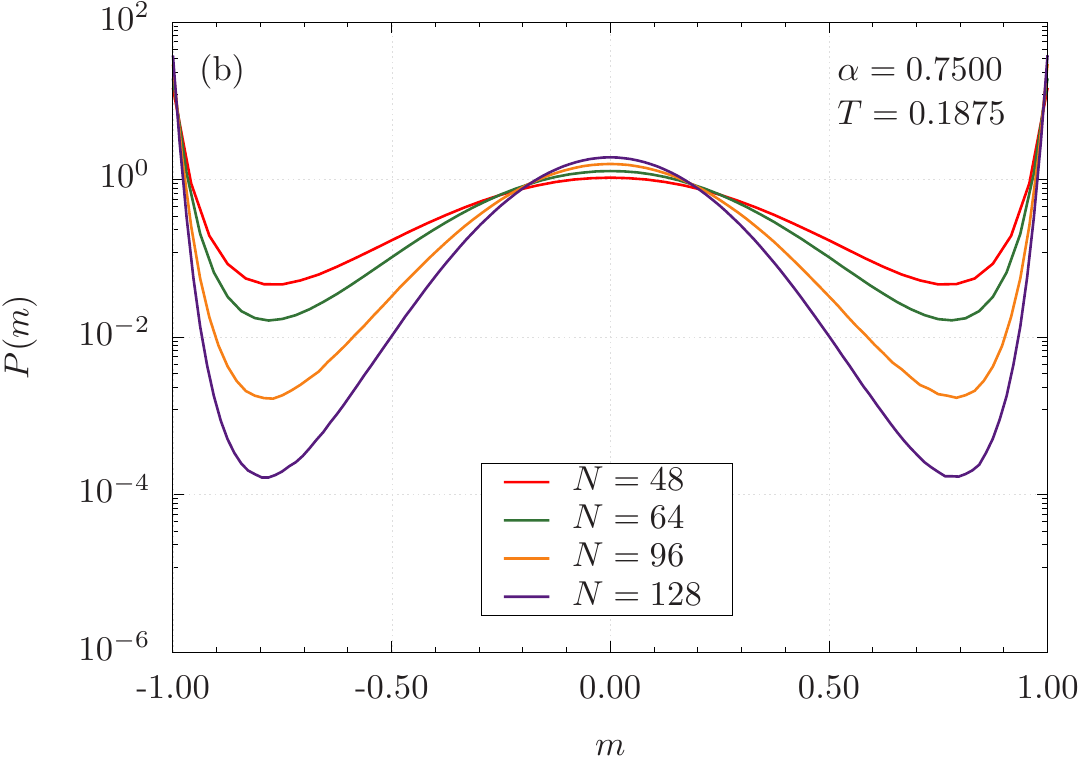}\\[2em]
\includegraphics[width=0.85\columnwidth]{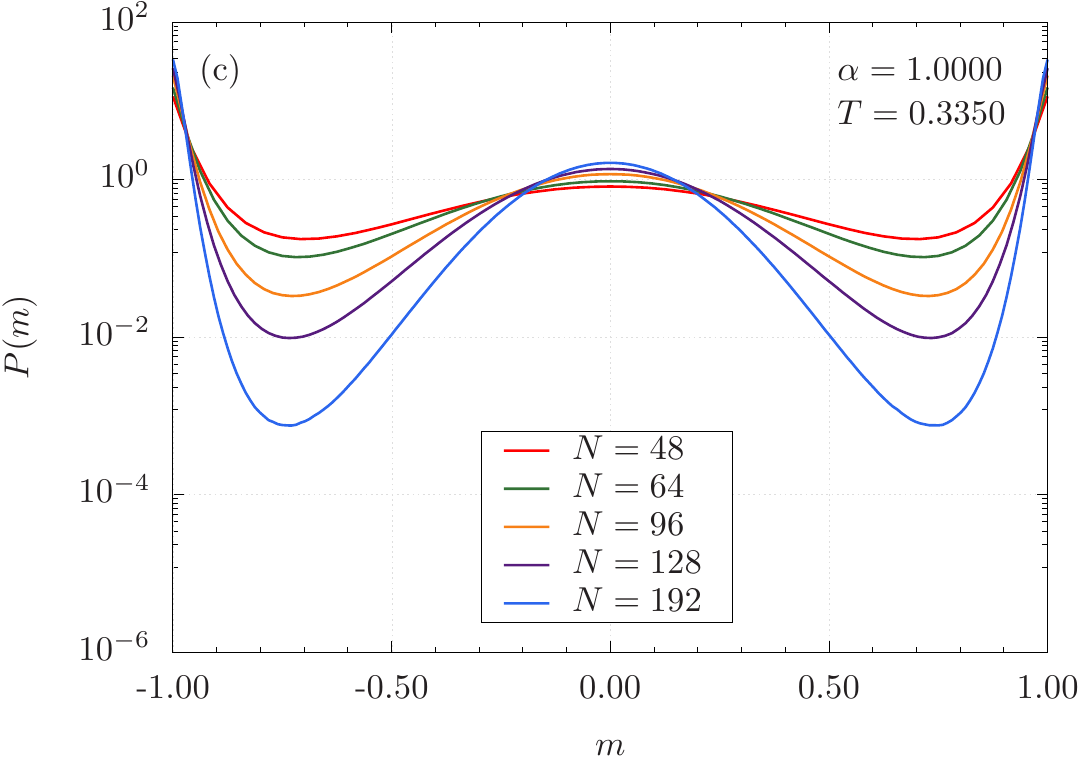}
\caption{
Order parameter distributions for  $\alpha = 0.50$, $0.75$, and $1.00$
in a linear-log scale. The data clearly show three peaks---two
close  to $|m| \to 1$ and one close to $m = 0$---for temperatures
close to the critical temperatures determined in Fig.~\ref{fig:binder}.
Close to the transition there are competing phases, i.e., a multivalue
order parameter. 
}
\label{fig:pm}
\end{figure}

In Fig.~\ref{fig:pm} data for the magnetization distribution are shown for 
$\alpha = 0.50$, $0.75$, and $1.00$ and $T \approx T_c$. In all three cases
peaks at $|m| \to 1$ are visible, signaling ferromagnetic order. However, 
a third peak for $m = 0$ grows with increasing $N$, thus signaling a jump
in the magnetization close to the transition. Note that for decreasing
$\alpha$ the first-order transition is more pronounced.

\section{Hardness phase transition}
\label{sec:hardnessPhase}

\begin{figure}
  \centering
  \includegraphics[width=1.\columnwidth]{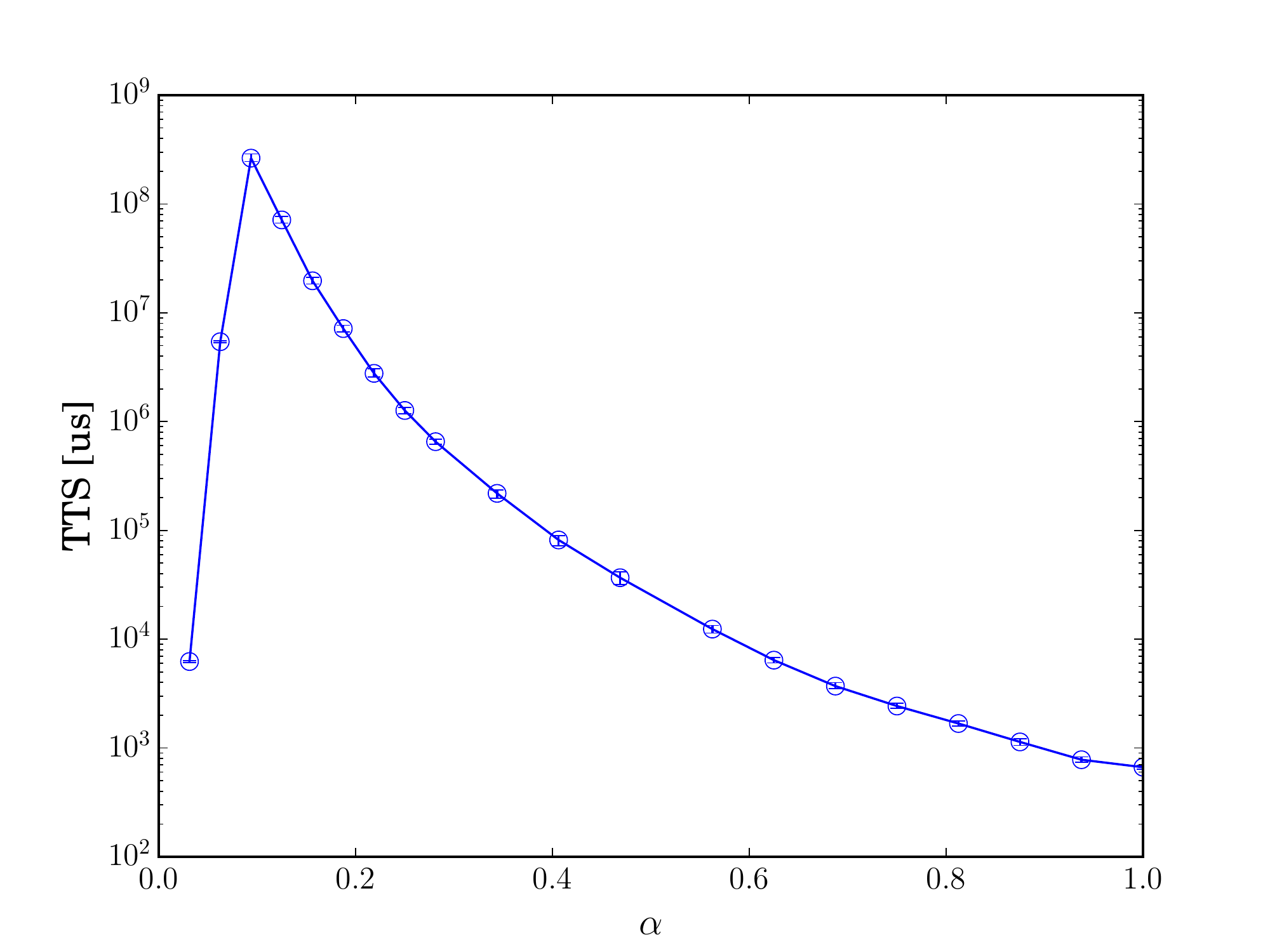}
  \caption{Optimized parallel tempering Monte Carlo time-to-solution
    for the WPE at $N=32$ for $20$ values of $1 \le M \le N$. This is a
    prototypical easy-hard-easy pattern with the peak occurring when
    $M=3$. Note that the vertical axis is logarithmic and spans many orders
    of magnitude. As discussed in Sec.~\ref{sec:hardnessPhase:hardnessTransition}, 
    a ``solution'' in this
    case is defined as finding a state whose energy lies within
    $\epsilon = 10^{-7}$ of the planted ground-state energy.}
  \label{fig:TTS_N_32}
\end{figure}

Having studied the phase behavior of the WPE and elicited properties
that would plausibly correlate with algorithmic difficulty, we now
proceed to examine some empirical results in which we probe the typical
time to find the ground state using a parallel tempering (PT)
Monte Carlo method. This algorithm is widely used to
simulate complex physical and biological systems; contemporary
implementations routinely examine spin glasses with several thousand
variables and are quite competitive as optimization algorithms as
well~\cite{moreno:03,zhu:16a}. In essence, it is a careful stochastic
local search~\cite{hoos:04} method such that configurations are
allowed to escape metastable states by traveling to high temperatures
and return afresh to probe the low-energy landscape while ensuring
asymptotically correct equilibrium sampling at each temperature.

We seek to check and account for the existence of an algorithmic
hardness peak, a sharp increase in the time required to find the
ground state as $\alpha$ is varied. Unfortunately, our numerical
studies were severely hampered by the extreme difficulty of the WPE in
its hard regime, which disallowed us from empirically localizing the
peak for sizes that would otherwise be considered modest by the
standards of other Ising ensembles (e.g., the SK model). We
systematically reduced the sizes in which PT failed, within our
computational resources, to approach the planted ground-state energy
for a range of $\alpha < 1$ values. Reliable statistics were finally
obtained for $N=32$; the plot of median time to solve the problem, or
more precisely and as justified in
Sec.~\ref{sec:hardnessPhase:hardnessTransition}, to find a state with
energy within $\epsilon=10^{-7}$ of that of the planted ground state,
is shown in Fig.~\ref{fig:TTS_N_32}. Other choices for $\epsilon$ are
certainly sensible, and in
Sec.~\ref{sec:hardnessPhase:hardnessTransition} we consider and
analyze the less stringent values of $\epsilon = 10^{-5}$ and
$\epsilon=10^{-3}$. Details of the simulation protocol are found in
Appendix \ref{sec:appendix:TTSMeasurements}.

The hardness peak is evident at $\alpha= 3/N = 3/32$, where the problems
typically took around $5$ minutes to solve. As expected, the instances
get steadily easier for the larger $\alpha$ values, but the
observation that $M=1$ is easy may initially seem surprising. As
alluded to previously, this is due to the fully frustrated nature of
$M=1$, giving rise to a tremendous number of low-energy states,
coupled with the fact that standard double-precision floating point
arithmetic was used for the computations. Consequently, the potential
exists for several states that under unbounded precision would have
had close but nonetheless distinct energies to be mapped to the same
value. Under the low-energy degeneracy associated with $M=1$, one thus
observes a huge and exponentially increasing (in $N$) number of
numerically indistinguishable ``ground states''; finding one such
acceptable state turns out to be relatively
easy~\cite{comment:groundStatePrecision}.

Yet, the number of solutions alone cannot be expected to predict
problem difficulty; in the computationally easy large-$M$ regime,
the Hamiltonian starts to look increasingly like a ferromagnet, and so is
essentially assured of having a unique solution. Thus, in contrast to
uncorrelated problems like number partitioning in which an easy-hard
transition is observed at the parameter value such that ``perfect''
solutions to the problem disappear~\cite{gent:96,mertens:06}, the correlations resulting
from the planting procedure give rise to another factor influencing
difficulty and which leads to the observed easy-hard-easy pattern.
We conjecture that hardness at a 
given $\alpha$ is determined by two competing factors:
First, the number of solutions satisfying the integer program and,
second, the \emph{intrinsic} search space size of the integer program.
It seems plausible that the regime in which the ratio of these
two quantities is minimized signals the hardness transition as this
would be analogous to having the fewest needles in the biggest
haystack. Our aim is to analytically predict the hardness peak for the
$N$-variable WPE at given $\alpha$ and numerical tolerance
$\epsilon$.

To determine the number of satisfying solutions under precision
$\epsilon$, we first derive the ensemble-averaged energy histogram for
the WPE, discussed in Sec.~\ref{sec:hardnessPhase:energyHist}. We
then formalize our notion of the intrinsic search space in Sec.~\ref{sec:hardnessPhase:IntrinsicSSSize}; as anticipated in Sec.~\ref{sec:WPE:construction}, the size of the nullspace of $\Wv^T$ plays
a key role. Finally, we make our quantitative conjecture regarding the
transition location in Sec.~\ref{sec:hardnessPhase:hardnessTransition} and show that it precisely
predicts the location of the peak in Fig.~\ref{fig:TTS_N_32}. In
Sec.~\ref{sec:hardnessPhase:localOpt}, we present some preliminary
numerically obtained features of the WPE energy landscape to complement the main analysis performed here.

\subsection{Energy histogram}
\label{sec:hardnessPhase:energyHist}

We consider an ensemble of problems over $N$ spins with parameter $M$
such that \[
\Jtildev = -\frac{1}{N}\Wv\Wv^T
\] with the columns
$\{ \wv^\mu \}$ of $\Wv$ simulated as $\Ncal(\ZeroVec, \Sigmav)$ and
Hamiltonian
\[
H(\sv) = -\frac{1}{2}\sv^T\Jtildev\sv
\]
for $\sv \in \Scube^N$. Note that the diagonal elements are included
in this formulation as it is presently more convenient to take the
ground-state energy to be zero. Without loss of generality, we assume
the ferromagnetic ground state is planted so that for all $\mu$
\[\sum_{i=1}^Nw_i^\mu=0.\]
We seek the distribution of energies marginalized over the problem
ensemble
\begin{equation}
  p_E(e) \triangleq \int_{\{\wv^\mu\}} p_E(e | \{\wv^\mu\} )
  f(\wv^1,\ldots,\wv^M) \dd \{\wv^\mu\} ,
  \label{eq:EMarginal}
\end{equation}
where $p_E(e|\{ \wv^\mu\} )$ is the probability of drawing by
\emph{uniform} sampling a state with energy $E=e$ from the given problem.

\begin{figure}
  \centering
  \includegraphics[width=1.0\columnwidth]{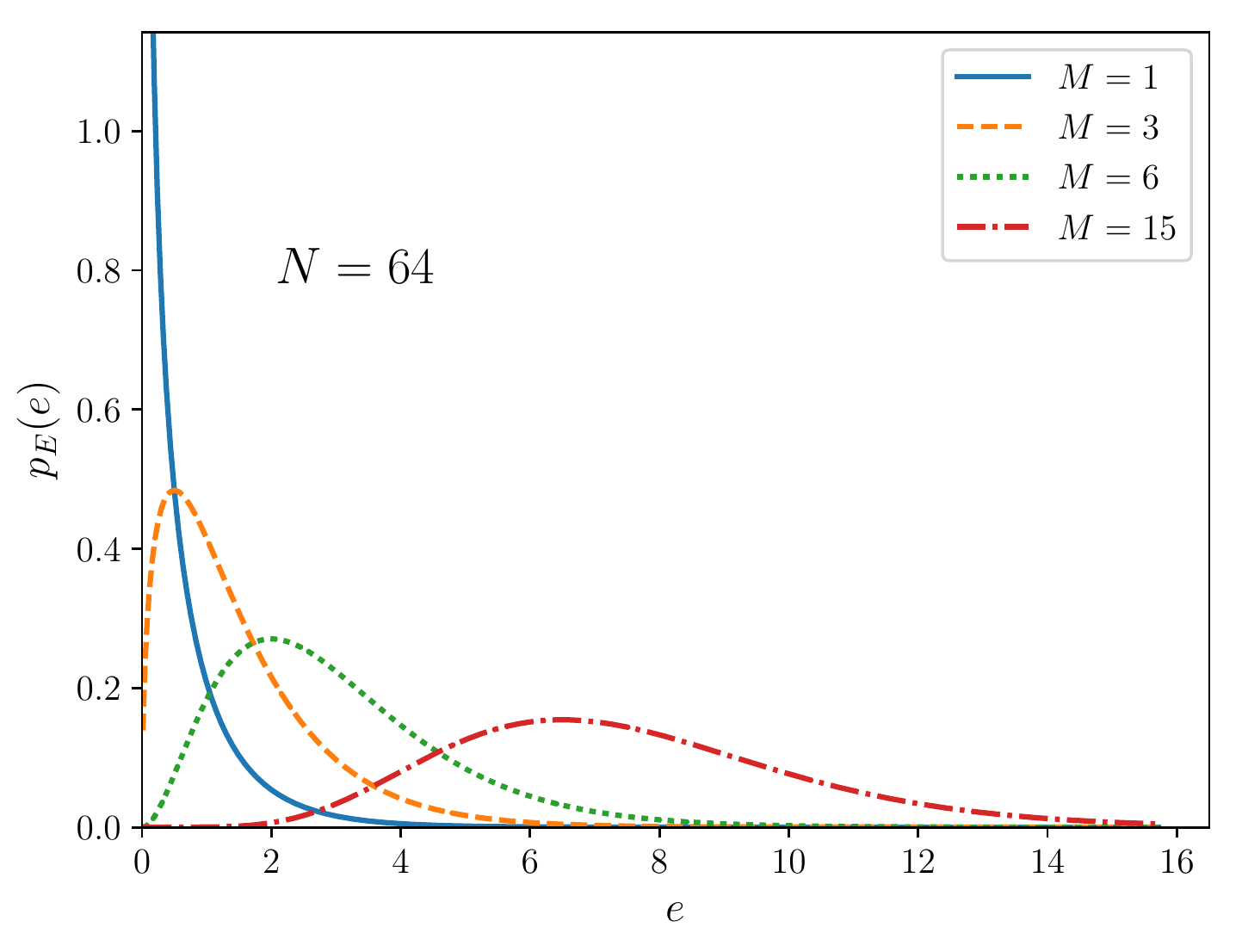}
  \caption{The energy distribution $p_E(e)$ of a WPE instance of size
    $N=64$ at different values of $M$. When $M=1$ (solid blue line),
    the density is
    overwhelmingly 
    concentrated on values very close to the ground-state energy of
    $E=0$. The energy mean and standard deviation (fluctuations) are $M/2$
    and $\sqrt{M/2}$, respectively; when $M=15$ (dash-dotted red line) there is
  vanishing probability of drawing a low-energy state by random
  chance.}
  \label{fig:pEsN_64}
\end{figure}

The result is surprisingly simple; the energies follow a gamma
distribution with density
\begin{equation}
  p_E(e) = \left \{
    \begin{array}{ll}
      \frac{1}{\Gamma(M/2)}e^{M/2-1}\exp( -e ) & \textrm{for $e \geq
                                                 0$}\\
      0 & \textrm{for $e < 0$} 
    \end{array}
  \right.
  \label{eq:EPDF}
\end{equation}
and cumulative distribution $P_E(e) \triangleq \Pr( E\leq e)$
\begin{equation}
  P_E(e) = \left \{
    \begin{array}{ll}
      \frac{1}{\Gamma(M/2)}  \gamma\Big(\frac{M}{2},e \Big) & \textrm{for $e \geq
                                                      0$}\\
      0 & \textrm{for $e < 0$}
    \end{array}
  \right. ,
  \label{eq:ECDF}
\end{equation}
where $\Gamma(x)$ and $\gamma(x,y)$ are the gamma and incomplete gamma
functions~\cite{abramowitz:64}, respectively. The calculation of the
distribution is shown in Appendix \ref{sec:appendix:energyHist}. It is
quite similar to Mertens'~\cite{mertens:01} analysis of the number
partitioning problem cost density, but here the energy is a more
complicated quadratic function with a correlated coupler matrix rather
than a rank-1 absolute value discrepancy over independent
coefficients. Figure \ref{fig:pEsN_64} displays $p_E(e)$ for the WPE
when $N=64$ for a few values of $M$. When $M=1$, the density is
overwhelmingly concentrated on low-energy values; combined with
bounded precision, this degeneracy is responsible for the problems
being easy. By the properties of the gamma distribution, the mean and
standard deviation of $E$ are $M/2$ and $\sqrt{M/2}$ respectively;
consequently (nonplanted) low-energy states become exponentially less
likely as $M$ increases.

If the coupler matrix $\Jv$ rather than $\Jtildev$ were used, the
density would simply be translated by the ground-state energy of
$-\alpha N/2$, i.e., 
\[
p_E(e) \gets p_E(e + \alpha N/2 ) .
\]

We point out that Eq.~(\ref{eq:EPDF}) in fact describes the
energy density of non-planted states, which as is shown in the Appendix,
is dominated by paramagnetic states ($m=0$) far from the planted
solution. In fact, the true density should be regarded as a mixture of
a $\delta$ function at zero energy with weight $2^{-(N-1)}$, representing the
probability of sampling the planted solution, and of the gamma
distribution (\ref{eq:EPDF}) with weight $1-2^{-(N-1)}$.

The distribution enables evaluation of the ensemble-averaged number of
states whose energies are at most $\epsilon$ and which are not related
by a global spin flip~\cite{comment:spinFlip}, which turns out to be
highly relevant to computational properties. This can be shown to be
\begin{align}
  \Expect{}\Big[ \#[E \leq\epsilon] \Big] = & 1 +
                                                 (2^{N-1}-1)P_E(\epsilon)
                                                 \nonumber \\
  =& 1 + (2^{N-1}-1)
     \frac{1}{\Gamma(M/2)}
     \gamma\Big(
     \frac{M}{2}, \epsilon
     \Big)
     \label{eq:expectNumberSolns}
\end{align}
The leading ``1'' is due to the persistence of the planted solution;
its presence in the expectation may initially seem strange and for
small values of $\alpha$ it is indeed irrelevant, but as $\alpha$
increases, the paramagnetically contributed solutions start to vanish
and its influence in fact becomes dominant.

\subsection{Intrinsic search space cardinality}
\label{sec:hardnessPhase:IntrinsicSSSize} 

We have speculated that
because we aim to solve an Ising-constrained linear program
or, equivalently, to solve for a state $\sv^* \in \Scube^N$ lying in
the nullspace of $\Wv^T$, the dimensionality of the nullspace plays
some role in problem complexity. In this section, we clarify the
notion and extract an $M$-dependent intrinsic search space. This is
defined as a discrete set of reduced-dimensional states over which one
would need to exhaustively search to solve the problem without
reference to any information about the values of $\Wv$. In some sense,
this is analogous to the reduced complexity enjoyed by various
optimization problems on graphs of low treewidth. The
result is that when $\Wv^T$ has $M$ independent rows, one only needs
to consider $O(2^{N-M-1})$ states in brute-force search; the
subtraction by one is simply to remind that the problem is invariant
to global spin flips. We note that the following procedure is not
meant for literal implementation because we are disregarding, for example,
numerical issues that may be important in practice, but simply to show
that up to a polynomial prefactor the exhaustive search complexity is
exponential in $K$, where
\[ K \triangleq \textrm{dim}( \textrm{null} (\Wv^T) ) = N-M . \]
We remind the reader that the integer programs are always feasible
due to the planting construction. Let the vectors
\[ \Vcal \triangleq \{ \vv_1, \vv_2, \ldots, \vv_K \} \] span
$\textrm{null}(\Wv^T)$. Such a basis can be obtained in polynomial
time using, for example, the singular value decomposition of $\Wv$.
The integer program can be equivalently phrased as
\begin{align*}
  & \textrm{find } \sv^* \in \Scube^N \\
  & \textrm{such that } \Vv \xv = \sv^* \textrm{ is consistent} ,
\end{align*}
where $\Vv$ is the $N\times K$ matrix whose columns are $\{ \vv_i\}$
and $\xv \in \Reals^K$. For any state $\sv$, one can in polynomial
time either solve for satisfying coefficients $\xv$ or show that no
such coefficients exist using standard linear algebra techniques.
While a solution will eventually be found when the method is repeated 
for all possible states, we now
show that not all $\sv \in \Scube^N$ need be checked.

We first obtain a $K\times K$ matrix $\Vtildev$ composed of
a subset of $K$ independent rows from $\Vv$. Such a matrix can be
determined using for example the QR decomposition with
pivoting~\cite{golub:12}.
We apply this decomposition to the columns of $\Vv^T$:
\[
\Vv^T\vc{P} = \vc{QR} ,
\]
where $\vc{P}$ is an $N\times N$ column-permuting matrix, $Q$ is
$K\times K$ and orthogonal, and $R$ is $K\times N$ and upper-triangular. This ensures
that the first $K$ columns of $\Vv^T\vc{P}$ are independent, so we take
\[
\Vtildev = \vc{P}_K^T\Vv ,
\]
where $\vc{P}_K$ is an $N\times K$ submatrix composed of the first
$K$ columns of $\vc{P}$ and encodes which rows of $\Vv$ were selected
for $\Vtildev$. These steps are computed only once for a given
problem.

Now let $\svtilde$ be an Ising state of length $K$. The system
\[
\Vtildev \xvtilde = \svtilde
\]
with respect  to partial state $\svtilde$ always has a solution
because $\Vtildev$ is full rank. After solving
for $\xvtilde$, we verify whether this generalizes to a feasible solution
over \emph{all} Ising variables by computing
\[
\yv = \vc{P}^T\Vv \xvtilde .
\]
The first $K$ components of $\yv$ will be $\svtilde$ by construction;
if the rest are also in $\{\pm 1\}$, then $\yv$ is a feasible solution.
If not, then we choose another $\svtilde$ and repeat the process starting
with solving for $\xvtilde$. This shows that up to the polynomial
overhead involved in the various steps, one only needs to consider the
number of states in $\Scube^K$; in fact due to the spin-flip symmetry,
the intrinsic search space contains $2^{K-1}$ elements. In the case
where $M\geq N$ and so the nullspace is one dimensional, only a single
state need be checked.

\subsection{Predicting the hardness transition}
\label{sec:hardnessPhase:hardnessTransition}

We now synthesize the two preceding antithetical factors into an
expression that predicts when problems are likely to become difficult.
For illustration, we first consider the case of degenerate ground
states, in which the reasoning does not in fact make any assumptions
about the specific generative process or family from which problems
are drawn. The number of solutions not related by global flip symmetry
to a specific problem is denoted by $N_G$, a random variable whose
distribution may correspond to some parametric setting.

In the exhaustive algorithm described in the previous section, a set of no
more than $2^{N-M-1}$ partial states $\{\svtilde\}$ need be checked
for feasibility. Given a traversal order over the partial states, the
expected time to locate a solution to a problem is simply the average
number of iterations until a specific $\svtilde$ mapping to a feasible
completion is found. Noting that each partial state corresponds to at
most one solution and assuming that an adversary uniformly allocates
the solutions among the partial states, by searching among
$\{\svtilde\}$ in arbitrary order, the solution time can be identified
with the number of draws of a partial state (without replacement) from
a population of $2^{N-M-1}$ such states, $N_G$ of which are solutions,
until a single solution is found. This quantity follows a specific
type of negative hypergeometric distribution, where for convenience we
obtain the inverse of its mean $\tau$ for our particular values as
\begin{equation}
  \frac{1}{\tau} = \frac{N_G + 1}{2^{N-M-1}-N_G} \approx \frac{N_G}{2^{N-M-1}}
  \label{eq:hypergeomMean}
\end{equation}
When averaged over all problems in the ensemble, the expected inverse solution
time is thus approximately
\begin{equation}
  \Expect{} \Big[ \frac{1}{\tau} \Big] =
  \frac{\Expect{}\Big[N_G\Big]}{2^{N-M-1}}
  \label{eq:ExpectInvSolTime}
\end{equation}
Equation~(\ref{eq:ExpectInvSolTime}) makes explicit the competing effects
of the number of solutions (numerator) and search space size
(denominator) on problem difficulty.

Returning our attention to the WPE, we presently determine how to
compute the expected number of solutions, or equivalently the mean
number of ground states in the Ising formulation, defined as
\[
\Expect{} \Big[ \#[E=0] \Big]
\]
where the degeneracy results from the
discreteness of the generator variables $\{\zv^\mu\}$. We recall from
Sec.~\ref{sec:WPE:representing} that in the WPE, all independent and
standardized $\{\zv^\mu\}$ yield the same limiting properties as those
resulting from using a Gaussian and presented in this work.
Consequently, a good estimate for the ground-state degeneracy can be
obtained for sufficiently large systems using the expected count in
Eq.~(\ref{eq:expectNumberSolns}). For example the number of ground
states in the Rademacher-discretized WPE presented in
Sec.~\ref{sec:WPE:representing} can be approximated by
\begin{align}
  \Expect{} \Big[ \#[E=0]\Big] = 1 + (2^{N-1}-1)
     \frac{1}{\Gamma(M/2)}
     \gamma\Big(
     \frac{M}{2}, \epsilon_1
     \Big)
\end{align}
where $\epsilon_1 = O(1/N^3)$ represents the smallest attainable
excited-state energy for the problem size.

As mentioned, the computational expense associated with simulating the
WPE in its hard regime forced us to simulate systems of relatively
small size. Such a restriction can, however, potentially introduce
finite-size effects; in other words, the properties of the discretized
systems may be far from those predicted by their Gaussian asymptotics.
To minimize such artifacts, we opted to sample the parameters using
Gaussian generators represented with the maximum available precision
and consequently, to contend with numerical errors. The task was thus
translated to that of approximate solution.

Bounded precision means we cannot represent all the ``true'' values of
the Gaussian variables $\{\zv^\mu\}$ used to generate $\Wv$ nor can we
in general exactly represent the values of $\Wv$ resulting from the
planting procedure using the rounded $\{\zv\}$ variables or even the
exact energy of a state relative to the rounded $\Wv$. Rounding and
discretization cause errors in the representation of some energies; in
particular, states that had distinct energies under arbitrary
precision may be mapped to numerically indistinguishable values, and
the relative ordering among closely spaced true energies may change.
Let $\Ehat$ be the random variable resulting from finite-precision
representation of some true continuous-valued energy $E$; its
corresponding CDF
$P_{\Ehat}(\epsilon) \triangleq \Pr(\Ehat\leq \epsilon)$ now has step
discontinuities. If we interpret the mapping of a state's exact energy
$E$ to its representable approximation $\Ehat$ as a pseudo-random walk
process leaving invariant the probability of having energy of at most
$\epsilon$, then provided $\epsilon$ is large enough to be
representable we can approximate the finite-precision energy CDF with
the continuous distribution introduced in
Sec.~\ref{sec:hardnessPhase:energyHist}:
\[ \Pr( \Ehat \leq \epsilon ) \approx \int_{0}^\epsilon p_E(e)\dd e =
P_E(\epsilon), \] where the latter equality follows from the
nonnegativity of $E$. We can then approximate the expected number of
states with observed energy of at most $\epsilon$ using
Eq.~(\ref{eq:expectNumberSolns}):
\begin{align}
  \Expect{}\Big[ \#[\Ehat\leq\epsilon] \Big] = 1 + (2^{N-1}-1)
     \frac{1}{\Gamma(M/2)}
     \gamma\Big(
     \frac{M}{2}, \epsilon
     \Big) 
\end{align}

The value of $\epsilon$ is in principle arbitrary but must be large
enough to reflect the approximation error of the planted ground-state
energy. We observe that under the aggregated discretization effects of
the construction procedure, the true planted ground-state energy of
zero is typically distorted to a value on the order of the so-called
machine epsilon $\approx 10^{-16}$ for double-precision arithmetic. A
sensible strategy is thus to take $\epsilon$ to be some value not much
larger than this, which would conservatively ensure that no planted
ground states are missed while accepting a certain number of ``false
positives,'' i.e., that some nonplanted states may be considered
successful solutions. Excessively small values of $\epsilon$, however,
resulted in prohibitively long computational times for the problems in
the hard regimes. Our methodology to analyze the empirical solution
time settled on using three target energy values:
$\epsilon \in \{10^{-7}, 10^{-5}, 10^{-3}\}$.
We hence strive to yield accurate predictions of the hardness
transition for the important task of approximate (or relaxed)
solution to the problem.

\begin{figure*}
  \centering
  \includegraphics[width=0.85\columnwidth]{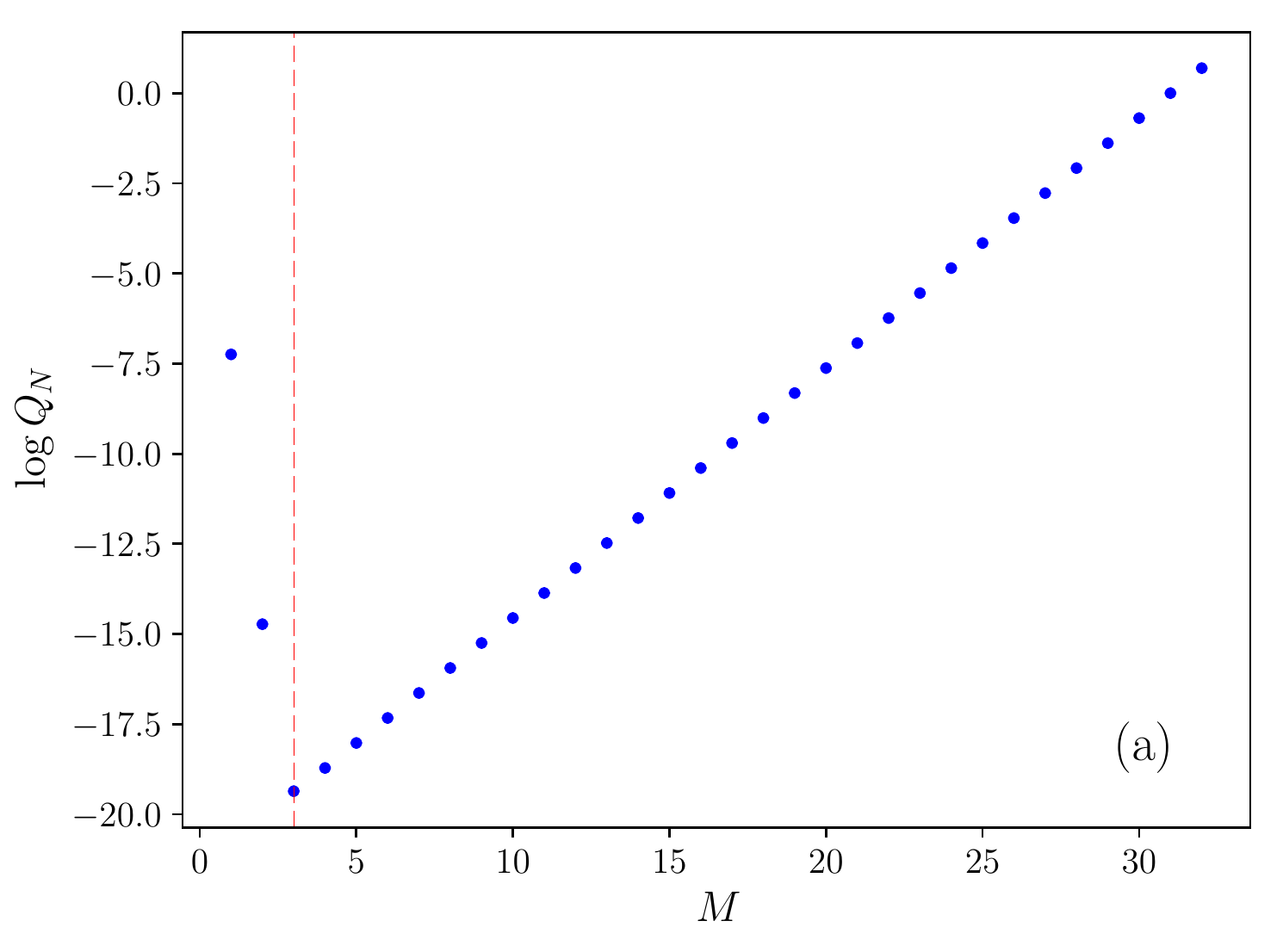}
  \hspace{0.5cm}
  \includegraphics[width=0.85\columnwidth]{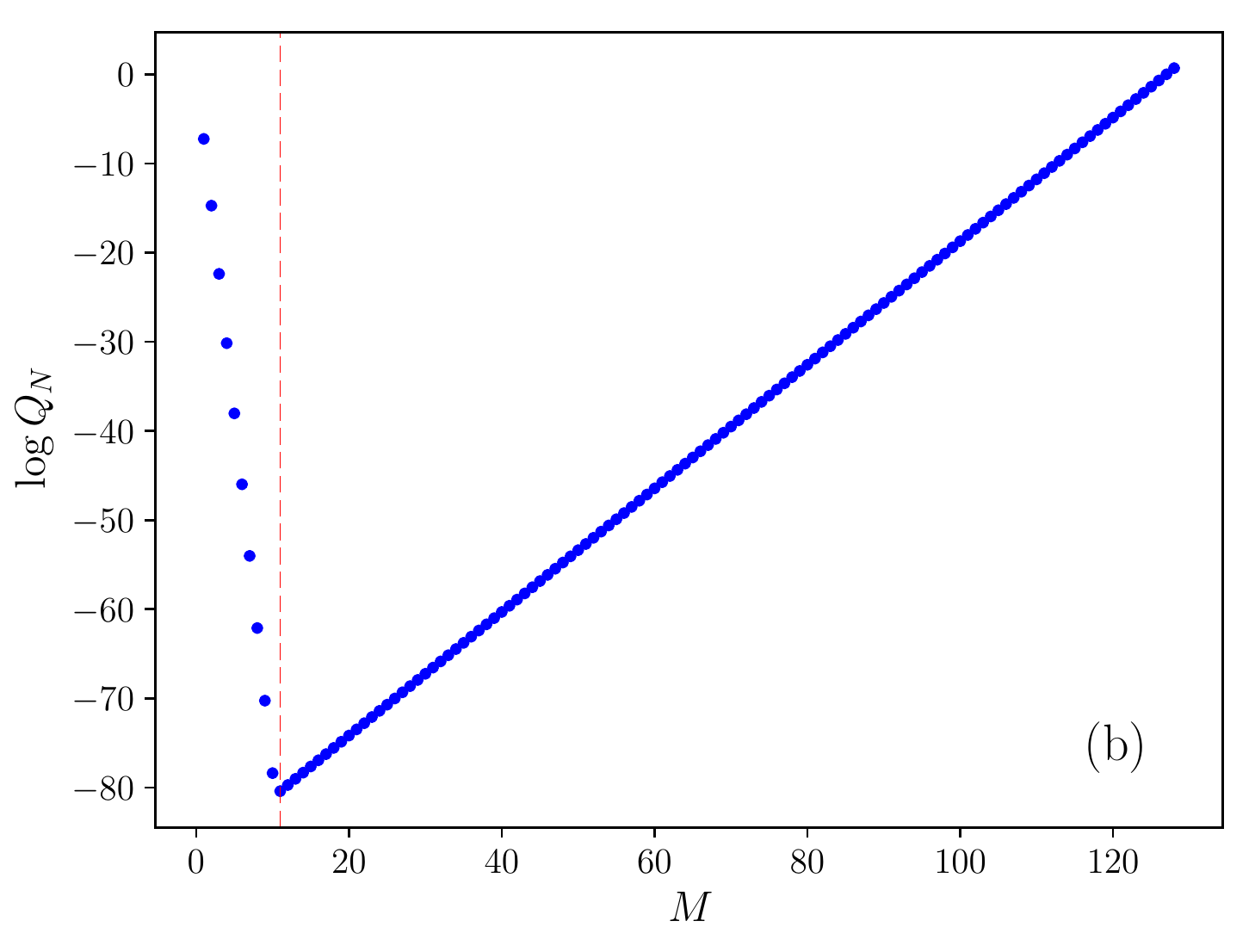}
  \caption{ $\log Q_N(M)$ whose minimum, denoted by the dashed red line,
    is conjectured to predict the WPE
    hardness transition at precision $\epsilon=10^{-7}$
    for sizes $N=32$ (left) and $N=128$ (right). For $N=32$, the minimizer occurring at
    $M=3$ coincides with the maximum parallel tempering
    time-to-solution observed in Fig.~\ref{fig:TTS_N_32}. When
    $N=128$, maximum difficulty is predicted to occur when $M=11$,
    suggesting that at constant precision, generating the hardest problems
    requires scaling $M$ with $N$.}
  \label{fig:logQNs}
\end{figure*}

Finally, we define the function trading off number of solutions to
effective search space size \begin{equation}
  Q_N(M) \triangleq \frac{\Expect{}\big[ \#[\Ehat\leq\epsilon] \big]}
  {2^{N-M-1}}
  \label{eq:QNM}
\end{equation}
and anticipate that problems become most difficult at
\[
M^* = \argmin\limits_{M\in \{1,\ldots,N\}} Q_N(M)
\]
as this coincides with the fewest number of solutions relative to the
effectively constrained search space. In Fig.~\ref{fig:logQNs}, we
plot $\log Q_N(M)$ for system sizes $N=32$ and $N=128$, when
$\epsilon \approx 10^{-7}$ in accordance with our most stringent
target. When $N=32$, $M^*=3$ in perfect accordance with the location
of the hardness peak shown in Fig.~\ref{fig:TTS_N_32}. The latter size
was far too large to solve in the hard regime within our constraints,
but the plot suggests that to obtain the most difficult problems for
this larger size, $M$ needs to increase. It is instructive to predict
the required scaling of $M$ with $N$ for maximally hard problems at
this level of precision.

\begin{figure}
  \centering
  \includegraphics[width=1.\columnwidth]{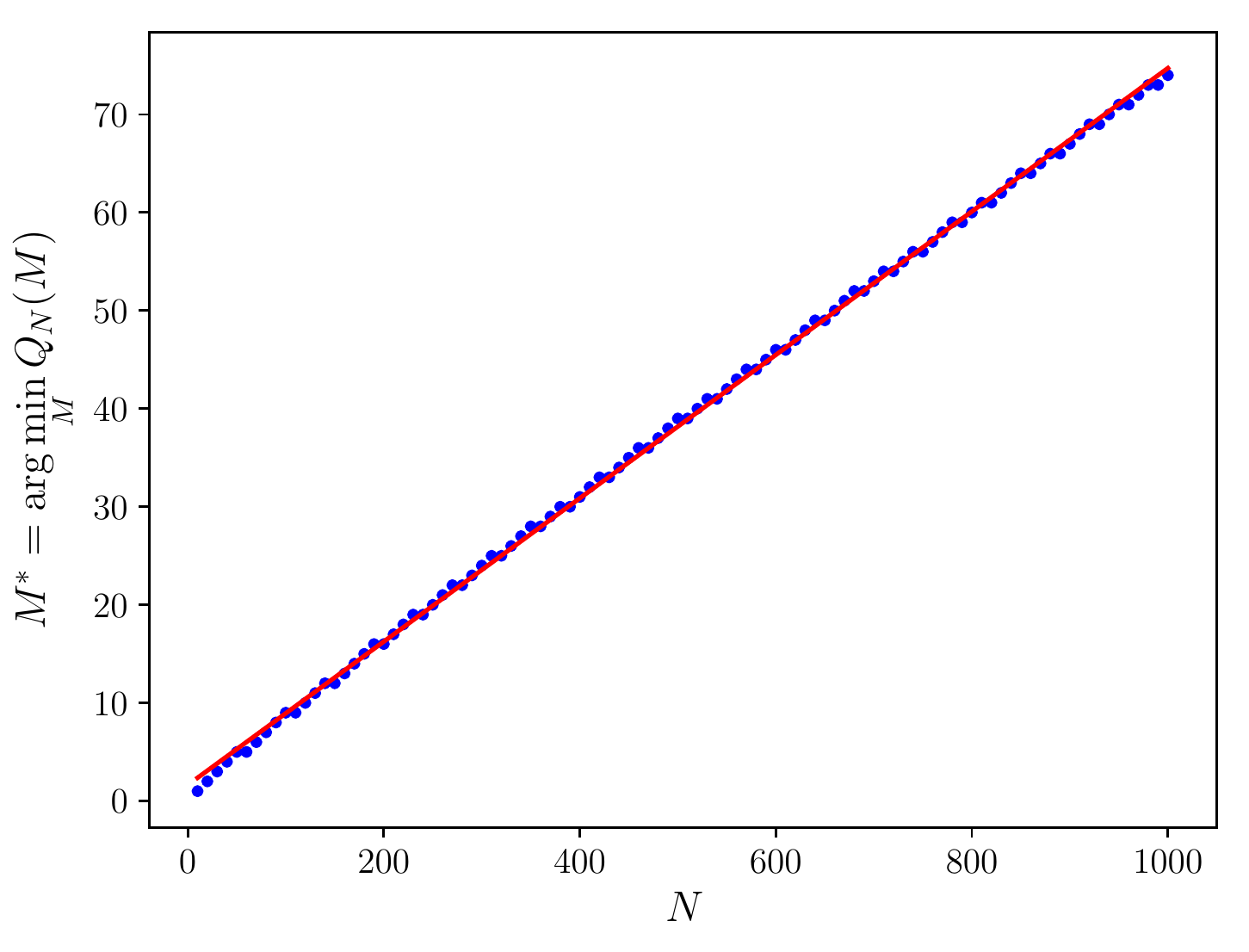}
  \caption{Predicted scaling of $M$ with $N$ to maximize WPE hardness
    at precision $\epsilon = 10^{-7}$. True values of
    $M^* = \argmin\limits_{M} Q_N(M)$ (blue points) clearly follow
    a linear relation. Regression approximates this relation as $M^*
    \approx 1.63 + 0.073N$ (red line), suggesting that the hardest
    problems occur for this precision restriction at $\alpha\approx 0.073$.
  }
  \label{fig:minQNScaling}
\end{figure}

Figure \ref{fig:minQNScaling} shows $M^*$ as a function of $N$, where
a clear linear scaling is apparent. By linear regression, we find this
relation to be
\[
M^* \approx 1.63 + 0.073N ,
\]
suggesting that the hardest problems at $\epsilon = 10^{-7}$ occur
when $\alpha \approx 0.073$. The general message is clear: To have
truly difficult problems under precision constraints, the number of
equations $M$ in the integer program cannot be constant.

The reader may notice that the definition of $Q_N$ resembles that of
the mean inverse solution time (\ref{eq:ExpectInvSolTime}) with
respect to the exact partial state traversal algorithm. Directly
relating $Q_N$, which considers all energies in a specified range, to
the solution time of an exact solver is not straightforward however.
The issue is that a partial state $\svtilde$ may now have several
completions to full Ising states whose energies match the target but
which are not ground states. While we have shown that finding the
ground-state completion of a given $\svtilde$ or verifying that no such
completion exists is straightforward, locating a completion guaranteed
to match a generic target $\epsilon>0$ is nontrivial. We may
nonetheless heuristically justify $Q_N$ as follows. Suppose that
fixing the $N-M$ partial state variables $\svtilde$ sufficiently
constrains the remaining free variables that locating the extension of
minimum energy can typically be done rapidly and with high probability
with a heuristic algorithm. An ``exhaustive-approximate'' algorithm
for finding a state whose energy is bounded by $\epsilon$ can thus
proceed as follows: Traverse the $\{\svtilde\}$ in some order, where
for each $\svtilde$, the minimizing extension is heuristically
obtained, and stop if a resultant state's energy is less than
$\epsilon$. Under the assumption that the target states are uniformly
distributed among the partial state ``bins,'' $Q_N$ can once again
have an interpretation as an expected inverse solution time. To
justify our conjecture, we now show that $Q_N(M)$ can serve to
localize the hardness peak for generic values of
 $\epsilon$.

\begin{figure}
  \centering
  \includegraphics[width=1.\columnwidth]{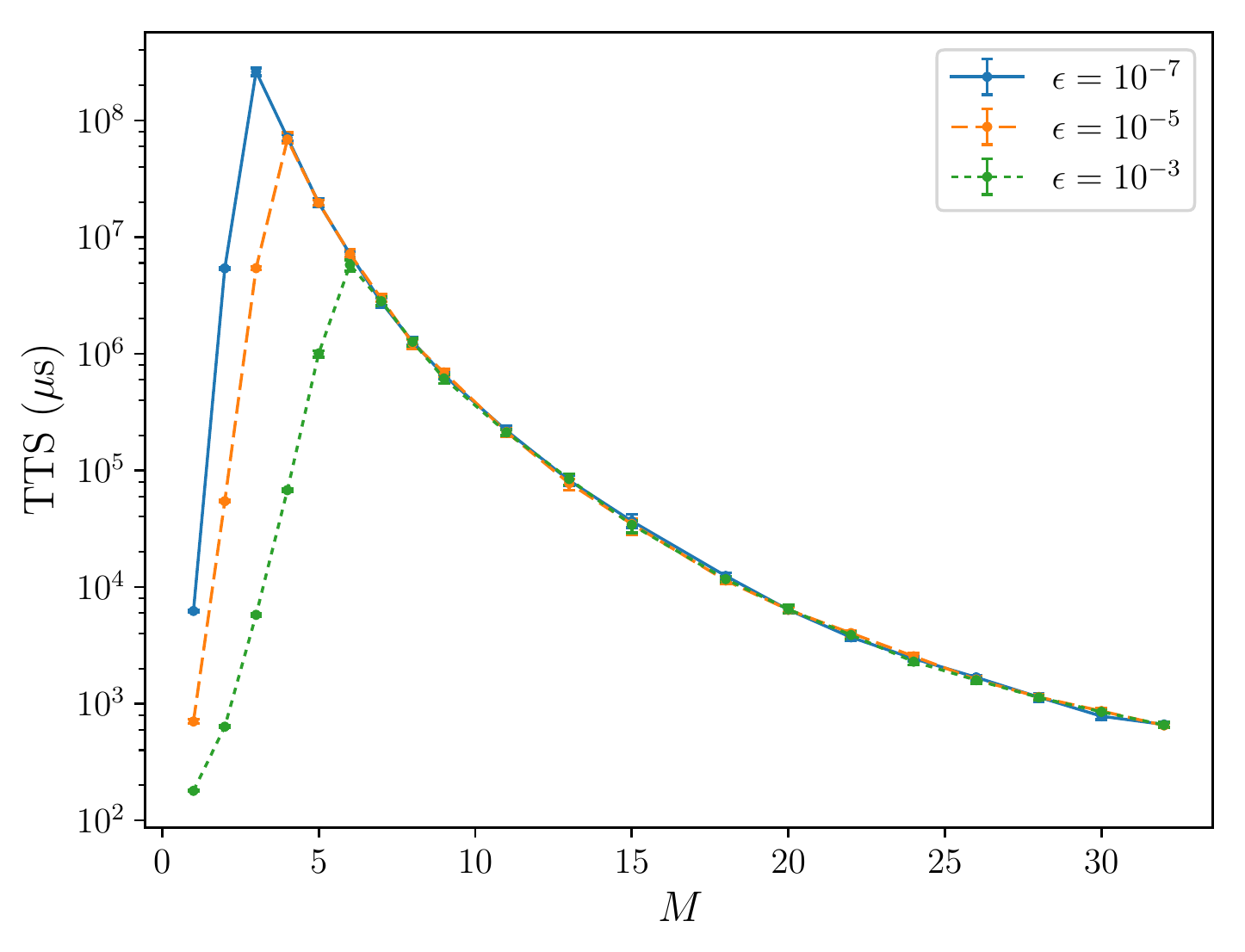}
  \caption{Optimized parallel tempering time to approximate solution for the
    $N=32$ WPE as $\epsilon$, the acceptable excess energy over that
    of the ground state, is varied. As expected, the peak difficulty
    decreases as $\epsilon$ increases, as this corresponds to making
    the objective more permissive. Interestingly, the
    hardness peak occurs at larger values of $M$ as $\epsilon$ grows.
    This phenomenon is discussed in the text, where we make analytical
    predictions of the peak location by reference to the function
    $Q_N(M)$, and illustrated in Fig.~\ref{fig:TTS_logQNs}.}
  \label{fig:TTS_N_32_epsilons}
\end{figure}

\begin{figure}
  \centering\includegraphics[width=0.85\columnwidth]{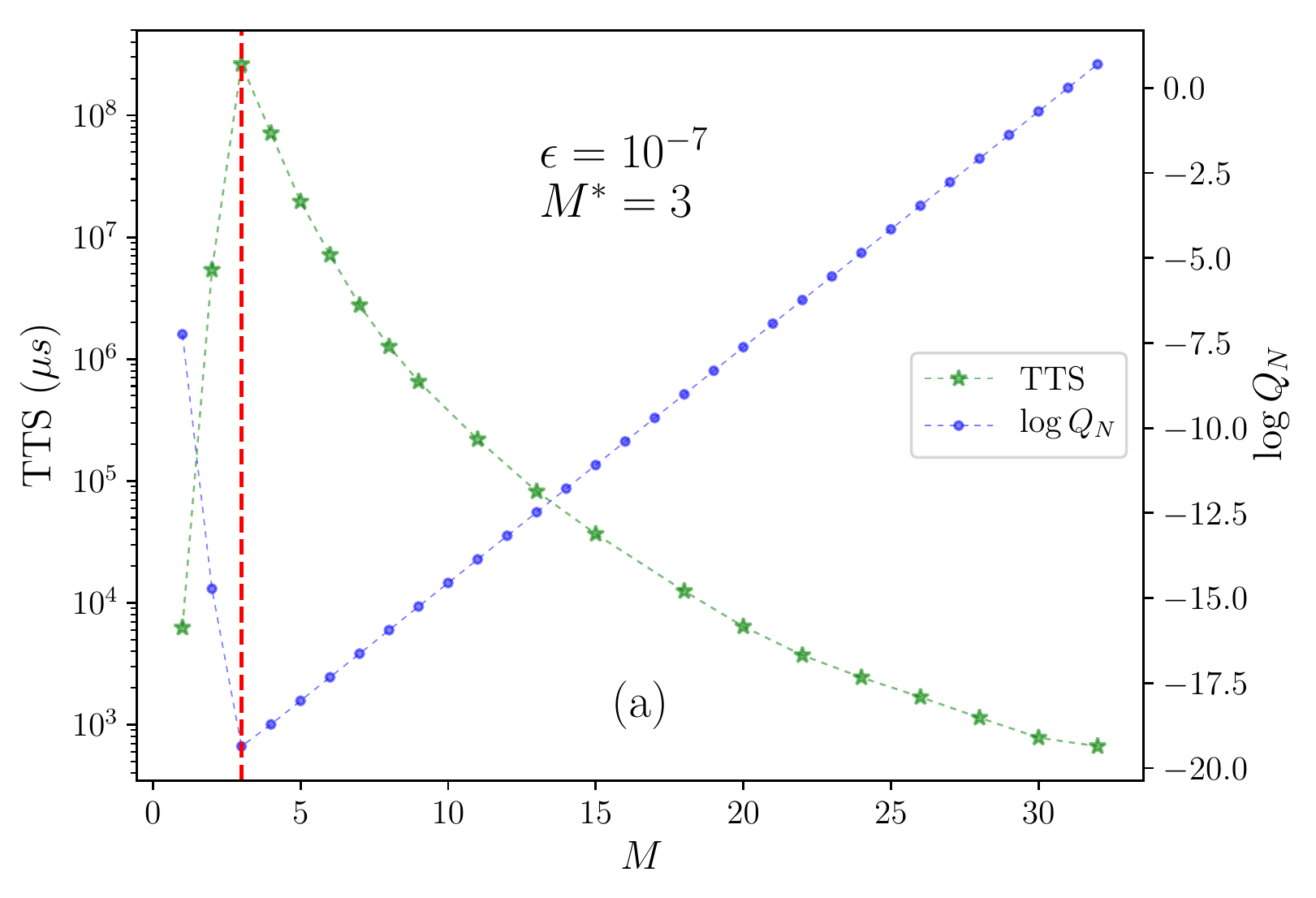}\\
  \centering\includegraphics[width=0.85\columnwidth]{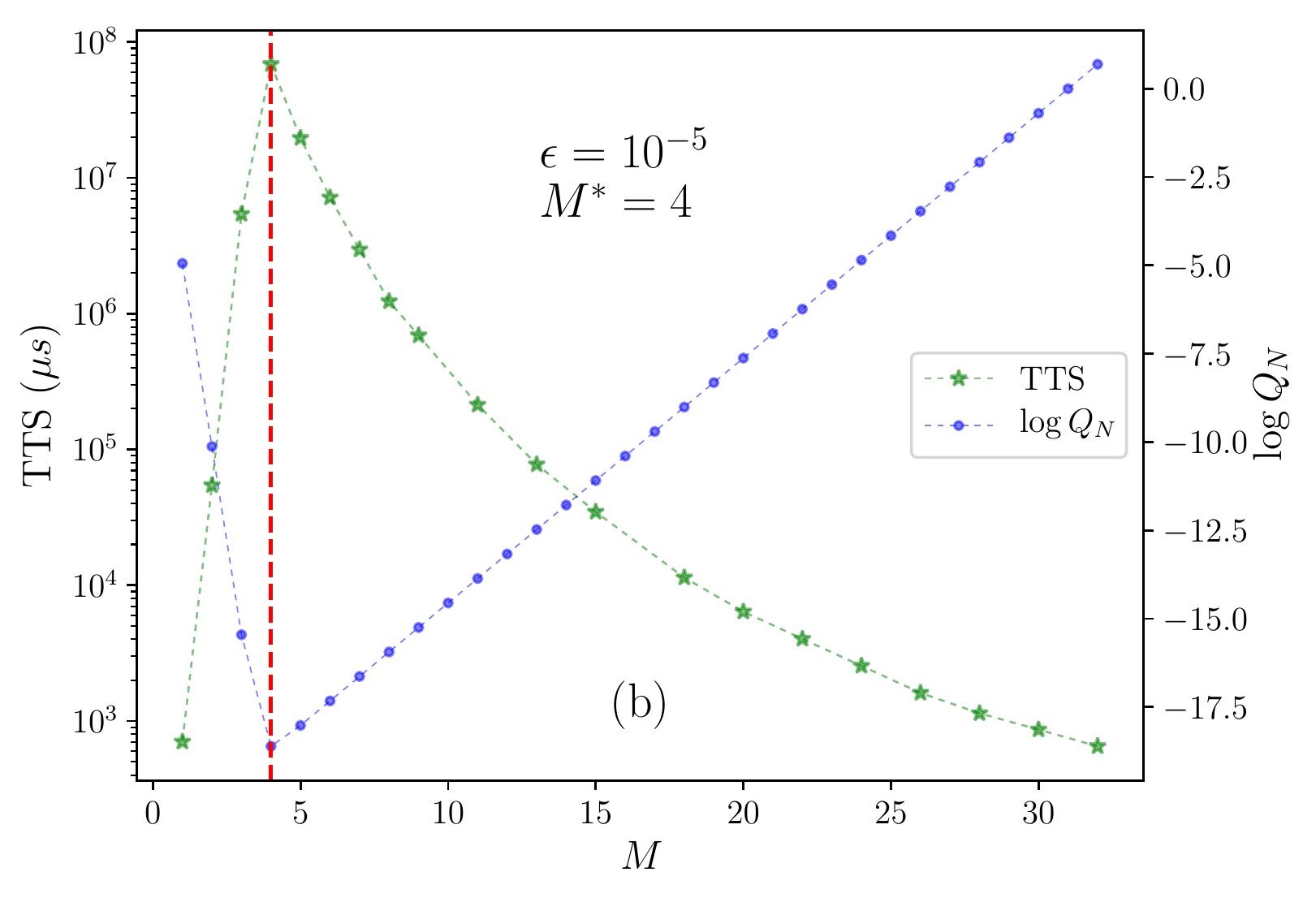}\\
  \centering\includegraphics[width=0.85\columnwidth]{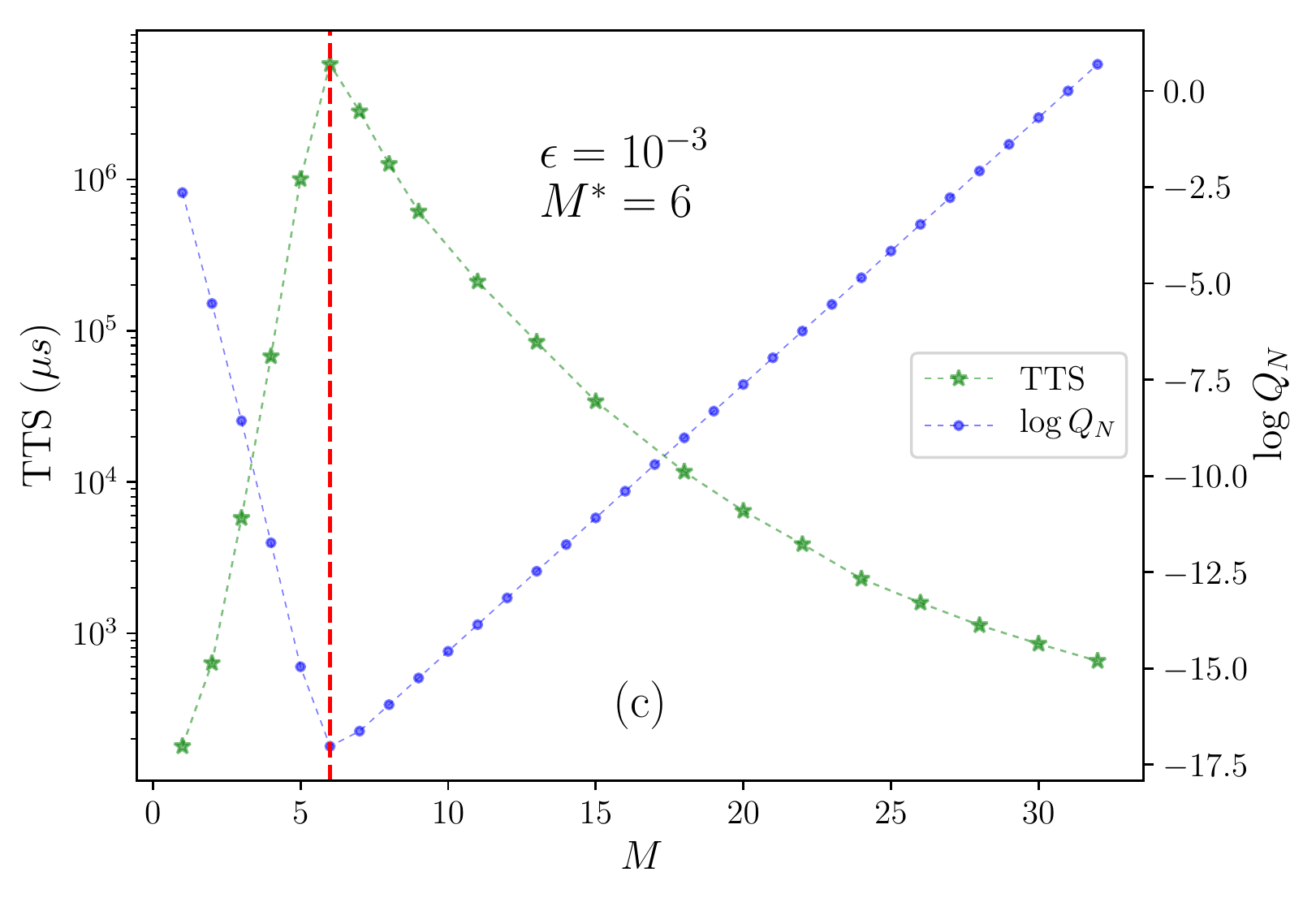}
  \caption{Illustration of the predictive power of
    $M^* = \argmin\limits_{M} Q_N(M)$ in localizing the algorithmic
    hardness peak for a range of target $\epsilon$ values: (a)
    $\epsilon=10^{-7}$, (b) $\epsilon=10^{-5}$, (c)
    $\epsilon=10^{-3}$. In each panel, we plot the parallel tempering
    solution times (green, left $y$-axis) for the $N=32$ WPE and
    $\log Q_N(M)$ (blue, right $y$-axis) for the corresponding
    $\epsilon$ values. The dashed red line displays the value of
    $M^*$, which agrees perfectly with the empirically measured
    hardness peak.}
  \label{fig:TTS_logQNs}
\end{figure}

Figure \ref{fig:TTS_N_32_epsilons} shows the PT median times to
solution for the same $N=32$ WPE ensemble considered so far, but for
three values of $\epsilon$ defining an acceptable solution:
$\epsilon=10^{-7}$, $10^{-5}$, and $10^{-3}$. Unsurprisingly, the
typical solution times decrease as the energy criterion becomes more
permissive. Additionally, we note that the location of the hardness
peak shifts to larger values of $M$ as $\epsilon$ is relaxed. Most
importantly, in Fig.~\ref{fig:TTS_logQNs} we observe the predictive
power of $Q_N(M)$ at these values of $\epsilon$: In all three cases,
the minimizer $M^*$ of $Q_N(M)$ at the respective values of $\epsilon$
precisely corresponds with empirically observed PT solution time. We
have hence proposed a robust, theoretically motivated framework for
generating tunably difficult problems over a wide range of approximate
solution targets.

\subsection{Properties of locally optimal states}
\label{sec:hardnessPhase:localOpt}

\begin{figure}
    \centering
    \includegraphics[width=0.85\columnwidth]{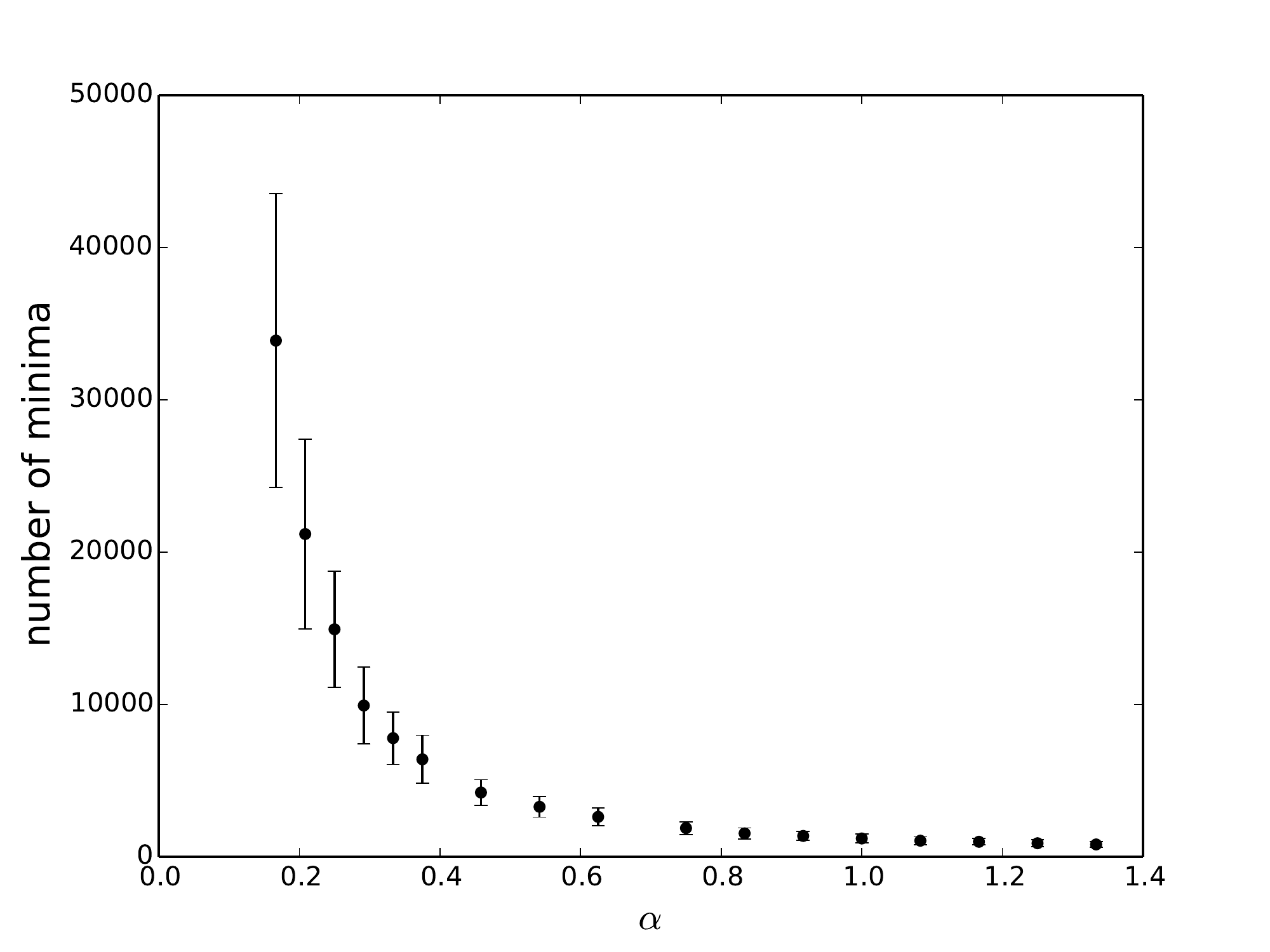}
    \caption{Expected number of local optima for WPE instances of size $N=24$ as a function of $\alpha$.}
    \label{fig:NumberOfMinima}
\end{figure}

\begin{figure}
    \centering
    \includegraphics[width=1.00\columnwidth]{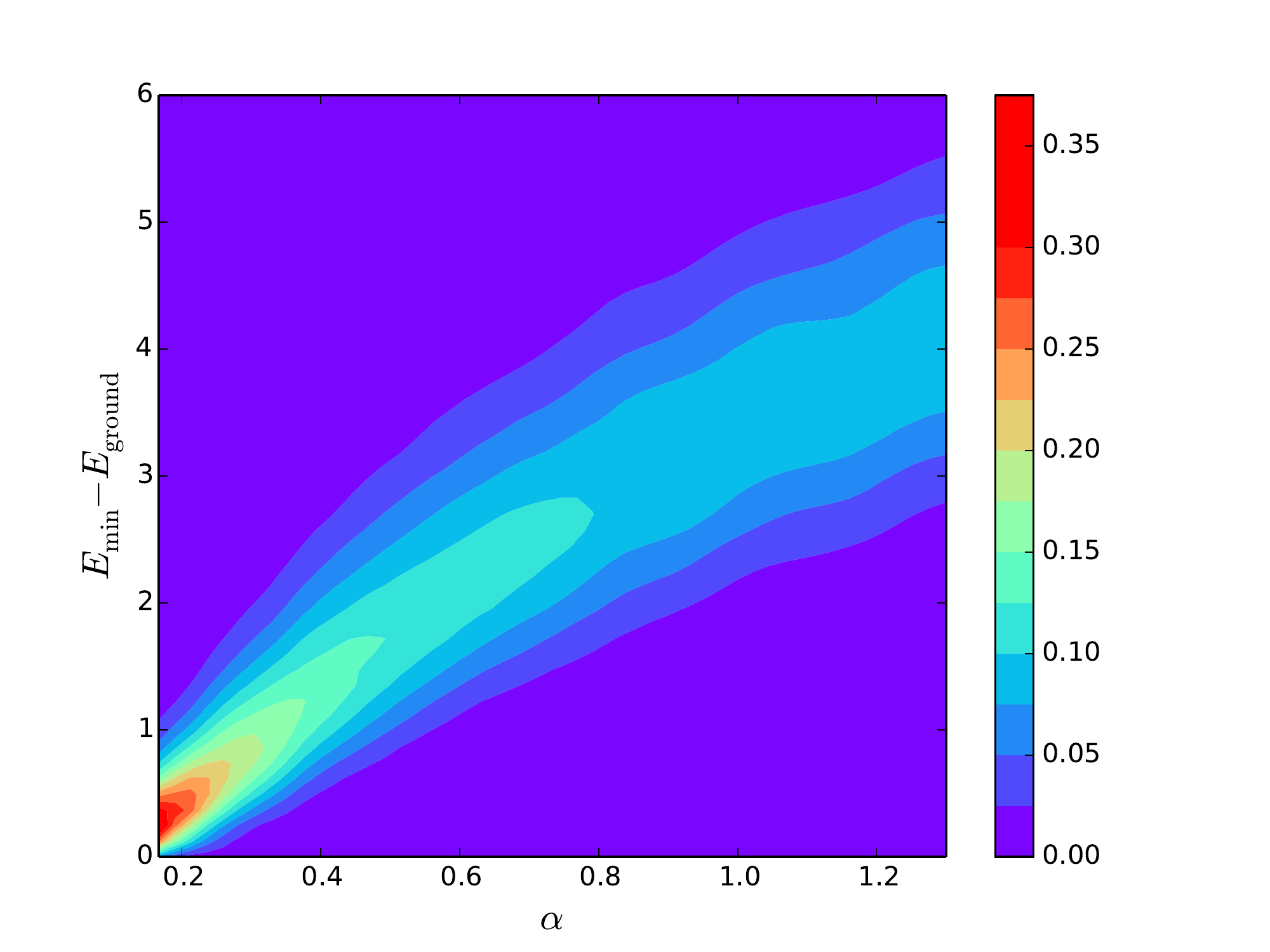}
    \caption{Residual locally optimal energy distributions for WPE instances of size $N=24$ as a function of $\alpha$. For small $\alpha$, the distributions are concentrated on low residual energy values.}
    \label{fig:resEMinimaDist}
\end{figure}

As we have seen, the energy histogram derived in Sec.~\ref{sec:hardnessPhase:energyHist} has been useful in predicting the algorithmic properties of WPE instances. Nonetheless, this distribution does not provide information about topological aspects of the local minima, i.e., states that are energetically stable with respect to a single spin flip.

In this section, we briefly probe the properties of local minima using exhaustive search on small instances; we save analytic examination of these properties, along the lines of Bray and Moore's~\cite{bray:80c} analysis for the SK model, for later work.

A natural statistic to analyze is the expected number of local minima as $\alpha$ is varied. Furthermore, it is instructive to define a residual energy histogram restricted to stable states. These are shown for a system with $N=24$ variables in Figs.~\ref{fig:NumberOfMinima} and \ref{fig:resEMinimaDist}, respectively. As expected, the number of minima decreases monotonically in $\alpha$ as the ensemble tends toward a ferromagnet. At small $\alpha$, we observe a large number of minima, but also that the residual energy itself is likely to be small. This is consistent with the observation that at small $\alpha$, the problems with restricted precision are easy.

\begin{figure*}
  \centering
  \includegraphics[width=0.85\columnwidth,trim={0cm 6cm 0cm 6cm},clip]{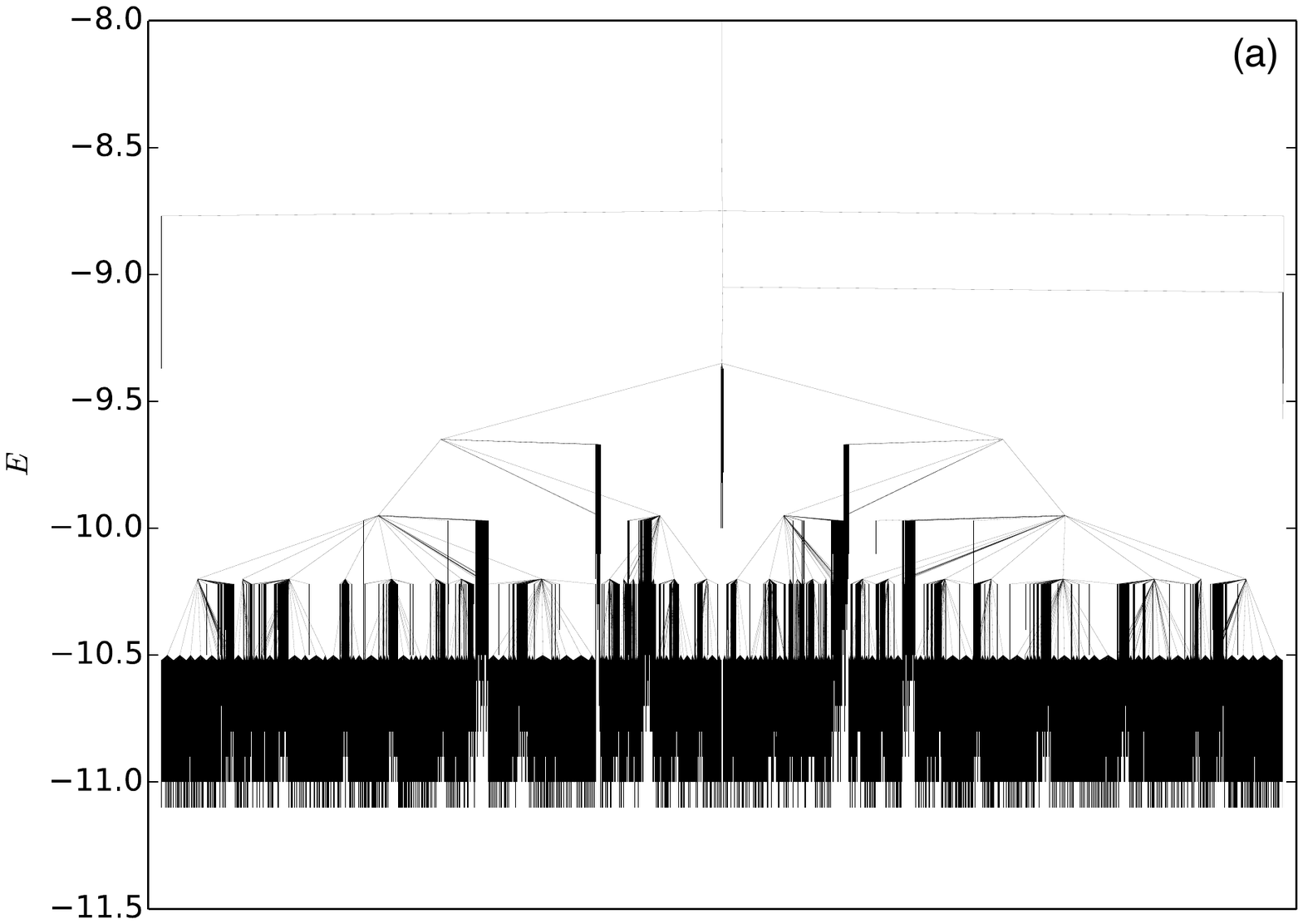}
  \includegraphics[width=0.85\columnwidth,trim={0cm 6cm 0cm 6cm},clip]{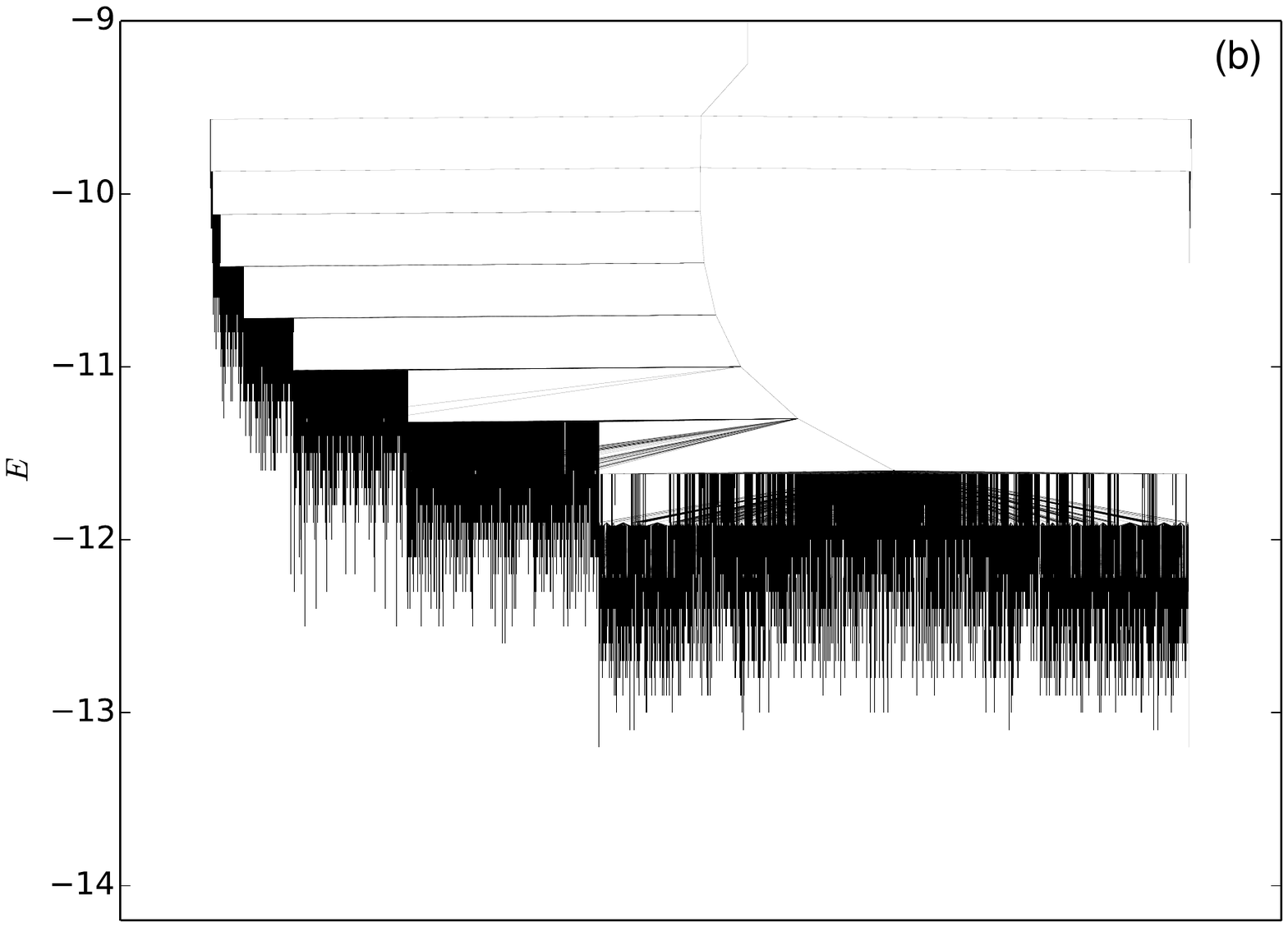}\\
  \includegraphics[width=0.85\columnwidth,trim={0cm 6cm 0cm 6cm},clip]{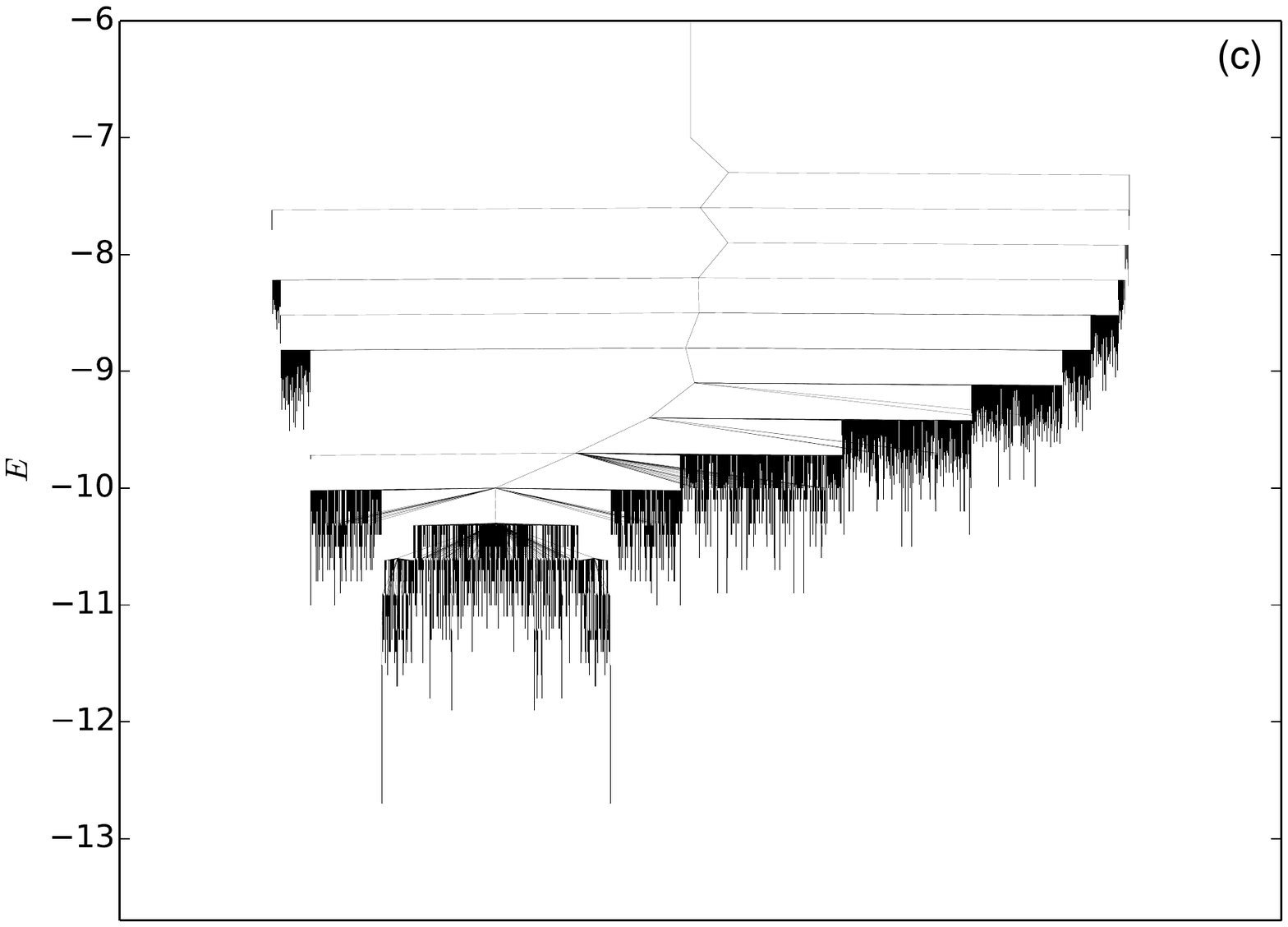}
  \includegraphics[width=0.85\columnwidth,trim={0cm 6cm 0cm 6cm},clip]{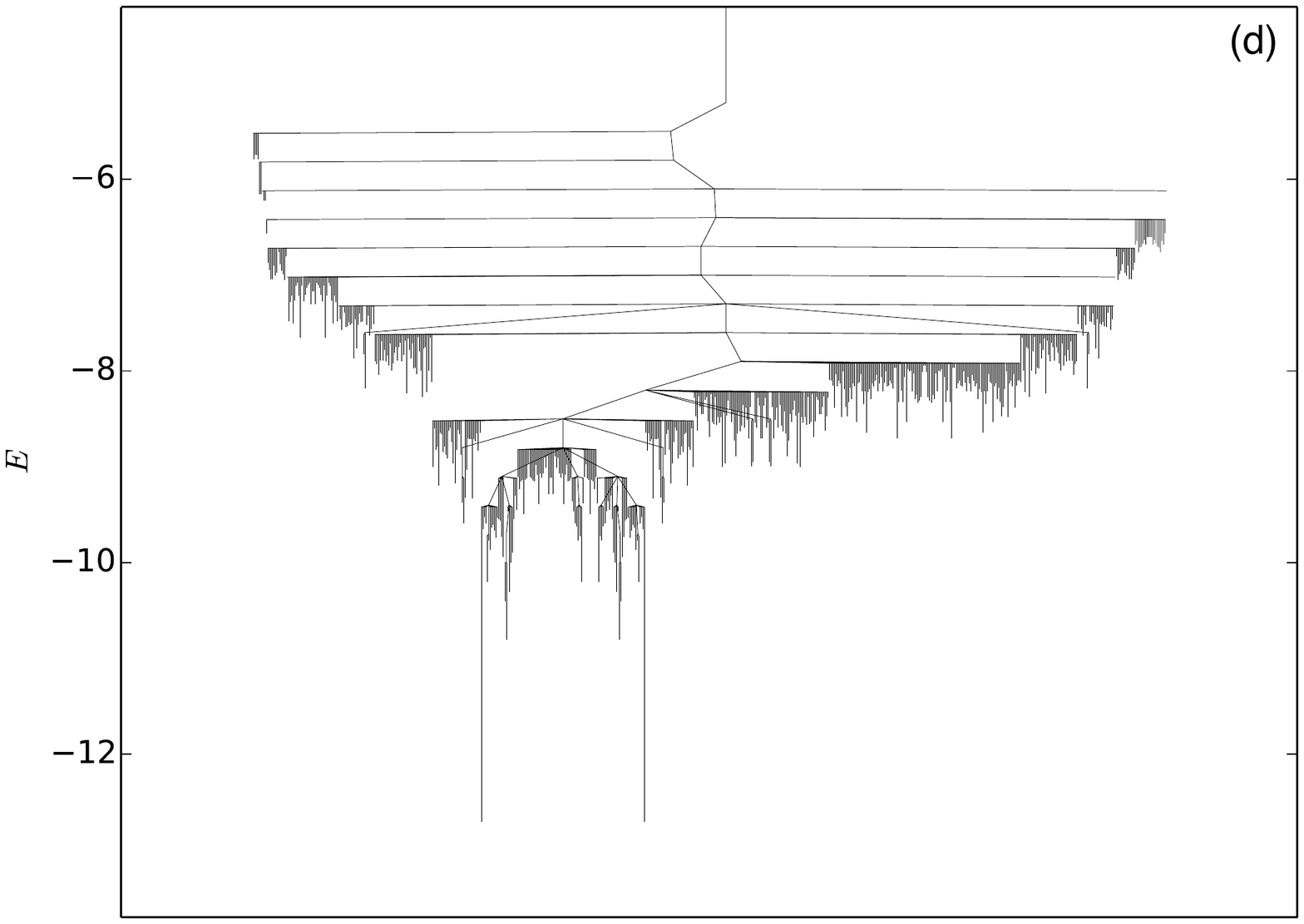}
  \caption{Disconnectivity graphs (described in the text) representing the energy landscape of four specific $N=24$ WPE instances at various values of $M$. Leaf nodes of the tree structures depict local minima, and internal nodes represent barrier states. (a) When $M=4$ (top-left), we observe a tremendous near-degeneracy of a large number of metastable states with energy very close to that of the planted solution. (b) When $M=8$ (top-right), the degeneracy begins to lift, and the number of minima starts to decrease. As $M$ increases to $15$ (c) and $32$ (d), the ground state becomes increasingly dominant and the problems
  computationally easier.}
  \label{fig:N24_DisconnectivityGraphs}
\end{figure*}  

To further illustrate properties of the WPE,
Fig.~\ref{fig:N24_DisconnectivityGraphs} shows examples of
disconnectivity graphs for four specific instances at their respective
values of $M$. Disconnectivity graphs are two-dimensional representations of high-dimensional energy landscapes. In their simplest form they depict the minima of the system and the lowest energy barrier connecting any two minima, where an energy barrier is defined as the highest energy value encountered along a specific pathway. The barriers represent the minimum increase in energy necessary to transition from one minimum to another. In this work, the minima of the system were obtained via complete enumeration and the barriers were calculated using a search over all possible pathways identified with flipping spins that are misaligned between pairs of minima. This method was outlined by Garstecki, Hoang and Cieplak~\cite{garstecki:99}. To deal with the large number of minima in a computationally efficient manner, we further made use of an approximation based on the relative proximity of minima; specifically, for each minimum only the barriers to the $50$ minima closest to it in Hamming distance were obtained. This approximation is based on the fact that transitions between two minima can also happen through basins of intermediate minima, and hence if two minima are separated by a large Hamming distance it is likely that the lowest energy barrier between these two minima will be already represented via transitions between intermediate minima and their corresponding barriers.  

In Fig.~\ref{fig:N24_DisconnectivityGraphs} the minima are represented
by vertical bars whose lowest points denote their energies. On top,
they are connected by lines converging to a common point representing
the height of the barrier that needs to be crossed in order to
transition between the connected minima. Due to the continuous nature
of the energy values, we sort the minima into a hierarchical cluster
structure whose end points comprise the intervals
$[ E_b - \Delta \ell , E_b + \Delta \ell )$, where $E_b$ is the energy
of the barrier and $2\Delta \ell$ denotes the length of the
interval. Minima whose connecting barriers fall within the same energy
interval are sorted into a common cluster. Within an individual
cluster the minima are arranged based on the number of spins in the up
or corresponding down states. Minima which have a high number of
up-state spins are sorted toward the left and, correspondingly, minima
with a high number of down-state spins are sorted toward the
right. Note, that this order strictly only applies within the
individual cluster; the order of the individual clusters relative to
each other is determined by the hierarchical structure. In this work
we set $\Delta \ell = 0.075$. Figure
\ref{fig:N24_DisconnectivityGraphs} shows a clear progression in the
energy landscape for which small values of $M$, e.g., $M=4$ in
Fig.~\ref{fig:N24_DisconnectivityGraphs} (a), are characterized by a
very large number of almost degenerate metastable states, and larger
values of $M$ [Figs.~\ref{fig:N24_DisconnectivityGraphs}(b)
--\ref{fig:N24_DisconnectivityGraphs}(d)] tend to break the
degeneracy, emphasize the planted ground state, and make the landscape
more funnel-like.

\begin{figure}
    \centering
    \includegraphics[width=1.00\columnwidth]{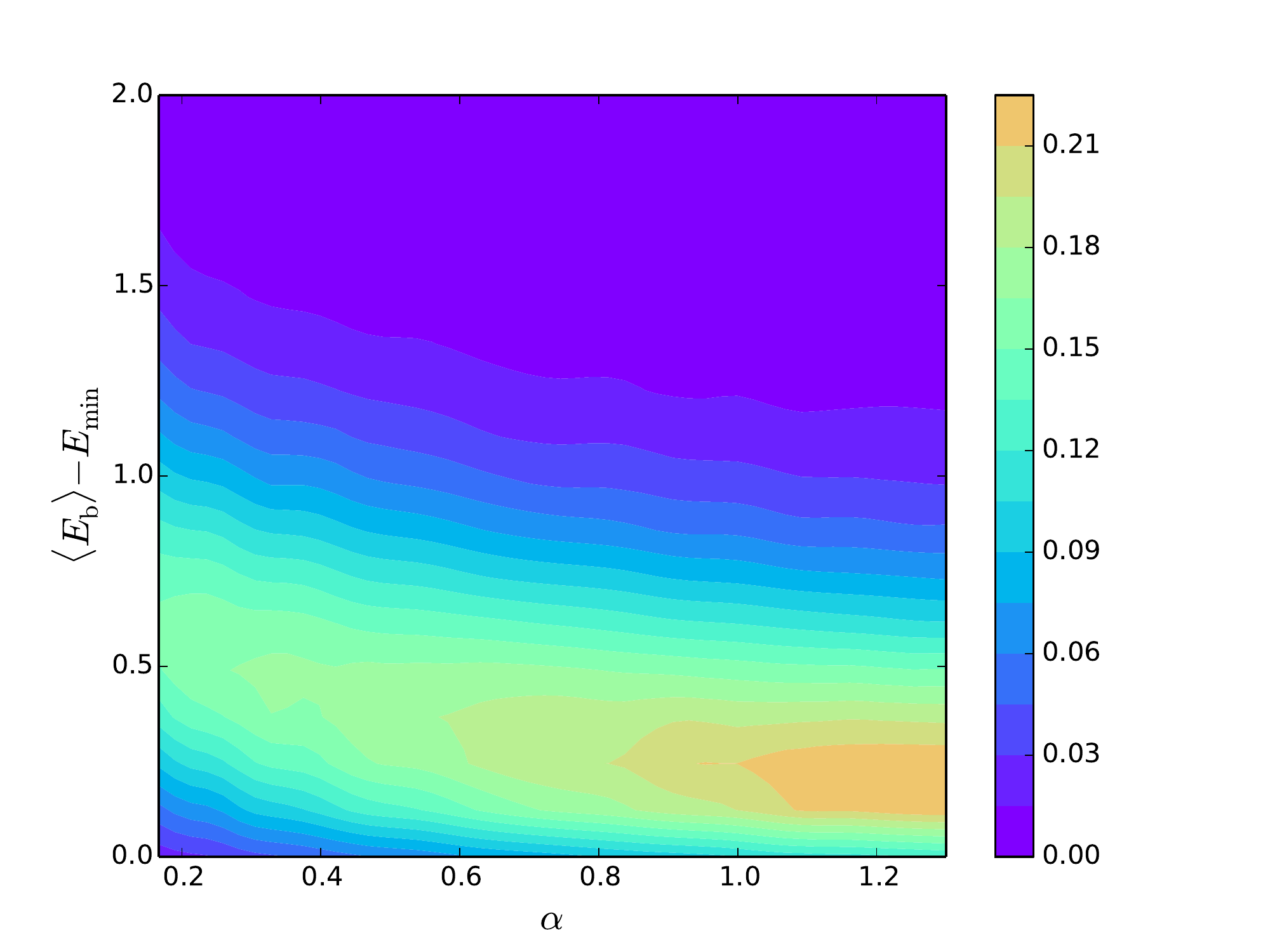}
    \caption{Average barrier height distribution (described in the
      text) for the WPE ($N = 24)$. This quantity represents the
      average increase in energy incurred in transitioning from one
      minimum to a closely lying neighbor. The distribution is more dispersed at small $\alpha$, where the barrier heights are furthermore larger in expectation than at larger $\alpha$. This indicates that transitions from the minima at large $\alpha$ are more likely to be achievable with less energy cost and therefore might be more probable than at small values of $\alpha$.}
    \label{fig:barrierDist}
\end{figure}

Figure \ref{fig:barrierDist} shows the distribution of the average barrier height within the first Hamming distance of its individual minima states. It represents the distribution of the average increase in energy necessary for the system to escape its minima via the shortest route, i.e., to transition to adjacent minima. For a given instance, the average $\langle \cdots \rangle$ was taken over the barriers to the minima which lie within the shortest Hamming distance to a given minimum, say minimum $i$. $E_{b}$ are the energies of the individual barriers and $E_{\mathrm{min}}$ is the energy of minimum $i$. The overall distribution is then obtained by sampling over all minima $i$ of $N=100$ sample systems for each of the values of $\alpha$.

As can be seen in Fig.~\ref{fig:barrierDist}, at small values of $\alpha$ the distribution is more spread than at large values, where additionally it is dominated by relatively small values of the average energy barrier. From an energy landscape perspective this indicates that transitions from the minima at large $\alpha$ are more likely to be achievable with less energy cost and therefore might be more probable than at small values of $\alpha$~\cite{comment:landscapeCaveat}.

\section{Discussion}
\label{sec:Discussion}

We have proposed a planted Ising ensemble with several
noteworthy physical and algorithmic properties. The model exhibits a
first-order temperature transition, a persistent locally stable
paramagnetic state when $\alpha < 1$, and when represented with finite
precision, an easy-hard-easy algorithmic difficulty profile. Its
physical properties are consistent with the observed hardness of
finding its ground state; moderately sized problems are extremely
difficult in the hard regime. This meshes well with the intuition that
the transition and paramagnetic stability give rise to a golf-course-like energy landscape. 
After deriving the instance-averaged
energy distribution---which turns out to follow a gamma law---we compare
the expected number of states matching a solution criterion
with a quantity we introduced quantifying the intrinsic search space
size at given $\alpha$ to analytically predict the location of the
hardness peak. The prediction is validated using solution times
obtained with a highly optimized implementation of parallel tempering
Monte Carlo.

The first-order transition between the planted and paramagnetic phases
is furthermore established by the TAP analysis in Sec.~\ref{sec:Thermodynamics:TAP}, 
with alternative derivations supported
by the replica method (Appendix \ref{sec:appendix:replicaMethod}) and
annealed approximations (Appendix
\ref{sec:appendix:annealedApproximation}). Careful Monte Carlo
simulations demonstrate the correctness of our analytically predicted
transition temperatures.

A connection is made with the anti-Hopfield model and this is
developed analytically in Appendix
\ref{sec:appendix:replicaMethod}~\cite{nokura:98}. This analysis
indicates that we might expect to see the impact of replica symmetry
breaking if we focus on the region \emph{orthogonal} to the planting
space. In this paper we have not focused on this for two reasons.
First, the emphasis has been on intermediate scale problems, where it
is difficult to numerically establish such phenomena, finite size
effects may dominate, and practical issues such as finite-precision
representation may have greater impact. Second, at large $N$ replica
symmetry breaking has little impact on the free-energy barrier
separating the bulk of the space from the planted solution, which is
the primary driver of hardness for heuristic optimization in the
planted case.

Nevertheless, we believe the qualitative description of the space
orthogonal to the planted solution, that of a rough energy landscape
with deep solutions almost orthogonal to the planted solution, is in
effect. Roughness is apparent at the small scales we have worked with
empirically. With this in mind, a modification of the planting
procedure is pursued in Appendices
\ref{sec:appendix:annealedApproximation} and
\ref{sec:appendix:replicaMethod} whereby the planted solution is
partially penalized. Its leading-order energy is then tunable at fixed
$\alpha$, at the cost of losing strict guarantees that it is a ground
state. This modification, it is believed, might be the basis for
interesting tests of dynamics (escape into or out of the planted
solution). In particular, an interesting suggestion has been made that for
this type of problem in the presence of a transverse field, 
quantum dynamics may be differentiated from classical
counterparts~\cite{smelyanskiy:18}. The features of our model with
this modified planting procedure represent a practical realization of
many of the abstract model features underlying the population transfer
hypothesis.

Future work will explore this direction more deeply, and consider as
well whether the features of the WPE energy landscape lend themselves
to the demonstration of fundamental speedup using emerging quantum
annealing devices, in particular those whose classical simulation is
known to be intractable. It is our hope that the insights into the
ensemble's physical and algorithmic classical properties presented in
this work will solidly underpin these future directions.

In addition, we are interested in pursuing the relation between the
WPE and the important class of low-rank estimation
problems~\cite{lesieur:17} used in unsupervised feature extraction and
dimensionality reduction, in further examining whether any deeper
connections can be made between the ensemble and biological
unlearning~\cite{nokura:98}, and in studying potential connections
with models for analyzing large-system code-division multiple-access
(CDMA) multiuser detectors~\cite{tanaka:02}.

\section{Acknowledgements}

We acknowledge useful discussions with Bill Macready, Federico
Ricci-Tersenghi, Hidetoshi Nishimori, Jon Machta, Cris Moore, Koji
Hukushima, Brad Lackey, Lenka Zdeborov{\'a}, Salvatore Mandr{\`a},
Mohammad Amin, Andrew King, Cathy McGeogh, and Aidan Roy.
H.G.K. thanks the Department of Poultry Science at
Texas A\&M University for providing support. H.G.K. and
C.P. are supported in part by the Office of the Director of National
Intelligence (ODNI), Intelligence Advanced Research Projects Activity
(IARPA), via MIT Lincoln Laboratory Air Force Contract
No.~FA8721-05-C-0002. The views and conclusions contained herein are
those of the authors and should not be interpreted as necessarily
representing the official policies or endorsements, either expressed
or implied, of ODNI, IARPA, or the U.S.~Government. The
U.S.~Government is authorized to reproduce and distribute reprints for
Governmental purpose notwithstanding any copyright annotation thereon.

\section{Appendix}
\label{sec:appendix}

\subsection{Cavity calculation of the TAP equations}
\label{sec:appendix:cavityTAP}

For notational convenience, we redefine the Hamiltonian of the
WPE to include the constant terms corresponding to the diagonal
elements of $\Jtildev$; this will not affect the final result. Further,
without loss of generality we assume that the ferromagnetic solution
$\tv = (1,1,\ldots,1)$ and its reversal are to be planted as the ground
state.  The Hamiltonian of the WPE is
\begin{equation*}
H(\sv) = -\frac{1}{2}\sum_{i,j=1}^N \Jtilde_{ij} s_is_j ,
\end{equation*}
where
\begin{equation*}
  \Jtilde_{ij} = -\frac{1}{N} \sum_{\mu=1}^M w_i^\mu w_j^\mu
\end{equation*}
and $\wv^\mu \sim \Ncal( \ZeroVec, \Sigmav )$. As before,
$M = \alpha N$ for $\alpha > 0$ and $N \to \infty$. The elements of
the covariance matrix $\Sigmav$ are as follows:
\begin{equation*}
  \Sigma_{ij} = \left \{
    \begin{array}{cc}
      1 & \textrm{for $i = j$}\\
      -\frac{1}{N-1} & \textrm{for $i \neq j$}
    \end{array}
    \right. .
\end{equation*}
This covariance structure implies that the Gaussian process generating
$\wv$ is not only stationary but \emph{exchangeable}. As discussed in
Sec.~\ref{sec:Thermodynamics}, this model bears many resemblances
to the Hopfield network, but in addition to the negation of the
interactions, we have the additional complication of the
``antipatterns'' $\{ \wv^\mu \}$ consisting of correlated elements.
Nonetheless, by closely following the two-step cavity-based method
presented by Shamir and Sompolinksi~\cite{shamir:00} for
the Hopfield model, taking special care to account for the
correlations, we may derive the TAP equations for the WPE.

The cavity approach~\cite{mezard:87} derives the self-consistent relation
for each local spin magnetization by first considering removal
of the spin from the $N+1$-spin system and defining a state
distribution on the $N$-spin subsystem. Remarkably, the joint
distribution of the field and spin in the original system can be
expressed in terms of the field distribution resulting from the
subsystem, allowing the spin and field statistics for the full system
to be related to those of the subsystem. On their own, these exact
relations do not give much insight because they are intractable to
compute, but when the subsystem field distribution can be justified to
be Gaussian, the TAP equations for the magnetizations may be obtained.
A substantial amount of the work is in deriving the correct parameters
for the field distribution. The details for the WPE follow.

\subsubsection{Cavitating a spin}

Consider an $(N+1)\times(N+1)$ WPE matrix $\Jtildev$ coupling spins
$\{0,\ldots,N\}$ through $M$ vectors $\{ \wv^\mu \}$,
i.e., $\Jtilde_{ij} = -\frac{1}{N+1}\sum_{\mu=1}^M w_i^\mu w_j^\mu$. We
can decompose this Hamiltonian into a sub-Hamiltonian $H^{(N)}$
consisting only of interactions among spins $\{1,\ldots,N\}$, denoted
here by $s_{1:N}$, and a term accounting for the interaction between
spin zero and the others:
\[
H^{(N+1)}(\sv) = H^{(N)}(s_{1:N}) - h_0(s_{1:N} ) s_0 + \frac{1}{2}\Jtilde_{00}
\]
with
\begin{align*}
  H^{(N)}(s_{1:N}) & = -\frac{1}{2}\sum_{i,j=1}^N\Jtilde_{ij}s_is_j
\end{align*}
and
\begin{align*}
  h_0(s_{1:N}) & = \sum_{j=1}^N\Jtilde_{0j}s_j .
\end{align*}
The final constant in $H^{(N+1)}$ is irrelevant and will be
dropped. The exact joint distribution over $(s_0,h_0)$ can be shown to
be
\[
P^{(N+1)}(h_0,s_0) = \frac{1}{\xi} \exp( \beta h_0 s_0) P^{(N)}(h_0) ,
\]
where
\[
\xi = \frac{Z_{N+1}}{Z_N} = \physExpect{2\cosh \beta h_0}_N
\]
and $\physExpect{\cdots}_N$ refers to thermal averaging with respect to
$P^{(N)}(s_{1:N}) = \frac{1}{Z_N} \exp( -\beta H^{(N)}(s_{1:N}))$. 
From this, we obtain the (intractable) relations
\begin{align}
  \physExpect{s_0}_{N+1} = & \sum_{s_0  } \int s_0 P^{(N+1)}(
                             h_0, s_0) \dd h_0 \nonumber \\
  = & \frac{ \int  \big[ \exp( \beta h_0) -
      \exp(-\beta h_0) \big] P^{(N)}(h_0) \dd h_0 }
      { \physExpect{2\cosh \beta h_0}_N} \nonumber \\
  = & \frac{\physExpect{\sinh(\beta h_0)}_N}{\physExpect{\cosh(\beta
      h_0)}_N}
      \label{eq:Es0Exact}
\end{align}
and
\begin{align}
  \physExpect{h_0}_{N+1} = & \sum_{s_0  } \int h_0 P^{(N+1)}(
                             h_0, s_0) \dd h_0 \nonumber \\
  = & \frac{ \int  h_0 \big[ \exp( \beta h_0) +
      \exp(-\beta h_0) \big] P^{(N)}(h_0) \dd h_0 }
      { \physExpect{2\cosh \beta h_0}_N} \nonumber \\
  = & \frac{\physExpect{h_0\cosh(\beta
      h_0)}_N}{\physExpect{\cosh(\beta h_0)}_N} .
      \label{eq:Eh0Exact}
\end{align}
The next step is to compute the field statistics to be used
following a Gaussian assumption for $P^{(N)}(h_0)$.

As for the Hopfield model, define the field mean and variance as
\begin{align*}
  \physExpect{ h_0}_N & = \sum_{j=1}^N \Jtilde_{0j} \physExpect{s_j}_N ,
  \\
  \physExpect{ (\delta h_0)^2}_N & = \physExpect{ h_0^2}_N - \physExpect{h_0}^2_N .
\end{align*}
Note that
\begin{align*}
  \physExpect{ (\delta h_0)^2}_N & = \sum_{i=1}^N\sum_{j=1}^N
                          \Jtilde_{0i}\Jtilde_{0j} \chi_{ij}^{(N)} ,
\end{align*}
where
\begin{align*}
  \chi_{ij}^{(N)}  \triangleq \physExpect{ \delta s_i\delta s_j} = \physExpect{s_is_j}_N - \physExpect{s_i}_N\physExpect{s_j}_N .
\end{align*}
Using the definition of $\Jtilde_{0i}$, we obtain
\begin{align}
  \physExpect{ (\delta h_0)^2}_N & = \frac{1}{(N+1)^2}
                                   \sum_{\mu=1}^M\sum_{\nu=1}^M w_0^\mu
                                   w_0^\nu
                                   \sum_{i=1}^N\sum_{j=1}^N w_i^\mu
                                   w_j^\nu \chi_{ij}^{(N)} .
                                   \label{eq:varh0}
\end{align}
If we define the overlap of spins $\{1,\ldots,N\}$ with the last $N$
components of $\wv^\mu$ as
\[
\eta_\mu \triangleq \frac{1}{N} \sum_{i=1}^Nw_i^\mu s_i
\]
and its covariance under the cavitated spin distribution
\[
\physExpect{\delta \eta_\mu \delta \eta_\nu }_N \triangleq \physExpect{\eta_\mu
  \eta_\nu}_N - \physExpect{\eta_\mu}_N\physExpect{\eta_\nu}_N ,
\]
then we obtain 
\begin{align*}
  \physExpect{\delta \eta_\mu \delta \eta_\nu }_N = \frac{1}{N^2}
  \sum_{i=1}^N\sum_{j=1}^N w_i^\mu w_j^\nu\chi_{ij}^{(N)}
\end{align*}
and hence
\begin{align*}
  \physExpect{ (\delta h_0)^2}_N & \approx 
                                   \sum_{\mu=1}^M\sum_{\nu=1}^M w_0^\mu
                                   w_0^\nu \physExpect{\delta \eta_\mu \delta \eta_\nu }_N .
\end{align*}

Noting $\chi_{ii}^{(N)} = 1 - \physExpect{s_i^2}_N = O(1)$ while
$\chi_{ij}^{(N)} = O(\frac{1}{\sqrt{N}})$ for $i\neq j$, we proceed to
determine the magnitude of $\physExpect{\delta \eta_\mu \delta \eta_\nu }_N$
in order to simplify the field variance, bearing in mind that while
$\wv^\mu$ is independent of $\wv^\nu$, there are componentwise
correlations within each vector not present in the Hopfield model. The
conclusion will be that just as for the Hopfield model,
$\physExpect{\delta \eta_\mu \delta \eta_\nu }_N$ is $O(\frac{1}{N^{3/2}})$
when $\mu \neq \nu$ and $O(\frac{1}{N})$ and when $\mu = \nu$.

Define
\[
S_{\mu\nu} = \frac{1}{N^2} \sum_{i=1}^N\sum_{j=1}^N w_i^\mu
w_j^\nu\chi_{ij} ,
\]
where the superscript on $\chi$ has been dropped. We seek
$\Expect{}[ S_{\mu\nu}] $ and the fluctuations
$\sqrt{ \Expect{}[ S_{\mu\nu}^2] - \Expect{}[ S_{\mu\nu}]^2 }$, where
the expectations are taken over $\{ \wv \}$.

The linear expectations are straightforward to compute; when $\mu \neq
\nu$, we have
\[
\Expect{}[ S_{\mu\nu}] = \frac{1}{N^2} \sum_{i=1}^N\sum_{j=1}^N
\Expect{}[w_i^\mu] \Expect{}[w_j^\nu]\chi_{ij} = 0
\]
while, recalling the covariance structure of $\wv$, when $\mu = \nu$,
\begin{align*}
  \Expect{}[ S_{\mu\mu}] & = \frac{1}{N^2} \Big [ \sum_{i=1}^N \Expect{}[
                       (w_i^\mu)^2 \chi_{ii} ] + \sum_{i\neq j} \Expect{} [ w_i^\mu
                       w_j^\mu\chi_{ij} ]
                       \Big] \\
                     & = \frac{1}{N^2} \Big[ O(N) - O\Big(
                       \frac{N(N-1)}{N\sqrt{N}}\Big) \Big] \\
                     & = O\Big( \frac{1}{N} \Big) ,
\end{align*}
which is different from the Hopfield model, in which this quantity is
zero.

Direct computation of the quadratic expectation $\Expect{}[
S_{\mu\nu}^2]$ is tedious but straightforward. In the expansion, there
will be a total of $N^4$ terms of the form $\chi_{ij}\chi_{kl}w_i^\mu
w_j^\nu w_k^\mu w_l^\nu$. We first make a relevant partitioning of
these terms for generic $\{\mu, \nu\}$ and then compute the expectations
for the cases where they are equal and different.
\vspace{0.5cm}
\begin{itemize}
\item 
  \textbf{Case $i=k$, $j=l$}

  If $i=j$, then there are $N$ terms of the form
  $\chi_{ii}^2( w_i^\mu)^2(w_i^\nu)^2$
  
  If $i \neq j$, then there are
  $N(N-1)$ terms like $\chi_{ij}^2( w_i^\mu)^2(w_j^\nu)^2$
  \vspace{0.5cm}
\item
  \textbf{Case $i=k$, $j\neq l$}
  
  If $i=j$, then there are $N(N-1)$ terms as $\chi_{ii}\chi_{il}
  (w_i^\mu)^2w_i^\nu w_l^\nu$
  
  If $i \neq j$, then we have $N$ terms like
  $\chi_{ij} \chi_{ii} (w_i^\mu)^2w_i^\nu w_j^\nu$ and $N(N-2)$ terms as
  $\chi_{ij}\chi_{il}(w_i^\mu)^2 w_j^\nu w_l^\nu$
  \vspace{0.5cm}
\item
  \textbf{Case $i\neq k$, $j=l$}
  
  This is a rotated version of the previous case. When $i=j$, there are
  $N(N-1)$ terms of the form
  $\chi_{jj}\chi_{kj}w_j^\mu w_k^\mu ( w_j^\nu)^2$

  When $i\neq j$, there are $N$ terms like
  $\chi_{ij}\chi_{jj}w_i^\mu w_j^\mu( w_j^\nu)^2$ and $N(N-2)$ terms as
  $\chi_{ij}\chi_{kj} w_i^\mu w_k^\mu ( w_j^\nu)^2$
  \vspace{0.5cm}
\item
  \textbf{Case $i\neq k$, $j\neq l$}
  
  When $i=j$, there are $N(N-1)$ terms like $\chi_{ii} \chi_{kk} w_i^\mu
  w_k^\mu w_i^\nu w_k^\nu $ and $N(N-1)(N-2)$ terms
  as $\chi_{ii} \chi_{kl} w_i^\mu w_k^\mu w_i^\nu w_l^\nu $
  
  When $i \neq j$, there are $N(N-1)(N-2)$ terms as $\chi_{ij}\chi_{kk}
  w_i^\mu w_k^\mu w_j^\nu w_l^\nu$ and $N(N-1)(N^2-3N+3)$ terms like
  $\chi_{ij} \chi_{kl} w_i^\mu w_k^\mu w_j^\nu w_l^\nu$
  \vspace{0.5cm}
\end{itemize}

Consider now the case of $\mu \neq \nu$. Adding the terms in the
expansion, recalling that $\wv^\mu$ is independent of $\wv^\nu$ and
that $\Expect{}[ w_i w_j] = -\frac{1}{N}$ when $i \neq j$, we find
that
\[
\Expect{}[ S_{\mu\nu}^2] = O\Big( \frac{1}{N^3} \Big) ,
\]
implying that $S_{\mu\nu}$ is $O\Big( \frac{1}{N^{3/2}} \Big) $. This quantity
is of identical order in the Hopfield model despite the correlations
in $\{ \wv \}$.

When $\mu = \nu$ on the other hand, we find that
\begin{align}
  \Expect{}[&S_{\mu\mu}] = \frac{1}{N^4} \Big[ N O(1) \Expect{}[ w_i^4] +
    N(N-1)O(\frac{1}{N})\ldots \nonumber \\
  & \Expect{}[ w_i^2w_j^2] + 2N(N-1)O(\frac{1}{\sqrt{N}}) \Expect{}[ w_i^3w_l] +
    2N \ldots \nonumber \\
  & O(\frac{1}{\sqrt{N}})\Expect{}[ w_i^3w_j] + 2N(N-2)
    O(\frac{1}{N}) \Expect{}[ w_i^2 w_j w_l] + \ldots \nonumber \\
  & N(N-1)O(1)\Expect{}[ w_i^2 w_k^2] + N(N-1)(N-2)\ldots \nonumber \\
  & O(\frac{1}{\sqrt{N}}) \Expect{}[
    w_i^2w_kw_l] +  N(N-1)(N-2) O(\frac{1}{\sqrt{N}} )\ldots \nonumber
  \\
  & \Expect{}[w_i w_jw_k^2] + N(N-1)(N^2-3N+3) O( \frac{1}{\sqrt{N}} )
    \ldots \nonumber \\
  & \Expect{}[ w_i w_j w_k w_l]
    \Big] ,
    \label{eq:Smumu}
\end{align} 
where the $O(\cdots)$ terms in Eq.~(\ref{eq:Smumu}) refer to the effects of the $\{\chi_{ij}\}$
products. Using the properties of the higher-order moments of a
correlated Gaussian distribution (e.g., Ref.~\cite{tong:12}), we have by the exchangeability of
our distribution for any indices such that different letters refer to
different values,
\begin{align*}
  \Expect{}[ w_i ^4]  = & 3 = O(1) \\
  \Expect{}[ w_i^3 w_j] =&  -\frac{3}{N} = O \Big (\frac{1}{N}\Big) \\
  \Expect{}[ w_i^2 w_j w_k ] = & -\frac{1}{N} + \frac{2}{N^2} = O \Big (
  \frac{1}{N} \Big) \\
  \Expect{}[ w_i^2 w_j^2 ] = & 1 + \frac{2}{N^2} = O(1) \\
  \Expect{}[ w_i w_j w_k w_l ] = & -\frac{3}{N^2} = O \Big ( \frac{1}{N^2}\Big) .
\end{align*}
Substituting these into Eq.~(\ref{eq:Smumu}), we obtain the result that
$\Expect{}[ S_{\mu\mu}^2] = O(1/N^2)$ and hence that
$S_{\mu\mu}$ is typically 
\begin{equation}
    \sqrt{ O\Big(\frac{1}{N^2}\Big) -
  \frac{1}{N^2}  } \pm \frac{1}{N} , 
\end{equation}
namely $O\Big(1/N\Big)$.
These results imply that the field variance in Eq.~(\ref{eq:varh0})
\begin{align*}
  \physExpect{(\delta h_0)^2}_N = & \sum_{\mu=1}^M(w_0^\mu)^2
  \physExpect{(\delta \eta_\mu)^2}_N + \sum_{\mu\neq\nu} w_0^\mu
  w_0^\nu \physExpect{\delta\eta_\mu \delta\eta_\nu}_N
\end{align*}
is
\[
O \Big ( \frac{M}{N} \Big) + O \Big ( \frac{\sqrt{M(M-1)}}{N^{3/2}} \Big) = O(1) + O \Big (\frac{1}{\sqrt{N}} \Big) ,
\]
where the order of the second summation follows from the independence
of $\mu$ and $\nu$; by self-averaging it can hence be approximated by
\begin{align}
  \physExpect{(\delta h_0)^2}_N = \sum_{\mu=1}^M \physExpect{(\delta
  \eta_\mu)^2}_N \triangleq V_N .
  \label{eq:h0VarVN}
\end{align}

Finally, recall that
$h_0 = -\frac{1}{N+1} \sum_{\mu=1}^Mw_0^\mu \sum_{i=1}^N w_i^\mu s_i
\approx -\sum_{\mu=1}^M w_0^\mu \eta_\mu$.  The $\{ \eta_{\mu} \}$
decorrelate at the same rate as they do in the Hopfield model;
further, $w_0^\mu$ follows deterministically from $\{ w_{1:N}^\mu\}$
and adds no information about the state distribution on sites
$\{1,\ldots,N\}$ and hence on the $\{ \eta_\mu \}$. This suggests that
the field can be approximated by a limiting Gaussian distribution:
\begin{align*}
  P^{(N)}(h_0) = \frac{1}{\sqrt{2\pi V}} \exp \Big[ - \frac{ \big(h_0 - \physExpect{h_0}_N\big)^2}{2V} \Big],
\end{align*}
where $V \triangleq \lim\limits_{N\to\infty} V_N$. This Gaussian
approximation dramatically simplifies relations in Eqs.~(\ref{eq:Es0Exact})
and (\ref{eq:Eh0Exact}), which reduce to
\begin{align}
  \physExpect{s_0}_{N+1} = & \tanh( \beta \physExpect{h_0}_N )\nonumber \\
  \physExpect{h_0}_{N+1}  = & \physExpect{h_0}_N + \beta V \physExpect{s_0}_{N+1} .
\end{align}
Considering deletion of any spin $i$ rather than zero, the TAP relation
\begin{equation}
\physExpect{s_i} = \tanh\Big[ \beta\big(  \sum_{j\neq i} J_{ij}
\physExpect{s_j} - \beta V \physExpect{s_i} \big) \Big]
\end{equation}
follows. Note that the distinction between $\Jv$ and $\Jtildev$
disappears at this point; a consequence of the cavity method is that
$\{\Jtilde_{ii}\}$ terms are disregarded as one expects. To fully
specify the TAP relation, $V$ must be determined for the WPE; we turn
to this task next.

\subsubsection{Cavitating a $\wv$}

Consider now a system of $N$ spins but with $M+1$ $\{ \wv^\mu \}$. The
corresponding Hamiltonian can be related to that of a system with a
single $\wv^\mu$ removed. Specifically, if
\[
H^{(M)}(\sv) = \frac{1}{2N} \sum_{i,j=1}^N\sum_{\mu=1}^M w_i^\mu
w_j^\mu s_i s_j
\]
then the full Hamiltonian is
\begin{align*}
  H^{(M+1)}(\sv) = & H^{(M)}(\sv) + \frac{1}{2N}\Big( \sum_{i=1}^N
                     w_i^0s_i\Big) \Big( \sum_{j=1}^N w_j^0 s_j \Big) \\
  = &  H^{(M)}(\sv) + \frac{N}{2}( \eta_0)^2 ,
\end{align*}
where again an irrelevant constant has been dropped.
Now following the cavity procedure, we obtain the distribution of 
$\eta_0$ relative to the Boltzmann distribution corresponding to
$H^{(M+1)}$ in terms of that corresponding to $H^{(M)}$:
\begin{align}
  &P^{(M+1)}( \sv) = \nonumber \\
                    & \frac{1}{Z_{M+1}} \sum_{\sv} \exp\big[ -\beta
                      H^{(M+1)}(\sv) \big] \delta\Big( \eta_0 - \frac{1}{N} \sum_{i=1}^N w_i^0
                      s_i \Big) = \nonumber \\
                    &  \frac{1}{Z_{M+1}} \sum_{\sv} \exp\big[ -\beta
                      H^{(M)}(\sv) + \frac{N}{2}( \eta_0)^2 \big] \delta\Big( \eta_0 - \frac{1}{N} \sum_{i=1}^N w_i^0
                      s_i \Big) = \nonumber \\
                    &  \frac{Z_M}{Z_{M+1}} \exp\big[ -\beta \frac{N}{2}( \eta_0)^2
                      \big] \sum_{\sv} \Big[ P^{(M)}(\sv) \delta\Big( \eta_0 - \frac{1}{N} \sum_{i=1}^N w_i^0
                      s_i \Big) \Big] = \nonumber \\
                    & \frac{Z_M}{Z_{M+1}} \exp\big[ -\beta \frac{N}{2}( \eta_0)^2
                      \big] P^{(M)}( \eta_0) .
      \label{eq:PMPM+1}
\end{align}
The expectation $\physExpect{\eta_0}_M = \frac{1}{N} \sum_{i=1}^N
w_i^0 \physExpect{s_i}_M = 0$ by self-averaging due to
the stationarity of $\wv^0$. The variance $\physExpect{(\delta \eta_0)^2}_M$
was shown to be $O \Big (\frac{1}{N}\Big)$ previously; for large $N$, its
typical value is
\begin{align*}
  \physExpect{(\eta_0)^2}_M = & \frac{1}{N^2} \Big[ \sum_{i\neq j}
                                w_i^0 w_j^0 \chi_{ij}^{(M)} +
                                \sum_{i=1}^N\chi_{ii}^{(M)} \Big] \\
  \to & \frac{1}{N^2}\Big[ -\sum_{i\neq j}
        \frac{1}{N-1} \chi_{ij}^{(M)} + \sum_{i=1}^N ( 1 -
        \physExpect{s_i}^2_M ) \Big] \\
  \to & \frac{1}{N}( 1- q) ,
\end{align*}
where
\[
q = \frac{1}{N} \sum_{i=1}^N \physExpect{s_i}_M^2 .
\]
Finally, assuming that $P^{(M)}(\eta_0)$ is Gaussian and using the
relation in Eq.~(\ref{eq:PMPM+1}), we obtain
\begin{align*}
  P^{(M+1)}(\eta_0) \propto & \exp \Big[ -\frac{N}{2(1-q)} \eta_0^2 -
                              \frac{N\beta}{2} \eta_0^2  \Big] \\
  = & \exp \Big[ -\frac{1}{2} \Big[
      \frac{N}{1-q} + N\beta  \Big] \eta_0^2 ,
      \Big] \\
\end{align*}
which is a Gaussian with mean zero and variance
\[
\physExpect{(\delta\eta_0)^2}_{M+1} = \frac{1-q}{N(
  1+\beta(1-q) )} .
\]
Note that this value would result for any $\wv^\mu$ removed. To
obtain the field variance, we use Eq.~(\ref{eq:h0VarVN}) to obtain
\begin{align*}
  V = & \sum_{\mu=1}^M \physExpect{ (\delta\eta_\mu)^2}_N\\
  = & \alpha N \physExpect{(\delta \eta_0)^2}_{M+1} \\
  = & \frac{\alpha(1-q)}{1+\beta(1-q)} .
\end{align*}
This resembles that of the Hopfield model but with a changed sign in
the denominator.

\subsection{Limiting spectral distribution of $\Jv$}
\label{sec:appendix:LSDJ}

We determine the limiting eigenvalue distribution of the $\Jv$ matrix.
To do so, it suffices to determine the spectral distribution of
$\Jtildev$. To see why, first note that $\Jtilde_{ii} \to -\alpha$ for
all $i$ by the law of large numbers. Recalling that
$\Jv = \Jtildev - \textrm{diag}(\Jtildev)$, this implies that $\Jv$ is
asymptotically related to $\Jtildev$ by addition of a uniform quantity
to the diagonal, namely
\[
\Jv = \Jtildev + \alpha \Iv_{N} ,
\]
which in turn means that the eigenvalues of $\Jv$ are simply those of
$\Jtildev$ translated by $\alpha$.

For Wishart matrices of the form $\frac{1}{M} \Xv\Xv^T$, where $\Xv$
are $N\times M$ matrices whose elements are independent
zero-mean unit-variance Gaussian variates and $M = \alpha N$, the
limiting spectral distribution is known as the
Marchenko-Pastur~\cite{marchenko:67} law.  The fact that the columns of
$\Wv$ are correlated Gaussian variates seems at first to complicate
the determination for $\frac{1}{N}\Wv\Wv^T$. The structure of the
specific covariance matrix $\Sigmav$ considerably simplifies matters,
however. Recalling that
\[
\Sigmav = \frac{N}{N-1} \Big[ \Iv - \tv\tv^T \Big] ,
\]
it is apparent that first, $\tv$ is an eigenvector with null
eigenvalue, and second, that any vector in the subspace orthogonal to
$\tv$ is an eigenvector with eigenvalue $N/(N-1)\to 1$. Hence, any
orthonormal set of $N-1$ vectors orthogonal to $\tv$ can be used to
represent $\Sigmav$, and the variation of $\wv$ along each of these
eigenvectors is asymptotically of unit magnitude. The procedure for
generating $\wv$ is for large $N$ thus equivalent to first
generating vector $\xv$ whose first $N-1$ elements are independently
$\sim \Ncal(0,1)$ and whose last element is zero and next,
transforming $\xv$ by some unitary matrix $\vc{U}$ rotating the
$N^{\textrm{th}}$ coordinate vector $\ev_N = (0,\ldots,0,1)$ to
$\tv$. This implies that the spectral distribution of $\Wv\Wv^T$
approaches that of $\Xv\Xv^T$, where
\[
\Xv =
\begin{bmatrix}
  \widetilde{\Xv} \\
  \ZeroVec^T
\end{bmatrix} 
\]
consists of an $(N-1) \times M$ matrix $\widetilde{\Xv}$ composed of
iid normal variates and a final row of zeros; the eigenvalues of
$\Xv\Xv^T$ are thus those of $\widetilde{\Xv}\widetilde{\Xv}^T$ with
an extra zero added. The limiting spectral distribution of
$\frac{1}{N-1} \widetilde{\Xv}\widetilde{\Xv}^T$ can be
straightforwardly obtained from the Marchenko-Pastur law by
appropriate change of variables. We then obtain the limiting
eigenvalue distribution of the matrix
$ -\Jtildev = \frac{1}{N} \Wv\Wv^T$ to be
\begin{equation}
  \ftilde(\lambda) = \left \{
      \begin{array}{ll}
        (1-\alpha) \delta(\lambda) + \ftilde_+( \lambda) & \alpha < 1 \\
        \frac{1}{N}\delta(\lambda) + \frac{N-1}{N}  \ftilde_+(\lambda) & \alpha \geq 1
      \end{array}
      \right. ,
  \label{eq:JtildeLSD}
\end{equation}
where
\[
\ftilde_+( \lambda) = \frac{1}{2\pi} \frac{\sqrt{
    (\lambda_+-\lambda)(\lambda - \lambda_-)}}
{\lambda}
\OneVec[ \lambda \in
[\lambda_-, \lambda_+]
\]
and
\[
\begin{array}{l}
  \lambda_- = \alpha - 2\sqrt{\alpha} + 1 \\
  \lambda_+ = \alpha + 2\sqrt{\alpha} + 1 .
\end{array}
\]
Note that the $\delta$ spike at zero never disappears, a feature that
turns out to crucially influence the phase behavior for large
$\alpha$. Finally, the spectral distribution of
$\Jv = \Jtildev + \alpha \Iv$ follows by reflection and translation:
\begin{equation}
  f(\lambda) = \left \{
      \begin{array}{ll}
        (1-\alpha) \delta(\lambda-\alpha) + f_+( \lambda) & \alpha < 1 \\
        \frac{1}{N}\delta(\lambda-\alpha) + \frac{N-1}{N}  f_+(\lambda) & \alpha \geq 1
      \end{array}
      \right. ,
  \label{eq:JLSD}
\end{equation}
where 
\[
f_+( \lambda) = -\frac{1}{2\pi} \frac{ \sqrt{
    (\lambda_+ - \lambda) (\lambda- \lambda_-)}}
{\lambda-\alpha}
\OneVec[ \lambda \in
[\lambda_-, \lambda_+]
\]
and
\[
\begin{split}
  \lambda_-  &= -2\sqrt{\alpha} - 1\\
  \lambda_+  &= 2\sqrt{\alpha} - 1 .
\end{split}
\]
The distribution consists of a continuous component over the support
$[\lambda_-, \lambda_+]$ and a persistent $\delta$ function at $\alpha$,
where of course $\alpha \geq \lambda_+$.

\subsection{Energy histogram of the Wishart ensemble}
\label{sec:appendix:energyHist}

Define the length $M$ vector of normalized state overlaps with the $\{ \wv^\mu\}$
\[\etav(\sv) = \frac{1}{\sqrt{2N}} \Wv^T\sv
\]
so that
$H(\sv) = \etav^T\etav$
and
\begin{align*}
  p_E(e | \{ \wv^\mu\} ) = & \Pr( \etav^T\etav = e | \{ \wv^\mu\} )\\
  = & \frac{1}{2^N} \sum_{\sv} \delta \Big( e -
      \frac{1}{2N}\sv^T\Wv\Wv^T\sv \Big) .
\end{align*}
The derivation of the marginal energy
\[
p_E(e) \triangleq \int_{\{\wv^\mu\}} p_E(e | \{\wv^\mu\} )
f(\wv^1,\ldots,\wv^M) \dd \{\wv^\mu\} 
\]
is simplified by decomposing the integration into sums of
expectations over subsets
of states with a constant number of positive elements and exploiting the
exchangeability of $f(\wv)$. Let $N_+(\sv)$ be the number of
elements in $\sv$ with value $+1$; for a magnetization
$m \in [-1,1]$, $N_+ = \frac{(1+m)N}{2}$. On the $N_+$ constrained
subset, the marginal density of $E$ is
\begin{align*}
  p_E( e | N_+) = & \frac{1}{\binom{N}{N_+}} \Expect{\{\wv^i\}} \Bigg [
                    \sum_{\sv:N_+} \delta \Big (
                    e - \frac{1}{2N} \sv^T\Wv\Wv^T\sv
                    \Big )
                    \Bigg ] \\
  = & \frac{1}{\binom{N}{N_+}} \sum_{\sv:N_+} \Expect{\{\wv^i\}} \Bigg [
      \delta \Big (
      e - \frac{1}{2N} \sv^T\Wv\Wv^T\sv
      \Big )
      \Bigg ] ,
\end{align*}
where the sums are over states with $N_+$ positive entries. The joint distribution
\[
f(\wv^1,\ldots,\wv^M) = f(\wv^1)\ldots f(\wv^M)
\]
by the independence
of the columns, but the components of each $\wv^\mu$ are correlated.
They are however exchangeable, and since the same $\sv$ appears in
each term in the sum this implies that
\begin{equation}
  \Expect{\{\wv^i\}} \Bigg[ \delta\Big(e - \frac{1}{2N}\sv^T
  \Wv\Wv^T\sv\Big) \Bigg ]
  \label{eq:expectETerm}
\end{equation}
only depends on $N_+$. Consider the specific case of $\sv$ whose
first $N_+$ elements are 1. Define the sums (one for each column of
$\Wv$)
\[
A_+^\mu \triangleq \frac{1}{\sqrt{2N}} \sum_{j=1}^{N_+} w_j^\mu .
\]
Because $\sum_{i=1}^Nw_i^\mu=0$, the sum
\begin{align*}
  A_-^\mu & \triangleq \frac{1}{\sqrt{2N}} \sum_{j=N_++1}^{N} w_j^\mu\\
          & = -A_+^\mu
\end{align*}
deterministically.
By the independence of the $\wv^\mu$ the sums $\{ A_+^\mu\}$ are 
independent random variables and from the properties of linear
transformations of Gaussian variables
each is distributed
according to a zero-mean Gaussian with variance
\[
\sigma_{N_+}^2 = \frac{N_+(N-N_+)}{2N(N-1)} .
\]
When $N_+=0$ or $N_+=N$, the zero-variance Gaussian is defined to
be a $\delta$ function as we expect. We express the expectation
(\ref{eq:expectETerm}) at fixed $N_+$ as
\begin{align}
  & \Expect{\{\wv^i\}} \Bigg [ \delta \Big (
                               e - \frac{1}{2N}\sv^T\Wv\Wv^T\sv
                               \Big ) \Bigg ]
                               \nonumber \\
                             & =  \Expect{\{A_+^\mu,A_-^\mu\}} \Bigg [ \delta \Big ( e -\sum_{\mu=1}^M (A_+^\mu - A_-^\mu)^2
                               \Big ) \Bigg ] \nonumber \\
                             & = \Expect{\{A_+^\mu\}} \Bigg [ \delta \Big ( e -\sum_{\mu=1}^M
    (2A_+^\mu)^2 \Big ) \Bigg ] .
\end{align}
Now let
\[
S \triangleq \sum_{\mu=1}^M(A_+^\mu)^2
\]
with density $f_S$. We then have
\begin{align}
  & \Expect{\{A_+^i\}} \Bigg [ \delta \Big ( e -\sum_{i=1}^M
    (2A_+^i)^2 \Big ) \Bigg ] \nonumber \\
  & = \Expect{S} \Bigg [ \delta \Big ( e -4S \Big ) \Bigg ] \nonumber \\
  & = \frac{1}{4}f_S\Big(\frac{e}{4} \Big) .
\end{align}
Because $S$ is the sum of squares of $M$ zero-mean iid Gaussians with
variance $\sigma_{N_+}^2$, it is gamma-distributed, i.e., with density
\begin{equation*}
  f_S(s) = \left\{
    \begin{array}{ll}
      \frac{1}{\Gamma(k)\theta^k}
      s^{k-1}\exp\Big( -\frac{s}{\theta}\Big) & \textrm{for $s \geq 0$}\\
      0 & \textrm{for $s < 0$}
    \end{array}
    \right.
\end{equation*}
with parameters $(k,\theta) =(M/2,2\sigma_{N_+}^2)$. For large $N$ and
$m$ relating $N_+$ to $N$, $\sigma_{N_+}^2 \approx \frac{1-m^2}{8}$.
We then obtain the constrained-$m$ energy histogram 
\begin{align*}
  p_E(e|m) = & \frac{1}{4}f_S\Big(\frac{e}{4}\Big) \\
  = & \frac{1}{\Gamma(M/2)(1-m^2)^{M/2}} e^{M/2-1}\exp\Big(
      -\frac{e}{1-m^2} \Big) 
\end{align*}
for $e \geq 0$ and zero otherwise and the overall energy distribution
\begin{align}
  p_E(e) = \frac{1}{2^N} \sum_{m} \binom{N}{\frac{(1+m)N}{2}} p_E(e|m) ,
\end{align}
where the sum runs over the $N+1$ values of $m$ mapping to
$N_+ \in \{0,\ldots,N\}$. As described, for example, in Refs.~\cite{mertens:06,mezard:09}, applying
Stirling's large-$N$ approximation to
the binomial coefficients, replacing the sum with an integral and
evaluating it with Laplace's method, we obtain
\[
p_E(e) \approx p_E(e|m=0)
\]
or the marginal gamma energy density for the WPE
\begin{equation}
  p_E(e) = \left \{
    \begin{array}{ll}
      \frac{1}{\Gamma(M/2)}e^{M/2-1}\exp( -e ) & \textrm{for $e \geq
                                                 0$}\\
      0 & \textrm{for $e < 0$} .
    \end{array}
         \right.
\end{equation}

\subsection{Annealed approximation}
\label{sec:appendix:annealedApproximation}

The purpose of the annealed approximation is to provide a simple bound
on the typical case behavior of the free energy at leading order in
$N$.  The free-energy density is defined
\begin{equation}
  f = \lim_{N\rightarrow \infty} -\frac{1}{\beta N} \left\langle \log \sum_{\sv} \exp\left[-\beta H(\sv)\right] \right\rangle , \label{eq:freeEn}
\end{equation}
where $\langle \cdot \rangle$ denotes an average over
instances. Discontinuities in the free energy describe the phase
transitions, and derivatives describe the order parameter(s) and other
statistically significant quantities at thermal equilibrium.

It is convenient for this section to consider a model defined by the
Hamiltonian
\begin{equation}
  H(\sv) = \frac{1}{N} \sum_{\mu=1}^{\alpha N} \left(\sum_{i',i} Z_{i',\mu}\left[\delta_{i,i'} - \frac{\kappa}{N}\right]s_i\right)^2, 
  \label{app:Ham}
\end{equation}
where $Z_{i,\mu}$ are independent and normally distributed random
variables. For the case $\kappa=1$ this Hamiltonian is identical to the
main text Hamiltonian up to the choice of the embedding solution
($t_i=1$, $\forall i$), inclusion of a nonzero diagonal term in the
coupling matrix (which adds a constant offset to the energy), and
corrections of $O(1/N)$. These restrictions are for the convenience of
analysis, and have no significant impact on the analysis method or
conclusions of the section.

The parameter $\kappa$ is useful in making the connection to the
anti-Hopfield model (obtained for the case
$\kappa=0$)~\cite{nokura:98}, and in identifying a variation on the
principle of embedding discussed in the main text. Tuning of this
parameter allows one to control the energy level of the planted
solution at leading order in $N$, allowing an embedding of a ground
state with control over the gap at fixed $\alpha$, or implanting an
excited state well separated from other stable and metastable states.

The annealed approximation may be used to obtain a lower bound, $f \geq f_A$, on the free-energy density
\begin{equation}
  f_A = \lim_{N \rightarrow \infty} -\frac{1}{\beta N} \log \left\langle \sum_{\sv} \exp\left[-\beta H(\sv)\right] \right\rangle . 
  \label{eq:freeEnAnn}
\end{equation}
The physical interpretation for this approximation is that the quenched
degrees of freedom ($Z$) are treated on an equal footing with the
dynamical degrees of freedom ($\sv$). This means that models of lower
energy can be selected disproportionately, since spin and model
variables can become correlated to lower the free energy. By this
process it is possible that atypical models can dominate the free
energy so that typical case is not reflected. However, in this
ensemble we show that the free energy is correct through much of the
phase space in agreement with the TAP analysis of Sec.~\ref{sec:Thermodynamics:TAP}.

Owing to the quadratic form of the Hamiltonian in the order parameter,
it is rather straightforward to take the disorder average explicitly,
which yields:
\begin{equation}
  f_A = \lim_{N \rightarrow \infty} -\frac{1}{\beta N} \log \sum_{\sv} \exp \left\lbrace \alpha N \mathrm{Tr} \log[\Iv + \beta \Xv(\sv)] \right\rbrace ,
  \label{eq:freeEnAnn2}
\end{equation}
where
$X_{ij}(s) =\frac{1}{N}(s_i - \kappa \sum_{i'} s_{i'})(s_j - \kappa
\sum_{j'} s_{j'})$.  Following this, we notice that the eigenvalues of
$\Xv$ are a function only of the sum of spin variables, thus the trace
log can be simplified, defining $m = \frac{1}{N}\sum s_i$ we can write
\begin{multline}
      f_A = \lim_{N \rightarrow \infty} -\frac{1}{\beta N} \log \sum_{m} {N \choose (1+m)N/2} \\
  \exp \left\lbrace \alpha N \log\left[ 1 + \beta (1- m^2\kappa (2 - \kappa))\right]
  \right\rbrace .
  \label{eq:freeEn1}
\end{multline}
Using Stirlings's approximation, and using a continuum approximation
for $m$ we can write
\begin{multline}
  f_A = \lim_{N \rightarrow \infty} -\frac{1}{\beta N} \log \int dm
  \exp \left\lbrace \right. \\ - N \sum_{x=\pm 1} \frac{1 + m x}{2}\log\left(\frac{1+m x}{2}\right)
  \\ \left.+ \alpha N \log[1 + \beta (1- m^2\kappa
    (2 - \kappa))]
    \right\rbrace. \label{eq:freeEn2}
\end{multline}
Finally, applying Laplace's approximation yields the result
\begin{align}
  f \geq \max_m \Big[& \frac{1}{\beta} \sum_{x=\pm 1} \frac{1+m x}{2}\log\left(\frac{1+m x}{2}\right) + \ldots
                     \nonumber \\
                   & \alpha \log[1 + \beta (1- m^2\kappa (2 -
                     \kappa))\Big] .
                     \label{eq:freeEn3}
\end{align}
The quantity $m$ can be readily associated with the planted state
overlap (magnetization, for the case of $t_i=1\;\forall i$). The first
term is the standard mean-field entropy term, the latter term being an
energy term.

Maximizing this equation involves solving for $d f/dm =0$, also called
the saddle-point equation. At high temperature there is only the
solution $m=0$, whereas at low temperature (for large enough $\kappa$)
there are two additional solutions. For the case $\kappa=1$ of the
main text [Eq.~(\ref{eq:freeEn3})] is identical to the dominant minima of
the TAP equation free energy [Eq.~(\ref{eq:FTAP})].

At low temperature it is interesting to observe that the
crystallization transition occurs due to a competition between the
energy term and the entropy term, one dominating in each regime. For
the paramagnetic solution is defined by $m=0$, whereas for the planted
solution $m\approx 1$. Equating these two terms for the case
$\kappa=1$ we can find a simple approximation to the first-order
transition, the approximation in Eq.~(\ref{eq:TcApprox}) proves a very accurate
lower bound for the critical temperature shown in Fig.~\ref{fig:transitionTemps} 
for small values of $\alpha$.

In this section we have found that the annealed approximation recovers
the more general TAP result derived in Sec.~\ref{sec:Thermodynamics:TAP}. 
For the special case of $O(N)$ planting
$\kappa<1$ we find that the embedded state remains thermodynamically
dominant at low-enough temperature for large-enough $\kappa$, at a
diminished critical temperature (and increased ground-state
energy). For the case $\kappa \approx 0$, and at small $\alpha$ the
annealed approximation is insufficient to predict all features of the
phase diagram, for this we can use the replica method of Sec.~\ref{sec:appendix:replicaMethod}.

\subsection{The replica method}
\label{sec:appendix:replicaMethod}

The annealed approximation of Sec.~\ref{sec:appendix:annealedApproximation},
and TAP analysis presented (Sec.~\ref{sec:Thermodynamics:TAP}), are
insufficient to describe all the features of our model phase space. To
go beyond these we here introduce a replica method.  The replica
method when solved is able to capture non trivial properties of the
exact free energy, at leading order in $N$, in a variety of disordered
models closely related to our proposed model~\cite{amit:87,parisi:95a,nokura:98}. 
As presented here, the replica
method is a nonrigorous method for purposes of obtaining insight
through related models.

The replica method for the anti-Hopfield model ($\kappa=0$) is
demonstrated by Nokura \emph{et al.}~\cite{nokura:98}, and may be derived along
similar lines as the Hopfield model~\cite{amit:87}. Only minor
variations to the form of the Hopfield replica method (free energy)
are necessary to analyze the WPE, but these changes
have significant consequences.

The free-energy density [Eq.~(\ref{eq:freeEn})] can be rewritten for
purposes of the replica method in terms of a replicated partition
function
\begin{equation}
  f = \lim_{N\rightarrow \infty} -\frac{1}{\beta N} \lim_{n \rightarrow 0}\frac{\partial}{\partial n} \langle \mathcal{Z}^n \rangle\;,
\end{equation}
where $\mathcal{Z}$ is the partition function, and the trick is to
solve for the case of positive integer $n$ and analytically continue
to real $n$. For integer $n$ we can write
\begin{equation}
  \mathcal{Z}^n = \sum_\sv \prod_{\rho=1}^n \exp(H(\sv^\rho))
\end{equation}
using Eq.~(\ref{app:Ham}), $\sv^\rho$ being a vector of dimension $N$.
From here we can follow closely the method in Section 2 of Ref.~\cite{amit:87}, 
which is considered a standard approach. This being
understood, we are sparse in our derivation. The main difference in the
method is the introduction of an additional order parameter $m$
through the identity
\begin{align}
  & \int d m^\rho \delta(m^\rho - \frac{1}{N}\sum_i s^\rho_i ) 
    \nonumber \\
  & = \int d m^\rho d{\hat m}^\rho \exp\left({\hat m}^\rho [ N m^\rho  - \sum_{i} s^\rho_i]  \right) .
\end{align}
The integral identity, and scaling with $N$, being appropriate for the
large system limit saddle-point, to be later identified. A similar
identity is introduced for the overlap
$q^{\rho,\rho'} = \frac{1}{N}\sum_i s_i^\rho s_i^{\rho'}$.

The replicated partition function following these manipulations be
written
\begin{align}
  \langle \mathcal{Z}^n & \rangle = \exp \Big\lbrace \frac{\alpha N}{2}
                                  \mathrm{Tr} \log\big( [(1-\beta)\Iv
                                  + \beta \mathbf{q} - \ldots \nonumber \\
                                & \beta
                                  \kappa(2-\kappa)\mv \mv^T]\big) - \frac{N}{2}\sum_{\rho \neq
                                  \rho'} q_{\rho,\rho'}{\hat
                                  q}_{\rho,\rho'} - \ldots \nonumber \\
                                & N \sum_{\rho}
                                  m_\rho {\hat m}_\rho + \log \mathrm{Tr}_s \exp\left(\frac{1}{2}
                                  \sv^T \mathbf{{\hat q}} \sv + {\hat \mv}^T \sv \right) \Big\rbrace ,
\end{align}
where $\Iv$ is the $n \times n$ identity matrix, $\mathbf{q}$ and
$\mathbf{{\hat q}}$ are matrices of the same dimension with zero on
diagonal (overlap order parameters), and $\mv$, ${\hat \mv}$ and $\sv$ are
$n \times 1$ vectors (alignment with planted solution order
parameters).

For the special case $m^\rho=0$ (or $\kappa=0$) this free energy is
identical to that of the anti-Hopfield model~\cite{nokura:98}. The
interpretation of this result is as follows: since $m$ describes the
degree of alignment with the planted solution, we can argue that when
there is no extensive alignment with the planted solution, such as at
high temperature (above the first-order planting transition) or in
general for the space orthogonal to the planted solution, the model
phenomena will be identical to the anti-Hopfield model.  This is
interesting, since in the anti-Hopfield model, and in related classes
of problems, there is understood to be a replica symmetry breaking
phenomena~\cite{parisi:95a}. The space becomes divided into modes
separated by extensive barriers at sufficiently low temperature. We
thus expect this same phenomena to carry over to our model either as a
stable or metastable solution. To determine which we must analyze the
free energy within an approximation. In this Appendix we will develop
only the replica symmetric theory, this being understood as sufficient
for the paramagnetic phase, and the crystal phase discussed in the
main text.

In the {\em replica symmetric} solution we take
$m^\rho=m, \forall \rho$ and that $q^{\rho,\rho'}=q$ for
$\rho\neq\rho'$ (the diagonal term is 0). The free energy at leading
order can be determined from the saddle-point free energy
\begin{align}
  \beta & f(q,{\hat q},m,{\hat m}) = -\frac{\alpha}{2}\Big\lbrace
          \log\left(1+\beta(1-q)\right) +\ldots \nonumber \\
        & + \frac{\beta (q - (2 \kappa - \kappa^2) m^2) }{1 + \beta (1
          - q)} \Big\rbrace + {\hat m} m - \frac{1}{2} {\hat q} q +
          \frac{1}{2} {\hat q} - \ldots \nonumber \\
        & \int
          dz \exp\left(-\frac{z^2}{2}\right) \log \left[2
          \cosh\left(\sqrt{{\hat q}} z + {\hat m}\right)\right] .
\end{align}
We can attempt to solve this equation as in the annealed case by
setting derivatives with respect to $m$, ${\hat m}$, $q$ and
${\hat q}$ to zero. These equations can be written
\begin{align}
  {\hat m} = & \alpha\beta \frac{(2 \kappa - \kappa^2) m}{1+\beta(1-q)}\label{eq1}\\
  m = & \int d z \tanh(\sqrt{{\hat q}} z + {\hat m}) \label{eq2}\\
  {\hat q} = & \alpha\beta \frac{q - (2 \kappa - \kappa^2) m^2}{[1+\beta(1-q)]^2} \label{eq3}\\
  q = & \int d z \tanh^2(\sqrt{{\hat q}} z + {\hat m}) .
  \label{eq4}
\end{align}
First we should note the solution $m ={\hat m} = q = {\hat q} = 0$,
which is the same (paramagnetic) free energy as in the annealed and
TAP analysis, and describes the high-temperature solution. The local
stability of this solution at $\kappa=1$, which determines the
critical temperature, is identical to that determined by alternative
analysis methods (Secs.~\ref{sec:appendix:annealedApproximation}
and \ref{sec:Thermodynamics:TAP}).

Next consider the subspace with ${\hat q}=0$ and $m\neq 0$. In this
case we find the relationship $q=(2\kappa - \kappa^2)m^2$, and $q=0$,
and recover the planted phase with identical properties to the
annealed and TAP analysis. The energy of this solution being
controlled by $\kappa$. This solution, where it exists is again
locally stable.

Finally, for $\alpha>1$ a solution with nonzero ${\hat q}$ is
possible in the subspace with $m=0$, the critical temperature 
given by $T = -1 + \sqrt{\alpha}$. Nevertheless, for $\kappa=1$ the
embedding-aligned phase ($|m|>0$) is dominant. Thus the replica
symmetric solution is in agreement with the understanding presented in
the main text. For $\rho<1$ this solution can become relevant for a
limited region $\alpha \gtrsim 1$, and describes a transition similar
to that in the SK model.

However, it is known that the replica symmetric solution can be
unstable, and it is necessary to go beyond these approximations to
first and higher orders of replica symmetry breaking to describe the
phase space correctly, this being most important at small $\alpha$. We
have not probed the interesting consequences in this part of the phase
diagram, but the results of the anti-Hopfield model are understood to
apply and there is expected to be dynamical and static transitions in
the space orthogonal to the planted
solution~\cite{nokura:98,parisi:95}. It is the existence of these
solutions, whether they are stable or metastable, alongside the
planted one, that offers the interesting possibility to undertake a
{\em population transfer} measurement in the context of this
model~\cite{smelyanskiy:18}. Being able to plant a deep stable or
metastable solution may also have other interesting applications,
particularly in sampling and inference applications where one must
provide more than a single ground-state certificate.

The aim of this paper is to provide practical intermediate scale
benchmarks, the replica method describes only the typical case
properties at leading order in $N$. A tension therefore exists between
the results found in this section, and practical limitations of the
ensemble such as precision, and instance to instance fluctuations at
finite size. These facts must always be borne in mind when
considering such analysis results.

\subsection{Time to solution measurements}
\label{sec:appendix:TTSMeasurements}

In this work the problem hardness is quantified via time to solution (TTS) of parallel
tempering Monte Carlo. We do emphasize that we expect similar results using other 
heuristics. 
For a single
disorder realization, the time to solution is defined as the run time
such that there is a $99\%$ success probability to have found the
solution at the end of the run.

Many solvers have parameters that affect the success probability which
we denote by a set $\{\phi\}$.  For parallel tempering Monte Carlo, the parameters
considered are the lowest temperature and number of replicas.  Often,
it is faster to carry out many short attempts taking run time $R$ with
a lower success probability $p(R, \{\phi\})$ which motivates the
definition \cite{boixo:14,ronnow:14a,albash:18} 
\begin{equation}
  \mathrm{TTS}(R, \{\phi\}) \triangleq R ~\frac{\log \left(1 - 0.99\right)}{\log \left(1 - p(R, \{\phi\})\right)} .
  \label{eq:tts_def}
\end{equation}
The number of attempts is taken to be a real number for convenience.

In the case of an ensemble, averages may be poorly defined and so the
median is used~\cite{boixo:14}.  For the TTS to be well defined, the minimum
value with respect to run time and parameters must be computed~\cite{ronnow:14a}.  The
ensemble TTS is defined as
\[\mathrm{TTS} \triangleq \min\limits_{R, \{\phi\}} \median\limits_{i} \left\{ \mathrm{TTS}_i (R, \{\phi\})\right\} , \]
where the index $i$ refers to the TTS of individual disorder samples
as a function of run time and parameters.

Due to the nonsequential nature of parallel tempering Monte Carlo, the success
probability as a function of run time can be efficiently measured with
only a small number of runs~\cite{mandra:18}: The algorithm is run $W$ times
until the solution is found.  For a given run time $R$, the success
probability is estimated as the percentage of runs where the solution
is found in time less than $R$.

Due to finite numerical precision, solutions are considered to be any
state with energy $E < E_{\rm GS} + \epsilon$ where $E_{\rm GS}$ is the
planted solution energy and $\epsilon$ is a numerical constant.
$\epsilon = 10^{-7}$ was used for the TTS study in Fig.~\ref{fig:TTS_N_32}.  
The effects of different values of $\epsilon$
on the location of the TTS maximum are shown in Fig.~\ref{fig:TTS_N_32_epsilons}.

\bibliography{refs,comments}

\begin{thebibliography}{82}
\bibitem[{com({\natexlab{a}})}]{comment:fullyFrustrated}
\bibinfo{note}{{For all variable triples $\{i,j,k\}$, it is impossible to
  simultaneously satisfy the edge preferences $\{J_{ij}, J_{jk}, J_{ik} \}$.}}
\bibitem[{com({\natexlab{b}})}]{comment:ftildeMinima}
\bibinfo{note}{{In this and some subsequent plots of $\ftilde$, the minimizers
  occurring at $m\neq 0$ may visually appear to occur exactly at the end points
  ($m=\pm 1$) but they in fact lie inside the interval, though very close to
  the boundaries. Indeed, for any finite $\beta$ the end points are technically
  local maxima on $[-1,1]$.}}
\bibitem[{com({\natexlab{c}})}]{comment:groundStatePrecision}
\bibinfo{note}{{Indeed, due to rounding errors, it can happen that the planted
  solution's energy exceeds that of some other spurious state by a tiny
  amount!}}
\bibitem[{com({\natexlab{d}})}]{comment:spinFlip}
\bibinfo{note}{{When including states related by global flip symmetry, the full
  expected count is of course twice the stated value.}}
\bibitem[{com({\natexlab{e}})}]{comment:landscapeCaveat}
\bibinfo{note}{{In practice however, i.e., when different optimization
  algorithms are concerned, this argument might not be strictly applicable as
  different optimization routines use different methodologies to access the
  possible state space of the systems.}}
\end{thebibliography}

\end{document}